\documentclass[11pt,a4paper]{article}
\usepackage{jheppub,bm,booktabs,multirow}

\usepackage{amsmath}
\allowdisplaybreaks[4]
\usepackage{amssymb}
\usepackage{graphicx}
\usepackage{booktabs}
\usepackage{bm}
\usepackage{psfrag}
\usepackage{cancel}
\usepackage[normalem]{ulem}
\usepackage{subcaption}
\usepackage{overpic}
\usepackage[utf8x]{inputenc}

\usepackage{scalefnt,pstricks}
\usepackage{cancel}
\usepackage[normalem]{ulem}
\usepackage{braket}
\usepackage{xcolor}

\makeatletter
\def\@fpheader{~}
\makeatother

\usepackage[Q=yes,pverb-linebreak=no]{examplep}

\usepackage{mathtools}

\newcommand{\spac}{{\hspace{0.3mm}}}

\def\as{\alpha_s}
\def\e{\epsilon}
\def\nno{\nonumber}

\def\bT{\boldsymbol{T}}

\definecolor{darkgreen}{RGB}{0,100,0}

\definecolor{CBBlue}{RGB}{0,114,178}
\definecolor{CBOrange}{RGB}{230,159,0}
\definecolor{CBGreen}{RGB}{0,158,115}
\definecolor{CBPink}{RGB}{204,121,167}
\definecolor{CBYellow}{RGB}{240,228,66}

\newcommand{\Dqqq}[1]{ \int \hspace{-0.05cm} \left[d\Omega_{#1} \right]}

\newcommand{\Hfun}[1]{\bm{\mathcal{ H}}_{{#1}}}

\newcommand{\plist}{\{\underline{p}\}}

\newcommand{\nlist}{\{\underline{n}\}}
\newcommand{\xlist}{\{\underline{x}\}}
\newcommand{\F}{\mathcal{F}}

\newcommand{\bH}{\boldsymbol{\mathcal{H}}}

\newcommand{\bGam}{\boldsymbol{\Gamma}}
\newcommand{\bU}{\boldsymbol{U}}
\newcommand{\bP}{\mathbf{P}}

\newcommand{\thetain}{\ensuremath{\Theta_{\mathrm{in}}}}

\newcommand{\thetaout}{\ensuremath{\Theta_{\mathrm{out}}}}

\newcommand{\bS}{\boldsymbol{\mathcal{S}}}
\newcommand{\bone}{\ensuremath{\boldsymbol{1}}}

\newcommand{\Rm}[1]{\ensuremath{\boldsymbol{R}_{#1}}}
\newcommand{\Vm}[1]{\ensuremath{\boldsymbol{V}_{#1}}}
\newcommand{\vm}[1]{\ensuremath{\boldsymbol{v}_{#1}}}
\renewcommand{\rm}[1]{\ensuremath{\boldsymbol{r}_{#1}}}
\newcommand{\dm}[1]{\ensuremath{\boldsymbol{d}_{#1}}}
\newcommand{\RLSm}[1]{\ensuremath{\overline{\boldsymbol{R}}_{#1}}}
\newcommand{\VLSm}[1]{\ensuremath{\overline{\boldsymbol{V}}_{#1}}}
\newcommand{\vLSm}[1]{\ensuremath{\overline{\boldsymbol{v}}_{#1}}}
\newcommand{\rLSm}[1]{\ensuremath{\overline{\boldsymbol{r}}_{#1}}}

\newcommand{\Rclm}[1]{\ensuremath{{\boldsymbol{R}}_{#1}^{cl}}}
\newcommand{\Vclm}[1]{\ensuremath{{\boldsymbol{V}}_{#1}^{cl}}}
\newcommand{\vclm}[1]{\ensuremath{{\boldsymbol{v}}_{#1}^{cl}}}
\newcommand{\rclm}[1]{\ensuremath{{\boldsymbol{r}}_{#1}^{cl}}}
\newcommand{\dclm}[1]{\ensuremath{{\boldsymbol{d}}_{#1}^{cl}}}
\newcommand{\dclAm}[1]{\ensuremath{{\boldsymbol{d}}_{#1}^{cl,A}}}
\newcommand{\dclFm}[1]{\ensuremath{{\boldsymbol{d}}_{#1}^{cl,F}}}
\newcommand{\dclSm}[1]{\ensuremath{{\boldsymbol{d}}_{#1}^{cl,S}}}
\newcommand{\RLSclm}[1]{\ensuremath{\overline{\boldsymbol{R}}_{#1}^{cl}}}
\newcommand{\VLSclm}[1]{\ensuremath{\overline{\boldsymbol{V}}_{#1}^{cl}}}
\newcommand{\vLSclm}[1]{\ensuremath{\overline{\boldsymbol{v}}_{#1}^{cl}}}
\newcommand{\rLSclm}[1]{\ensuremath{\overline{\boldsymbol{r}}_{#1}^{cl}}}
\newcommand{\dLSclm}[1]{\ensuremath{\overline{\boldsymbol{d}}_{#1}^{cl}}}

\newcommand{\Eout}{\ensuremath{E_\mathrm{out}}}
\newcommand{\Etot}{\ensuremath{E_\mathrm{tot}}}
\newcommand{\ptot}{\ensuremath{\vec{p}_\mathrm{tot}}}
\newcommand{\ceps}{\tilde{c}^\epsilon}

\newcommand{\XS}{\ifmmode\overline{\mathrm{XS}}\else$\overline{\mathrm{XS}}$ \fi}
\newcommand{\MS}{\ifmmode\overline{\mathrm{MS}}\else$\overline{\mathrm{MS}}$ \fi}
\newcommand{\LS}{\ifmmode \overline{\mathrm{LS}}\else$\overline{\mathrm{LS}}$ \fi}
\newcommand{\BS}{\ifmmode\overline{\mathrm{BS}}\else$\overline{\mathrm{BS}}$ \fi}

\newcommand{\gamcuspm}[1]{\ensuremath{\gamma_{#1}^\mathrm{cusp}}}

\newcommand{\rd}{\mathrm{d}}

\newcommand{\tJ}{\tilde{J}}

\newcommand{\barxi}{\mathchoice
  {\overline{\mkern-1mu\xi}}
  {\overline{\mkern-1mu\xi}}
  {\overline{\mkern-1.5mu\xi}}
  {\overline{\mkern-1.5mu\xi}}
  \mathstrut
}

\title{Two-Loop Anomalous Dimension for Non-Global and Clustering Logarithms}
\author[a]{Thomas Becher,}
\author[a]{J\"urg Haag,}
\author[b]{Nicolas Schalch}
\affiliation[a]{Albert Einstein Center for Fundamental Physics, Institut f\"ur Theoretische Physik, Universit\"at Bern,
  Sidlerstrasse 5, CH-3012 Bern, Switzerland}
\newcommand{\OXaff}{Rudolf Peierls Centre for Theoretical Physics, Clarendon Laboratory, Parks Road,
  University of Oxford, Oxford OX1 3PU, UK}
\affiliation[b]{\OXaff}

\emailAdd{becher@itp.unibe.ch}
\emailAdd{juerg.haag@unibe.ch}
\emailAdd{nicolas.schalch@physics.ox.ac.uk}

\date{\today}

\abstract
{We derive the two-loop anomalous dimension relevant for the resummation of subleading non-global and, for the first time, also clustering logarithms within the framework of effective field theory. Special emphasis is put on the choice of the renormalization scheme. We show that the modified minimal subtraction scheme is problematic since it requires extra dimensions to set up the parton shower to perform the resummation. We discuss a set of modified dipole subtraction schemes which are free from this complication and can be used to obtain a frame-independent result for the renormalization-group evolution. We implement the leading-color limit of our result in the {\sc Marzili} framework and numerically verify the scheme independence for gap-between-jets cross sections.  
}

\begin{document}

\maketitle

\newpage

\section{Introduction}

Jet cross sections are obtained by grouping soft and collinear emissions together with the energetic partons produced in hard collisions. They are inclusive enough to be computable in perturbation theory, but exclusive enough to be a good proxy for the underlying hard scattering processes. For this reason, jet observables are the most commonly used hadronic observables at hadron colliders. 

The clustering of soft and collinear radiation into jets is either based on angular constraints (cone jets \cite{Sterman:1977wj}) or through a recombination procedure (sequential-recombination jets \cite{JADE:1986kta}). In both cases, it induces a complicated pattern of logarithmically enhanced higher-order corrections. For cone jets, this was discovered 25 years ago by Dasgupta and Salam who found that secondary emissions from partons inside a jet cone lead to a complicated pattern of logarithms \cite{Dasgupta:2001sh}. These non-global logarithms (NGLs) do not exponentiate into the simple Sudakov form familiar from global event shapes and were not captured by the methods used for analytic resummation \cite{Catani:1992ua}. The dependence of NGLs on the directions and colors of individual partons inside jets can be viewed as a breakdown of color coherence for the soft radiation in the presence of hard phase space constraints. For hadronic collisions, a second, more severe breakdown of coherence arises through complex low-energy phases in the scattering amplitudes. These Glauber phases impede the cancellation of soft-collinear terms in the cross section, which manifests itself through double-logarithmic contributions. Compared to the single-logarithmic NGLs, these are super-leading logarithms (SLLs) \cite{Forshaw:2006fk,Forshaw:2008cq}. Their perturbative structure is also different: NGLs generally start at two loops, whereas the first SLLs occur only at four loops, relative to the Born cross section.

For sequential-recombination jet cross sections, there is a third category of nontrivial logarithmic enhancements, the clustering logarithms (CLs). While NGLs and SLLs are characteristic of non-abelian dynamics, clustering logarithms have their origin in the non-trivial phase space constraints imposed by the clustering sequence. Since the clustering changes order by order, the associated soft logarithms do not exponentiate, even in the abelian case \cite{Kelley:2012kj}. CLs were first studied shortly after the discovery of NGLs \cite{Appleby:2002ke}, since then they have been analyzed in several different contexts \cite{Appleby:2002ke,Banfi:2005gj,Kelley:2012kj,Banfi:2010pa,Khelifa-Kerfa:2011quw,Khelifa-Kerfa:2015mma,Ziani:2021dxr,Khelifa-Kerfa:2024dut,Khelifa-Kerfa:2024roc,Khelifa-Kerfa:2024hwx,Khelifa-Kerfa:2025cdn,Khelifa-Kerfa:2026mbq,Khelifa-Kerfa:2025jzl,Khelifa-Kerfa:2025jev}. 

A standard tool for the analysis of multi-scale problems in quantum field theory are effective field theories, in particular soft-collinear effective theory (SCET)~\cite{Bauer:2000ew,Bauer:2000yr,Bauer:2001ct,Bauer:2001yt,Beneke:2002ph,Beneke:2002ni}  for processes involving energetic particles. Using effective theory methods, factorization theorems for jet processes can be derived. The simplest example is a theorem for exclusive fixed-cone cross sections which, for an $M$-jet cross section at a lepton collider, takes the form \cite{Becher:2015hka,Becher:2016mmh}
\begin{align}\label{eq:factSymb}
\sigma(Q, Q_0) &=  \sum_{m=M}^\infty \big\langle \bH_m(Q,\mu) \otimes \bS_m(Q_0,\mu)  \big\rangle\, ,
\end{align}
where $Q$ is the center-of-mass energy and $Q_0$ the veto on radiation outside the jets. In this framework the resummation of NGLs is obtained by solving the renormalization group (RG) equation for the renormalized hard functions $\bH_m(Q,\mu)$ and evolving them from $\mu \approx Q$ down to the low-energy scale $Q_0$. The soft functions $\bS_m(Q_0,\mu)$ are squared matrix elements of soft Wilson lines along the directions of the $m$ hard partons inside the jet described by $\bH_m(Q,\mu)$. The hard and soft functions must be computed for fixed directions of the hard partons, and the symbol $\otimes$ indicates the integral over the $m$ directions of the hard partons. The symbol $\langle \dots \rangle$ denotes the color trace, which can be taken after combining the hard and soft functions. 

The complexity of the NGLs is reflected by the fact that hard and soft functions of different multiplicity $m$ mix under renormalization. The anomalous dimension relevant for the evolution of the hard function increases the multiplicity during the evolution, i.e. it adds legs to the hard function, in analogy to parton shower evolution. Indeed, the leading logarithmic (LL) solution of the RG precisely maps to the parton shower introduced in \cite{Dasgupta:2001sh}, and the BMS equation \cite{Banfi:2002hw}, as explicitly shown in \cite{Becher:2015hka,Becher:2016mmh,Li:2026yrp}.
The structure and phenomenological impact of NGLs have been explored for jet masses and shapes, filtered jets, isolation cones, processes with massive particles, and at high fixed orders \cite{Rubin:2010fc,Hornig:2011iu,Kelley:2011aa,Dasgupta:2012hg,Kerfa:2012yae,vonManteuffel:2013vja,Schwartz:2014wha,Larkoski:2015zka,Becher:2016omr,Larkoski:2016zzc,Neill:2016stq,Becher:2017nof,Balsiger:2018ezi,Balsiger:2019tne,Balsiger:2020ogy,Larkoski:2025afg}. 
Recently, machine-learning methods have also been explored as a way to model NGLs~\cite{Li:2026yrp}. For many years, only LL resummations were available.
Going beyond LL requires incorporating higher-order corrections to soft radiation \cite{Catani:1998nv,Catani:1999ss,Catani:2000pi}, in particular correlated double-soft emission, as encoded in the next-to-leading soft-evolution kernel derived in~\cite{Caron-Huot:2015bja}. Practical all-order resummations of the next-to-leading logarithmic (NLL) tower have since been developed in three complementary numerical frameworks: a generating-functional approach implemented in {\sc Gnole}~\cite{Banfi:2021owj,Banfi:2021xzn}, effective-theory RG evolution implemented in {\sc Marzili}~\cite{Becher:2021urs,Becher:2023vrh}, and logarithmically accurate parton showers in {\sc PanScales}~\cite{Dasgupta:2020fwr,FerrarioRavasio:2023kyg}.

Within the RG formulation, NLL accuracy requires the one-loop matching corrections to the hard and soft functions \cite{Balsiger:2019tne}, as well as the two-loop anomalous dimension \cite{Becher:2021urs}. These ingredients were implemented into the NLL parton shower framework {\sc Marzili} \cite{Becher:2023vrh}. The factorization \eqref{eq:factSymb} can be generalized to hadron colliders \cite{Becher:2021zkk,Becher:2023mtx,Becher:2024nqc} and was used to perform resummations of the leading SLLs generated by the phase terms in the anomalous dimension \cite{Becher:2021zkk,Becher:2023mtx,Boer:2024hzh,Boer:2023ljq,Becher:2025igg}. In the hadron collider case, the low-energy matrix elements also involve collinear fields, as well as Glauber interactions between soft and collinear partons; remarkably the Glauber exchanges restore the PDF factorization below the scale $Q_0$ \cite{Becher:2024kmk,Becher:2025igg}. We note that coherence violating effects also affect global variables \cite{Banfi:2010xy,Gaunt:2014ska,Forshaw:2021fxs,Banfi:2025mra,Becher:2026kbr}.

An important part of the present work is to extend the resummation of subleading logarithms from fixed-cone jets to sequential-recombination jets. For these, the area of the jet emerges from the clustering sequence, which depends on the energy fractions carried by the individual hard partons. To account for this, the hard functions must be computed differentially in the energy fractions, 
and the angular integral in \eqref{eq:factSymb} is promoted to an integral over the angles and energy fractions of the hard partons \cite{Becher:2023znt}. However, at LL the clustering sequence simplifies considerably because the evolution generates strongly-ordered partons. The resulting simplification is most drastic for the case of the anti-$k_T$ algorithm \cite{Cacciari:2008gp} which clusters the most energetic partons first. This implies that every soft emission is immediately clustered back into the hard jets. At LL, anti-$k_T$ jets are therefore equivalent to cone jets. This does not mean that anti-$k_T$ clustering avoids clustering logarithms at LL --- even flagship LHC vector-boson-fusion type observables \cite{CMS:2017dmo,ATLAS:2020nzk} as well as dedicated dijet measurements with a veto on additional central jet activity \cite{ATLAS:2011yyh} define veto regions not as the complement of some tagging jets, but instead as a rapidity slice between two tagging jets, introducing CLs even at LL. Still, the clustering sequence in the anti-$k_T$ algorithm  trivializes at LL, which simplifies the calculation of CLs. In contrast, the $k_T$ algorithm \cite{Catani:1991hj,Catani:1993hr} first clusters the softest emission. Additional soft emissions thus grow the area of the jets. Nevertheless, the LL clustering is only sensitive to the energy ordering, not the particle energies themselves, and the clustering constraint can be solved recursively, one particle at a time \cite{Becher:2023znt}. This situation changes at NLL. The two-loop anomalous dimension needed for NLL accuracy describes two un-ordered soft emissions, i.e. emissions with comparable energies, and the clustering sequence can change according to the relative energy fraction of the two emitted partons. The two-loop anomalous dimension, differential in the relative energy fraction, will be derived in the second part of this paper and will be one of its main results.  An interesting difference between the fixed-cone and the clustering two-loop anomalous dimension is the presence of quadrupole color-correlations, which contribute already in the large-$N_c$ limit. 

However, before turning to clustering effects, we will revisit the two-loop anomalous dimension for the fixed-cone case and analyze its scheme dependence in detail. Its derivation was originally performed in a scheme where the angular integrals are kept $d$ dimensional, and only the energy integrals are expanded around $d=4$ \cite{Becher:2021urs}. This does not correspond to the standard minimal subtraction ($\MS$) scheme and leads to an anomalous dimension which is frame dependent. Indeed, since the full angular integral is subtracted, the associated soft function vanishes for the canonical scale choice and we therefore call this the maximal subtraction ($\XS$) scheme. The paper \cite{Becher:2021urs} presented the scheme change to $\MS$, but only in implicit form, as a difference between angular integrations in $d$ and $4$ dimensions. While the \MS scheme is standard for fixed-order computations and for the resummation of global event shapes, its use in the context of the factorization \eqref{eq:factSymb} for non-global observables involves some delicate issues:
\begin{enumerate}
\item As terms of all multiplicities contribute, even at LL accuracy, one needs to consider angular integrations over arbitrary many particle directions. In $d$ dimensions, we can introduce arbitrarily many angles in the $d-4$ unphysical dimensions. Do these evanescent angles need to be kept in the renormalized results?
\item How are the non-global cone (or clustering) constraints generalized to $d$-dimensions? Are the full $d$-dimensional momenta constrained or are the constraints formulated strictly in the physical subspace?
\end{enumerate}
Analyzing these questions in detail, we find that the standard \MS scheme is ill-suited for resummations based on \eqref{eq:factSymb}. To see this, we analyze the one-loop soft function and demonstrate that its \MS-renormalized version depends on an evanescent angle. The same angle is then necessarily also present in the two-loop anomalous dimensions and the one-loop hard functions. In the context of a Monte Carlo parton shower solution of the RG equation associated with \eqref{eq:factSymb}, this would induce significant complications. At NLL, one would need to set up the shower with 5-dimensional vectors in order to be able track the additional angle and then deal with distribution-valued angular functions. This evanescent-angle dependence was not accounted for in the original {\sc Marzili} implementation \cite{Becher:2023vrh}, which lead to deviations in higher-order terms that were observed in the numerical comparison to results obtained within the {\sc Gnole} \cite{Banfi:2021owj,Banfi:2021xzn} and {\sc PanScales} \cite{FerrarioRavasio:2023kyg} frameworks.\footnote{We thank Pier Francesco Monni for engaging in a detailed numerical comparison \cite{GnoleComparison}.} In the present paper, we introduce a set of renormalization schemes, based on subtractions with a modified dipole emitter, which are free of this complication. With a suitable choice of the dipole subtraction one can obtain an anomalous dimension which does not depend on evanescent angles and is frame independent. This Lorentz subtraction ($\LS$) scheme is optimally suited for a parton shower evolution, and was first introduced by Caron-Huot \cite{Caron-Huot:2015bja}. Our results for the two-loop anomalous dimensions are valid for arbitrary $N_c$ and we present them both in the color-space formalism and in the color-flow basis. We implement the leading-color two-loop anomalous dimension both in the \XS and \LS scheme into {\sc Marzili} and numerically verify that the physical results are scheme independent. This provides a stringent validation of the anomalous dimension and its implementation.

Partial results for the two-loop structures in the color-flow basis were given earlier in \cite{Platzer:2020lbr} and we confirm their result. 
The color-flow basis is useful since it provides a simple way to track factors of $N_c$ and has been used as the basis of the {\sc CVolver} framework designed to resum NGLs at finite $N_c$ \cite{Platzer:2013fha,Martinez:2018ffw,DeAngelis:2020rvq,Platzer:2022jny,Forshaw:2025fif,Forshaw:2025bmo}. 
Results at finite $N_c$ first became available in the pioneering works of \cite{Hatta:2013iba,Hagiwara:2015bia,Hatta:2020wre}, based on \cite{Weigert:2003mm}. The {\sc PanScales} NODS prescription \cite{Hamilton:2020rcu} provides a complementary treatment: it reproduces the full-color matrix element for sequences of energy-ordered, commensurate-angle emission pairs when these pairs are well separated in rapidity, and is in numerical agreement with \cite{Hatta:2013iba}.

Our paper is organized as follows. We start with a review of the factorization theorem relevant for sequential-recombination jets in Section \ref{sec:renschemedep}, and explain the renormalization of the ingredients and their scheme dependence. In Section \ref{sec:soft_function}, we then discuss angular integration in $d$ dimensions using the explicit example of the one-loop soft function. We introduce a set of modified dipole subtraction schemes and explain why these are superior to \MS renormalization in the context of a parton shower. After reviewing the two-loop anomalous dimension in the \XS scheme in Section \ref{sec:XSanomdim}, we discuss the general form of the scheme change and then provide the result in the \LS scheme. In Section \ref{sec:largeNc} we take the large-$N_c$ limit of the anomalous dimension, as relevant for the implementation in the {\sc Marzili} code, and present numerical results. The clustering anomalous dimension is derived in Section \ref{sec:clustering}. In a series of appendices, we collect the symmetry and color-conservation relations associated with different color structures, provide their color-flow basis representation, and summarize the Lorentz transformation properties of kinematic building blocks entering the anomalous dimensions. 

\pagebreak

\section{Renormalization and scheme dependence}
\label{sec:renschemedep}
Let us now discuss the factorization theorem \eqref{eq:factSymb} in more detail. It reads 
\cite{Becher:2015hka,Becher:2016mmh,Becher:2023znt}
\begin{align}\label{eq:fact}
\sigma(Q, Q_0) &=  \sum_{m=M}^\infty \big\langle \bH_m(\nlist,\{\underline{x}\},Q) \otimes_x \bS_m(\nlist,\{\underline{x}\},Q_0) \big\rangle\, ,
\end{align}
where we now explicitly indicate the directions $\nlist = \{ n_1,\dots, n_m\}$ of the individual hard partons, as well as their energy fractions $\xlist = \{ x_1,\dots, x_m\}$ as arguments of the hard and soft functions. 

The soft functions $\bS_m$ describe the soft emissions from $m$ hard partons inside the jets, along the directions $\nlist = \{ n_1,\dots, n_m\}$. The squared amplitudes for the production of the hard partons constitute $\bH_m$. The bare soft functions are given by the matrix elements
\begin{equation} \label{eq:Smdef}
    \bS_m(\nlist, \{\underline{x}\}, Q_0)=\int\limits_{X}\hspace{-0.55cm}\sum\bigl\langle 0\bigl|\bS_1^{\dagger}\left(n_1\right) \ldots \bS_m^{\dagger}\left(n_m\right)\bigr| X\bigr\rangle\bigl\langle X\bigl|\bS_1\left(n_1\right) \ldots \bS_m\left(n_m\right)\bigr| 0\bigr\rangle \theta\left(Q_0-2\Eout\right),
\end{equation}
where we impose the veto on the total energy outside the jets, but the definition can be generalized to other veto constraints. Below, we will also present results for the transverse energy outside dijets. Importantly, the Wilson lines in this function are only sensitive to the directions of the hard partons, and for simple geometric constraints, such as a cone around the thrust axis of the event, the soft function only depends on the directions $\nlist$ of the hard partons, and the cone opening angle. However, for the sequential-recombination jet cross sections with a gap analyzed in \cite{Becher:2023znt}, clustering effects change the jet region, and one has to take into account the energy fractions of the individual hard partons to obtain the veto region.  In this situation, the soft function has a complicated implicit dependence on the energy fractions through the precise definition of $\Eout$. In the factorization formula \eqref{eq:fact}, the symbol  $\otimes_x$ denotes the integral over the directions and energy fractions of the hard partons \cite{Becher:2023znt},
\begin{equation}\label{eq:otimesz}
  \bH_m \otimes_x \bS_m = \prod_{i=1}^m \Dqqq{i} \int_0^1\! \rd x_i\; \bH_m(\nlist,\xlist,Q)\, \bS_m(\nlist,\xlist,Q_0)\,.
\end{equation}
The angular integral in $d=4-2\e$ space-time dimensions is defined as
\begin{equation}\label{eq:dOmega}
     \Dqqq{q}\equiv\tilde{c}^\epsilon \int \!\frac{d^{d-2} \Omega_q}{2(2 \pi)^{d-3}}\left(v\cdot n_{q} \right)^{2\epsilon}\,,
\end{equation}
with $\tilde{c}=e^{\gamma_E}/\pi$ \cite{Becher:2021urs}. The velocity vector is $v^\mu = (1, \vec{0})$ in the lab frame, i.e.\ the center-of-mass frame of the collision. The factor $\left(v\cdot n_{q} \right)^{2\epsilon}$ arises when performing the energy integrations in a general frame and is useful to assess the frame independence of our results. For the thrust-cone case, the $x$-dependence is trivial and can immediately be integrated out, which leads back to \eqref{eq:factSymb} with the angular integral denoted by $\otimes$. 

The hard functions $\bH_m(\nlist,\{\underline{x}\},Q)$ are obtained by integrating the squared amplitudes 
\begin{equation}\label{eq:Hmtildedef}
 \widetilde{\bm{\mathcal{H}}}_m(\{\underline{p}\})  = |\mathcal{M}_m(\{\underline{p}\})\rangle \langle\mathcal{M}_m(\{\underline{p}\})| 
\end{equation}
over the energies, while keeping their directions and energy fractions fixed. More specifically, we define
\begin{align}\label{eq:Hm}
\bH_m(\nlist,\xlist,Q)&=
\frac{1}{2Q^2}\Biggl(\prod_{i=1}^m \int \frac{d E_i E_i^{d-3}}{\ceps(2 \pi)^2{(n_i\cdot v)^{2\epsilon}}} \Biggr) \widetilde{\bH}_m(\plist) \times \notag\\
&\quad
(2\pi)^d\delta(Q-\Etot)\delta^{d-1}(\ptot) \F(\plist)\thetain (\plist) \prod_{j=1}^m \delta\!\left(x_j-\frac{2E_j}{Q}\right) .
\end{align}
The condition $\thetain (\plist)$ defines the jet region, i.e., the part of phase space inside the $M$ jets in which hard radiation is allowed, while $\F(\plist)$ contains additional kinematic cuts one may want to impose on the jets. In \cite{Becher:2023znt}, we worked with energy-ordered hard functions and defined the energy fractions with respect to the next harder parton. While convenient at LL, the ordering obscures symmetries among generated emissions. We therefore work with the standard definition $x_i=2E_i/Q$ and impose the appropriate ordering through the anomalous dimension itself, see Section~\ref{sec:clustering}.

The factorization formula is based on a method-of-regions expansion of the cross section around the limit $Q_0\ll Q$, and the hard and the soft functions correspond to the contributions of the hard and the soft momentum regions in the phase-space and loop integrations. The cross section as a whole is infrared finite, but the individual hard and soft functions contain divergences regularized by working in $d=4-2\epsilon$ dimensions. In the context of the effective theory, we should view these hard functions as bare Wilson coefficients and the soft functions 
as bare matrix elements. Expanding the bare functions in coupling constant and suppressing the arguments, we write
\begin{equation}\label{eq:Sexpansion} 
    \bS_m=\bone + \frac{\alpha_0}{4\pi}\bS_m^{(1)}+ \left( \frac{\alpha_0}{4\pi} \right)^2\bS_m^{(2)} +\ldots \, ,
\end{equation}
and similarly for the hard functions $\bH_m$. We defined 
\begin{equation}\label{eq:baresoftcoupling}
  \frac{g_s^2}{4\pi} =\alpha_0 \left( \frac{\mu^{2}\tilde{c}}{4} \right)^\epsilon\, ,
\end{equation}
where $g_s$ is the bare coupling; $\alpha_0$ is dimensionless and includes the usual \MS factor. The renormalized $\MS$ coupling constant $\alpha_s(\mu)$ is obtained from $\alpha_0$ after renormalization, $\alpha_0=Z_\alpha \alpha_s$.

After coupling constant renormalization, the cross section \eqref{eq:fact} is finite, but the hard and soft functions individually contain divergences in the energy integrations, which cancel between the two. In addition, there are collinear divergences which cancel among the hard functions of different multiplicities after angular integration. The renormalization of $\bH_m$ and $\bS_m$ was discussed in \cite{Becher:2015hka}, and explicit two-loop expressions for the $Z$-factor were provided in \cite{Becher:2021urs}. The $Z$-factor is obtained from an anomalous dimension $\bm{\Gamma}$, which is a matrix in the space of parton multiplicities and can be used to perform resummation of large logarithms by solving the associated renormalization group equation. We expand the anomalous dimension as
\begin{equation}\label{eq:GammaHexp}
    \bGam^H =\frac{\alpha_s}{4 \pi}  \bGam^{(1)} + \left(\frac{\alpha_s}{4 \pi}\right)^2  \bGam^{(2)} +\ldots\, .
\end{equation}
For dijet production, $M=2$, the one and two-loop entries have the form
\begin{align}
\bGam^{(1)} &= \left(\begin{array}{ccccc}
\Vm{2} & \Rm{2} & 0 & 0 & \ldots \\
0 & \Vm{3} & \Rm{3} & 0 & \cdots \\
0 & 0 & \Vm{4} & \Rm{4} & \cdots \\
0 & 0 & 0 & \Vm{5} & \cdots \\
\vdots & \vdots & \vdots & \vdots & \ddots
\end{array}\right), & \bGam^{(2)} &= \left(\begin{array}{ccccc}
\vm{2} & \rm{2} & \dm{2} & 0 & \ldots \\
0 & \vm{3} & \rm{3} & \dm{2}  & \cdots \\
0 & 0 & \vm{4} & \rm{4} &  \cdots \\
0 & 0 & 0 & \vm{5} & \cdots \\
\vdots & \vdots & \vdots & \vdots & \ddots
\end{array}\right).
\label{eq:Gamma1matrix}
\end{align}
The diagonal entries $\Vm{m}$ and $\vm{m}$ contain the purely virtual contributions to the anomalous dimension which leave the multiplicity unchanged, while $\Rm{m}$ and $\rm{m}$ describe a single real emission which increases the parton multiplicity. The terms $\rm{m}$ correspond to real-virtual corrections. At the two-loop order also double-real emissions $\dm{m}$ arise. These describe two emissions with commensurate energy; energy-ordered emissions are obtained from iterating $\Rm{m}$. For the case where the veto region is independent of the energy fractions $\{\underline{x}\}$, the two-loop anomalous dimension was derived in \cite{Becher:2021urs}. In the present paper, we will extend this result to the general case.

The scale independence of the cross section implies that the functions $\bH_m$ and $\bS_m$ fulfill RG equations which can be used to resum the large logarithms of $Q/Q_0$. The formal solution for evolving the hard functions from a scale $\mu_h\approx Q$ down to the scale $\mu_s\approx Q_0$ is given by the $\mu$-ordered exponential
\begin{equation}\label{eq:evolop}
    \bU\left(\nlist,\xlist,  \mu_s, \mu_h\right)=\bP \exp \left[\int_{\mu_s}^{\mu_h} \frac{d \mu}{\mu}\bGam^H(\nlist,\xlist, \mu)\right].
\end{equation}
Expanding the anomalous dimension and the path-ordered exponential perturbatively, the evolution matrix up to two-loop order takes the form
\begin{equation}\label{eq:Uevol}
\bU = \bm{1} + \frac{\alpha_s}{4\pi}\, L\, \bGam^{(1)} + \left(\frac{\alpha_s}{4\pi}\right)^2 \, \left[  L^2 \, \left(\frac{1}{2}\bGam^{(1)} \, \bGam^{(1)} +\beta_0 \,\bGam^{(1)}\right)+ L \,\bGam^{(2)}  \right]
\end{equation}
with $\alpha_s \equiv \alpha_s(\mu_h)$, $L=\ln(\mu_h/\mu_s)$ and 
$\beta_0=\frac{11}{3} \spac C_A-\frac{4}{3} \spac n_F\spac T_F $. 
This equation explicitly displays the logarithms generated by RG evolution. At two-loop order, we get double logarithms from iterating the one-loop anomalous dimension, as well as single logarithms associated with the two-loop anomalous dimension. In addition, there are terms generated by the running of the coupling constant.

The paper \cite{Becher:2021urs} addressed the renormalization scheme dependence of the anomalous dimension. In particular, it showed that the two-loop anomalous dimension extracted from analyzing the divergences in the energy integrations of the hard functions, did not correspond to the usual $\MS$ prescription because the counter terms are not pure poles due to the $\e$-dependence of the angular integrals. The paper also gave the relevant scheme change terms to convert the original results to $\MS$, albeit only in implicit form. To discuss the scheme dependence in detail, it is helpful to write out the cross section order by order in perturbation theory and to evaluate the hard and soft functions at their natural scales $\mu_h=Q$ and $\mu_s=Q_0$. Using \eqref{eq:Uevol}, we then get explicit expressions for the coefficients of the logarithms in terms of one-loop coefficients and anomalous dimensions. Keeping only terms up to next-to-leading logarithmic (NLL) accuracy, we obtain
\begin{align}\label{eq:LogsSigmaNNLO}
\sigma &= \langle \bH_2^{(0)}\otimes \bm{1} \rangle + \frac{\alpha_s}{4\pi} \left( L \,\Big\langle \bH_2^{(0)} \otimes   \bGam^{(1)} \hat{\otimes} \bm{1}  \Big\rangle + \Big\langle \bH_2^{(1)} \otimes \bm{1}+  \bH_3^{(1)} \otimes \bm{1} +  \bH^{(0)}_2 \otimes \bS_2^{(1)} \Big\rangle \right) \nonumber\\
& \quad +\left(\frac{\alpha_s}{4\pi}\right)^2 \left( \frac{L^2}{2}  \Big\langle \bH^{(0)}_2 \otimes  \bGam^{(1)} \hat{\otimes } \bGam^{(1)}  \hat{\otimes} \bm{1}+  2 \beta_0 \bH^{(0)}_2 \otimes  \right. \bGam^{(1)} \hat{\otimes} \bm{1} \Big\rangle  \nonumber \\
&\hspace{2.3cm} \left. + L\, \Big\langle \bH^{(0)}_2 \otimes  \bGam^{(2)} \hat{\otimes} \bm{1}+ \bH_2^{(1)} \otimes \bGam^{(1)} \hat{\otimes} \bm{1}+  \bH_3^{(1)} \otimes \bGam^{(1)} \hat{\otimes} \bm{1} \right. \nonumber \\ 
&\hspace{4cm} \left.+  \bH^{(0)}_2 \otimes \bGam^{(1)} \hat{\otimes} \, \bS^{(1)} \Big\rangle + \dots \right) + \mathcal{O}(\alpha_s^3)\,.
\end{align}
The notation $\hat{\otimes}$ indicates the angular integral over additional directions added by the real-emission parts of the anomalous dimensions. Each angular bracket in the above equation is separately scheme independent, since they correspond to the coefficient of a logarithm in the perturbative expansion of the cross section and are physical quantities once a renormalization scheme for the coupling has been chosen. From this we observe, e.g., that the one-loop anomalous dimension is scheme independent for $\epsilon \to 0$.\footnote{Surprisingly, this statement is only true under certain conditions, see Section \ref{sec:msbarfailure}.} Similarly, the non-logarithmic one-loop corrections are scheme independent. They are the last term in the first line of \eqref{eq:LogsSigmaNNLO}, and given by the sum of hard and a soft one-loop corrections. Since the sum is scheme independent, any change in the renormalization scheme for the soft function translates to an equal and opposite change of the hard function. Such a change then also affects the terms in the third and fourth line of \eqref{eq:LogsSigmaNNLO} and induces a compensating change in the two-loop anomalous dimension. 

In perturbative computations in QCD, the default renormalization scheme is the $\MS$ prescription, and in \cite{Becher:2021urs} it was proposed to adapt this for the numerical solution of the RG evolution via a parton shower framework. However, it turns out that this scheme is impractical because the expansion of the bare soft functions around $d=4$ involves additional angles in the $(d-4)$ dimensions, if the observable is differential enough. In the following section, we will analyze the soft function and will show that even for the simplest cone constraint in dijet production, these unphysical angles arise already in the next-to-leading order (NLO) soft functions. While it is easily possible to work with the extra angles in a fixed-order problem, their implementation into a parton shower is problematic, especially since the required extra angular integrals need to be treated as distributions.

For the parton shower, it is much more efficient to perform the subtraction of the soft divergences in terms of $d$-dimensional angular integrals. While such subtractions are non-minimal, they allow us to subtract the dependence on the additional integrations and set up a shower purely in terms of angles in physical space, at least up to NLL. In the next section, we will analyze the soft function in detail and introduce a family of subtraction schemes suitable for shower implementation. 

\section{Soft functions and angular integrals in \texorpdfstring{$d$}{d} dimensions}
\label{sec:soft_function}

\subsection{Bare soft function and angular integrals}
\label{sec:baresoft}

The bare one-loop soft function for a fixed-cone gap-between-jets observable is given by a sum of dipole contributions
\begin{align}\label{softfundef}
	\frac{\alpha_0}{4\pi} &\bm{S}^{(1)}_m(\{\underline{n}\},Q_0)=\nno \\
& - g_s^2 \sum_{(ij)}\,\,\bm{T}_{i,L}\cdot\bm{T}_{j,R}  \,\int\frac{d^d q}{(2\pi)^{d-1}}\frac{n_i \cdot n_j}{n_i \cdot q \, n_j \cdot q}\delta(q^2)\theta(q^0)\,\theta(Q_0-2 \Eout)\Theta_{\mathrm{out}}(n_q)\, .
\end{align}
The radiated gluon with momentum $q$ is restricted to the veto region by $\thetaout(n_q)$, which equals one if the gluon is in the veto region and zero otherwise. Radiation in the in-region would be allowed but is scaleless and thus vanishes in dimensional regularization. The energy constraint within the veto region is imposed by the step function $\theta(Q_0-2\Eout)$. Up to NLL accuracy, the specific definition of the soft veto observable $\Eout$ only enters through the one-loop soft function, since the hard functions and the anomalous dimensions do not depend on it. At lepton colliders, one typically defines $\Eout$ as the total lab-frame energy in the gap, but it is natural to consider a more general class of veto variables, covering also the hadron-collider case where one typically measures transverse momentum. For a single emission, any such veto can be written as $\Eout = E_q f(n_q)$, where $E_q$ is the energy of the emitted gluon in the frame where we evaluate the soft function and $n_q$ is its direction. Possible choices for $f$ are
\begin{align*}
   &\text{energy in the lab frame:} & f(n_q) &= n_{q v} \,,\\
   &\text{$q_T$ with respect to $n_1$ and $n_2$:} & f(n_q) &= \sqrt{2}(W_{12}^q)^{-\frac{1}{2}}\,,
\end{align*}
where the reference vector $v$ defines energies in a Lorentz-invariant way---in the lab frame it reads $v=(1,0,0,0)$. The dipole radiator is
\begin{equation}\label{eq:Wdef}
    W_{ij}^q = \frac{n_{ij}}{n_{iq}\, n_{jq}} \,,
\end{equation}
with the shorthands $n_{ab} = n_a\cdot n_b$ and $n_{vq}=v\cdot n_q$.

Performing the energy integration, the bare soft function becomes
\begin{equation}
  \label{eq_S1bare_after_energy_integ}
  \bS_m^{(1)}\left(\{\underline{n}\}, Q_0 \right)=\frac{2}{\epsilon}\left( \frac{\mu}{Q_0} \right)^{2\epsilon}\sum_{(i j)} \bT_i \cdot \bT_j \, S_{ij}\, ,
\end{equation}
with the angular integral
\begin{equation}\label{eq:Sijdef}
   S_{ij}= \Dqqq{q} W_{i j}^q \, \thetaout(n_q)\left(\frac{f(n_q)}{n_{vq}}\right)^{2\epsilon} \, ,
\end{equation}
with the measure $\Dqqq{q}$ defined in \eqref{eq:dOmega}. We write the expansion of this integral around $d=4$ as
\begin{equation}\label{eq:Sijexpansion}
   S_{ij}= S^{[0]}_{ij} + \epsilon\, S^{[1]}_{ij}  + \dots\, .
\end{equation}
The factor $(n_{vq})^{2\epsilon}$ in the
measure \eqref{eq:dOmega} renders $\Dqqq{q}W_{ij}^q$ Lorentz invariant, allowing us to change frames for the calculation; the transformation properties of the angular measure and of the other building blocks of our results are summarized in Appendix~\ref{sec:lorentz}.

\subsection{Expansion and renormalization schemes}
\label{sec:schemes}

For this subsection and the next we make the simplifying assumption that the measurement and the emitting dipoles are four-dimensional. We assume the veto function $\thetaout$, the weight $f$ and the dipole directions $n_i$, $n_j$ reference at most four linearly independent physical vectors, which for the thrust-cone observable are $v$, $n_i$, $n_j$ and the thrust axis $n_T$. In the following, we compute using this setup, the fully general case will be analyzed in Section~\ref{sec:fourdlimit}.

The one-loop soft function in a given renormalization scheme (RS) is obtained as
\begin{equation}\label{eq:SRS}
  \bS^{\mathrm{RS}(1)}=\lim_{\epsilon\to 0} \left[\bS^{(1)}-\frac{1}{2 \epsilon} \bGam^{\mathrm{RS}(1)} \hat{\otimes} \bone \right] .
\end{equation}
The counterterm subtracts the divergent part of the bare soft function but is otherwise arbitrary. We consider the general class of schemes satisfying
\begin{equation}\label{eq:SijRS}
  -\frac{1}{2 \epsilon}  \bGam^{\mathrm{RS}(1)} \hat{\otimes} \bone=-\frac{2}{\epsilon}\sum_{(i j)} \bT_i \cdot \bT_j \, S^{\mathrm{RS}}_{ij}\,,\qquad
  S^{\mathrm{RS}}_{ij}=\Dqqq{q} \widetilde{W}_{i j}^q(\epsilon)  \,\Theta_{ij}^{\mathrm{RS}}(n_q)\, ,
\end{equation}
where $\widetilde W^q_{ij}(\epsilon)$ is an $\epsilon$-dependent modified dipole with $\widetilde W^q_{ij}(0)=W^q_{ij}$, and $\Theta^{\mathrm{RS}}_{ij}(n_q)$ is a modified angular constraint that agrees with $\thetaout(n_q)$ in four dimensions in a sense that will become clear below. The angular integration in \eqref{eq:SijRS} is kept $d$-dimensional and the limit $\epsilon\to0$ is only taken after the subtraction. After subtraction, the renormalized soft function is
\begin{equation}
  \label{eq_Sren_with_DeltaSij}
    \bS^{\mathrm{RS}(1)}=\sum_{(i j)} \bT_i \cdot \bT_j \left[ 4\ln\!\left( \frac{\mu}{Q_0} \right)S^{[0]}_{ij} + 2\Delta S_{ij} \right] ,\qquad
    \Delta S_{ij}=S^{[1]}_{ij}-S^{\mathrm{RS}[1]}_{ij}\, .
\end{equation}
The subtraction used in \cite{Becher:2021urs} is defined by $\widetilde W^q_{ij}=W^q_{ij}$ and $\Theta^{\mathrm{RS}}_{ij}=\thetaout$: for the gap energy, $f=n_{vq}$, it gives $\Delta S_{ij}=0$, since the full $d$-dimensional angular integral is subtracted. We call this the maximal subtraction ($\XS$) scheme. For $f=n_{vq}$ it yields a soft function that is a pure logarithm, at the price of explicit $v$-dependence in the two-loop anomalous dimension needed at NLL.

\paragraph{Adapted coordinates and the projected direction.}
To study different schemes, we need to expand the bare soft function and the counterterms in $\epsilon$. To this end, we introduce coordinates adapted to the dipole and to the measurement. Under our assumption, every reference vector is an ordinary four-dimensional momentum, so we can choose the axes such that
\begin{equation}\label{eq:physframe}
  v= (1,\vec{0},\vec{0}_{d-4} ) , \quad n_i = (1, \vec{n}_i,\vec{0}_{d-4}), \quad n_j = (1, \vec{n}_j,\vec{0}_{d-4})\,,\quad n_T = (1, \vec{n}_T,\vec{0}_{d-4})\, .
\end{equation}
We refer to the span of $v$ and the first three spatial axes as the \emph{physical space}: by construction it contains every direction the measurement can resolve. Beyond the dipole vectors, one further physical direction plays a role: the normal to the dipole plane. We define $n_\perp$ as the space-like unit vector ($n_\perp^2=-1$) in the physical space orthogonal to $v$, $n_i$ and $n_j$; it is unique up to a sign, and in the frame \eqref{eq:physframe} it reads $n_{\perp}=- (0, \vec{n}_i \times \vec{n}_j,\vec 0_{d-4})/\lvert \sin\theta_{ij}\rvert$.
It is convenient to evaluate the angular integrals in the center-of-mass (COM) frame of the dipole, in which $n_i$ and $n_j$ are back-to-back along the first spatial axis, $v$ lies in the plane of the first two, and $n_\perp$ points along the third.\footnote{Explicitly, $\Lambda n_i=\frac{M}{2}( 1,1,0,0,\vec 0)$, $\Lambda n_j=\frac{M}{2}( 1,-1,0,0,\vec 0)$, $\Lambda v=\frac{2}{M}( 1,0,\beta,0,\vec 0)$ and $\Lambda n_\perp=(0,0,0,1,\vec 0)$, with $M^2=2n_{ij}$ and $\beta=\sqrt{1-M^2/4}$, where $\Lambda$ is the Lorentz transformation to the dipole frame.} There we parametrize the emission as
\begin{equation}\label{eq:comparam}
  n_q(\theta,\phi,\chi,\hat{n}_{d-4})=\left( 1,\cos{\theta},\sin{\theta}\cos{\phi},\sin{\theta}\sin{\phi}\cos{\chi},{\sin\theta}\sin\phi\sin{\chi} \,\hat{n}_{d-4} \right) ,
\end{equation}
where $\hat{n}_{d-4}$ is a space-like unit vector in the orthogonal complement to the physical space.
The central geometric object of the scheme discussion is the \emph{projected direction} 
\begin{equation}\label{eq:nqtilde}
  \tilde{n}_q \equiv  n_q(\theta,\phi,0)=\left( 1,\cos{\theta},\sin{\theta}\cos{\phi},\sin{\theta}\sin{\phi},\vec{0}_{d-4} \right) ,
\end{equation}
where we removed $\hat{n}_{d-4}$ from the arguments since the vector $n_q$ is independent of it for $\chi=0$. We stress that $\tilde{n}_q $ is a genuinely dipole-dependent object --- the same direction vector $n_q$ gets mapped to a different $\tilde{n}_q $ if the emission is off a different dipole. Its role below is to provide a modified angular constraint $\Theta^{\mathrm{RS}}_{ij}(n_q)=\thetaout(\tilde n_q)$ for subtraction schemes such as \MS that are built from four-dimensional data alone; since $\tilde n_q$ is dipole dependent, so is this constraint.

\paragraph{Expansion of the angular integral.}
Under our assumption the veto function depends on the emission only through scalar products with physical vectors, and therefore only on the three angles of \eqref{eq:comparam}, 
\begin{equation}\label{eq:thetaoutangles}
  \thetaout(\theta,\phi,\chi)\;\equiv\;\thetaout\big(n_q(\theta,\phi,\chi,\hat n_{d-4})\big)\,,
\end{equation}
independent of $\hat{n}_{d-4}$. The integration over the extra-dimensional directions is then trivial and simply produces the volume $\Omega_{d-4}$. The direction \eqref{eq:comparam} is invariant under the combined replacement \mbox{$\phi\to-\phi$}, \mbox{$\chi\to\pi-\chi$}, \mbox{$\hat{n}_{d-4}\to-\hat{n}_{d-4}$}, and the veto function therefore obeys \mbox{$\thetaout(\theta,-\phi,\pi-\chi)=\thetaout(\theta,\phi,\chi)$}. This allows us to fold $\chi$ into $[0,\tfrac\pi2]$ while extending $\phi$ to $(-\pi,\pi)$ after which the angular integral becomes
\begin{equation}\label{eq:angularmeasure}
\begin{aligned}
    \int \! [d\Omega_{q}]\,  &= \frac{e^{\epsilon \gamma_E}\Omega_{d-4} }{(4\pi)^{1-\epsilon} }  \int_{0}^\pi d\theta  \left(\sin\theta\right)^{1-2\epsilon} \int_{-\pi}^{\pi} d\phi \left(\sin\phi\right)^{-2\epsilon}\times\\
    &\hspace{2.2cm} \int_{0}^{\frac{\pi}{2}} d\chi\left(\sin\chi\right)^{-1-2\epsilon}  \left(n_{vq} \right)^{2\epsilon}\, ,
\end{aligned}
\end{equation}
where the $\chi$-integration acts as a distribution,
\begin{equation}\label{eq:chidist}
 \int_{0}^{\frac{\pi}{2}}\! d\chi\, 4^\epsilon\left(\sin\chi\right)^{-1-2\epsilon} g(\chi) =   -\frac{g(0)}{2\epsilon}+ \int_{0}^{\frac{\pi}{2}} \frac{ d\chi}{\sin\chi} \left[ g(\chi)-g(0) \right] + \mathcal{O}(\epsilon) \, .
\end{equation}
At $\chi=0$ the emission lies in the physical space and $\thetaout(\theta,\phi,0)=\thetaout(\tilde{n}_q)$. With this we obtain
\begin{align}\label{eq:S1bare}  
      S_{ij}^{[0]}&= \frac{1}{4\pi}\int_{-1}^{1} d\!\cos\theta\int_{-\pi}^{\pi}d\phi \frac{2}{\sin^2\!\theta} \thetaout\left(\tilde{n}_q\right)\\
  S_{ij}^{[1]}&=\frac{1}{2\pi}\int_{-1}^{1}d\!\cos\theta\int_{-\pi}^{\pi}d\phi\int_{0}^{\frac{\pi}{2}}\frac{d\chi}{\sin\chi}\frac{2}{\sin^2\!\theta}\left( \thetaout(\theta,\phi,0)-\thetaout(\theta,\phi,\chi) \right) \nonumber\\
  &+\frac{1}{4\pi}\int_{-1}^{1}d\!\cos\theta\int_{-\pi}^{\pi}d\phi\frac{2}{\sin^2\!\theta}\thetaout(\tilde{n}_q)\left( \ln\left( \frac{f^2(\tilde{n}_q)}{n_{vq}^2} \right)+\ln\left( \frac{4 n_{vq}^2}{\sin^2\!\theta} \right)-\ln\left( 4\sin^2\phi \right) \right) . \nonumber
\end{align}
The veto constraint will in general induce a dependence on the angle $\chi$ through the plus-distribution term of $S^{[1]}_{ij}$. One might wonder whether the veto function can simply be chosen such that this dependence drops out --- after all, the measurement only fixes $\thetaout$ in four dimensions. However, $\tilde n_q$ depends on the dipole, so the $\chi$-dependence cannot be avoided for all dipoles simultaneously.

\paragraph{Scheme family.}
A scheme in the class \eqref{eq:SijRS} is specified by two choices: the modified constraint $\Theta^{\mathrm{RS}}_{ij}(n_q)$, for which we consider the full $d$-dimensional veto $\thetaout(n_q)$ or its projected version $\thetaout(\tilde n_q)$ introduced above, and the modified dipole, for which we take
\begin{equation}\label{eq:dipoledefinition}
 \widetilde{W}_{ij}^q(\epsilon)=  W_{i j}^q \left(\frac{ \sin^2\!\theta}{4 n_{vq}^2} \right)^{a \epsilon} \left(4 \sin^2{\phi}  \right)^{b\epsilon}
 = W_{i j}^q \left( 2 \hat{W}_{ij}^q \right)^{-a \epsilon}\left(4 S_{ijqv}  \right)^{b\epsilon}\, ,
\end{equation}
where the first form is valid in the COM frame \eqref{eq:comparam}, and the second, Lorentz invariant form uses $\sin^2\!\theta=2/W^q_{ij}$ and
\begin{equation}\label{eq:Sijqvdef}
 \sin^2\phi = S_{ijqv} = 1-\frac{(n_{vi} \,n_{jq}+n_{iq}\, n_{vj}-n_{ij} \,n_{vq})^2}{2 n_{iq}\, n_{jq} (2 n_{vi}\, n_{vj}-n_{ij})}\,,
\end{equation}
with $\hat W^q_{ij}=W^q_{ij}(n_{vq})^2$; both $\hat W^q_{ij}$ and $S_{ijqv}$ are invariant under rescalings of the light-cone vectors, see Appendix~\ref{sec:lorentz}. Different pairs $(a,b)$, combined with one of the two constraint choices, correspond to different schemes; for $a=b=1$ the dipole cancels the $\mathcal O(\epsilon)$ terms of the measure \eqref{eq:angularmeasure}, so that the subtraction is effectively performed with the four-dimensional angular measure. The expansion of the counterterm integral reads
\begin{equation}
  \label{eq_SRS_expansion}
  \begin{aligned}
    S^{\mathrm{RS}[1]}_{ij}&=\frac{1}{2\pi}\int_{-1}^{1}d\!\cos\theta\int_{-\pi}^{\pi}d\phi\int_{0}^{\frac{\pi}{2}}\frac{d\chi}{\sin\chi}\frac{2}{\sin^2\!\theta}\left( \thetaout(\theta,\phi,0)-\Theta_{ij}^{\mathrm{RS}}(\theta,\phi,\chi) \right)\\
  &+\frac{1}{4\pi}\int_{-1}^{1}d\!\cos\theta\int_{-\pi}^{\pi}d\phi\frac{2}{\sin^2\!\theta}\thetaout(\tilde{n}_q)\left( \left( 1-a \right)\ln\left( \frac{4 n_{vq}^2}{\sin^2\!\theta} \right)-\left( 1-b \right)\ln\left( 4\sin^2\phi \right) \right) ,
  \end{aligned}
\end{equation}
where $\Theta^{\mathrm{RS}}_{ij}(\theta,\phi,\chi)$ denotes the modified constraint in the angular variables and the relation $\Theta^{\mathrm{RS}}_{ij}(\theta,\phi,0)=\thetaout(\tilde n_q)$ is required for the pole to cancel. The finite part of the renormalized soft function in this class of schemes is
\begin{equation}
  \label{eq:DeltaSSchemes}
\begin{aligned}
    \Delta S_{ij}&=\frac{1}{2\pi}\int_{-1}^{1}d\!\cos\theta\int_{-\pi}^{\pi}d\phi\int_{0}^{\frac{\pi}{2}}\frac{d\chi}{\sin\chi}\frac{2}{\sin^2\!\theta}\left(\Theta_{ij}^{\mathrm{RS}}(\theta,\phi,\chi)-\thetaout(\theta,\phi,\chi) \right)\\
  &+\frac{1}{4\pi}\int_{-1}^{1}d\!\cos\theta\int_{-\pi}^{\pi}d\phi\frac{2}{\sin^2\!\theta}\thetaout(\tilde{n}_q)\left( \ln\left( \frac{f^2(\tilde{n}_q)}{n_{vq}^2} \right)+a\ln\left( 2 \hat{W}_{ij}^q \right)-b\ln\left( 4 S_{ijqv} \right) \right) .
\end{aligned}
\end{equation}
The scheme $a=b=1$ with $\Theta_{ij}^{\mathrm{RS}}(n_q)=\thetaout(\tilde{n}_q)$ corresponds to the \MS scheme: the counterterm is then a pure pole. In this scheme there is a mismatch between the gap definition in the bare result and in the subtraction, which induces a non-trivial $\chi$-dependence in the renormalized function (first line of \eqref{eq:DeltaSSchemes}); it arises already for the simple thrust-cone constraint, as we show at the end of this subsection. In contrast, choosing $\Theta_{ij}^{\mathrm{RS}}(n_q)=\thetaout(n_q)$ makes the $d$-dimensional veto constraint drop out of the renormalized soft function entirely. The maximal subtraction scheme $\XS$ introduced after \eqref{eq_Sren_with_DeltaSij} is of this type, with $a=b=0$: for the gap energy it gives $\Delta S_{ij}=0$ and the simplest formulas in the two-loop anomalous dimension, at the price of anomalous dimensions carrying explicit $v$-dependence. The Lorentz subtraction scheme ($\LS$, $a=1$, $b=0$, $\Theta^{\mathrm{RS}}=\thetaout(n_q)$) \cite{Caron-Huot:2015bja} is also of this type and in addition frame independent: the explicit $v$-dependence cancels between the measure and $\widetilde W$, and the $\phi$- and $\chi$-contributions are subtracted completely. Both $\XS$ and $\LS$ are well suited for the parton-shower implementation of the resummation, and we will use both: $\XS$ has slightly simpler formulas, while $\LS$ is frame independent and does not require evaluating the reference vector $v$ in the COM frames of the dipoles. The boosted subtraction scheme ($\BS$, $a=b=1$, $\Theta^{\mathrm{RS}}=\thetaout(n_q)$) of \cite{Balsiger:2019tne} is identical to \MS up to the treatment of the extra angles; its drawback is the dependence on the complicated invariant $S_{ijqv}$. The four schemes are summarized in Table~\ref{tab:schemes}.

\begin{table}[t]
\centering
\begin{tabular}{@{}lcccp{7cm}@{}}
\toprule
\textbf{Scheme} & $a$ & $b$ & $\Theta_{ij}^{\mathrm{RS}}(n_q)$ &  \textbf{Comment} \\
\midrule
$\XS$  & 0 & 0 & $ \thetaout(n_q)$  & Soft function for energy is a pure logarithm. $\bGam^{(2)}$ is frame-dependent.\\[5pt]
$\LS$  & 1 & 0 & $ \thetaout(n_q)$ & Simple and frame-independent $\bGam^{(2)}$. No azimuthal angle dependence in one-loop hard and soft functions. \\
$\BS$  & 1 & 1 & $ \thetaout(n_q)$ & Similar to $\MS$. $\bGam^{(2)}$ frame-independent but complicated.\\
$\MS$ & 1 & 1 & $ \thetaout(\tilde{n}_q)$ &  Dependence on $(d-4)$-dimensional angle in $\bS^{(1)}$ and  $\bGam^{(2)}$; convolutions must remain $d$-dimensional (Section~\ref{sec:msbarfailure}).\\
\bottomrule
\end{tabular}
\caption{Summary of renormalization schemes and their properties. \label{tab:schemes}}
\end{table}

\begin{figure}
    \centering
    \begin{subfigure}[b]{0.49\linewidth}
        \centering
        \includegraphics[width=\linewidth]{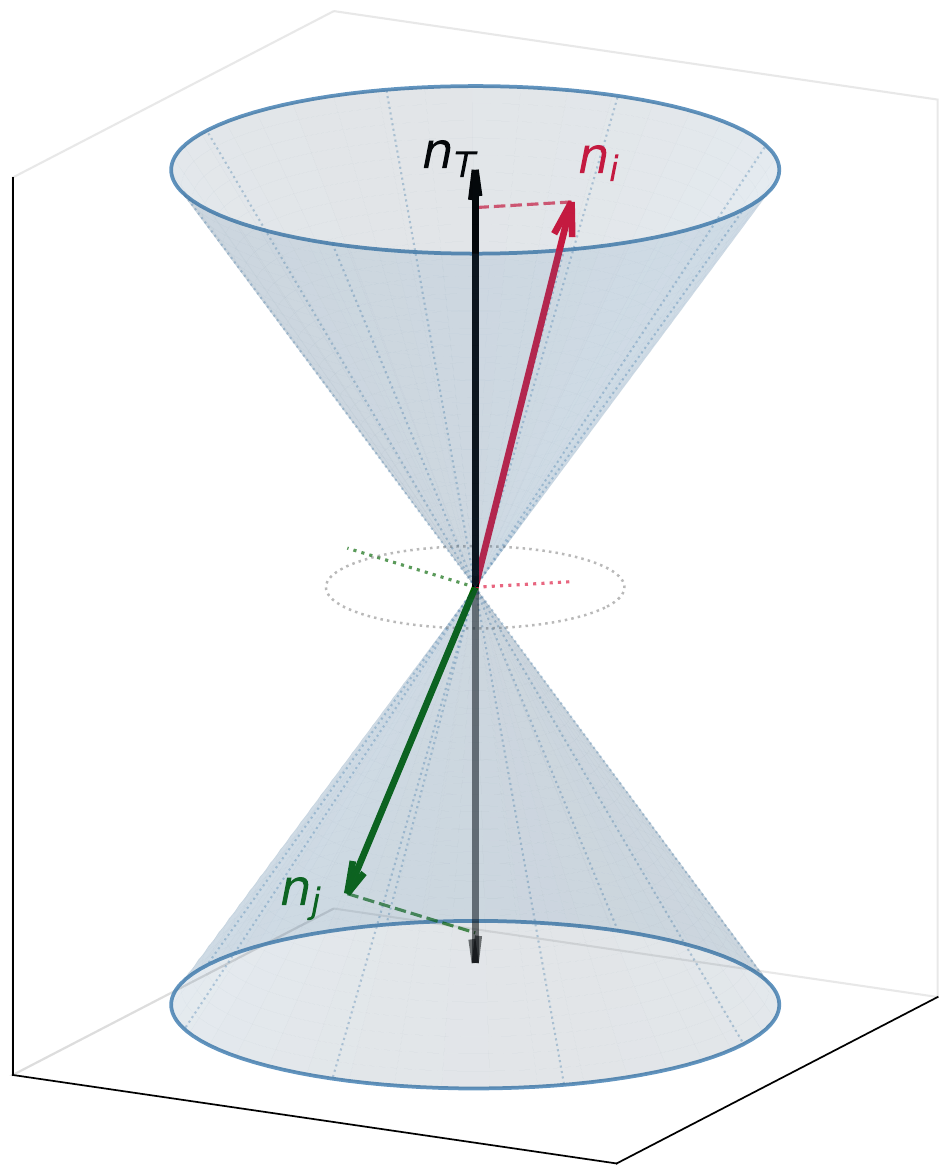}
        \caption{}
        \label{fig:cone_topbottom}
    \end{subfigure}
    \hfill
    \begin{subfigure}[b]{0.49\linewidth}
        \centering
        \includegraphics[width=\linewidth]{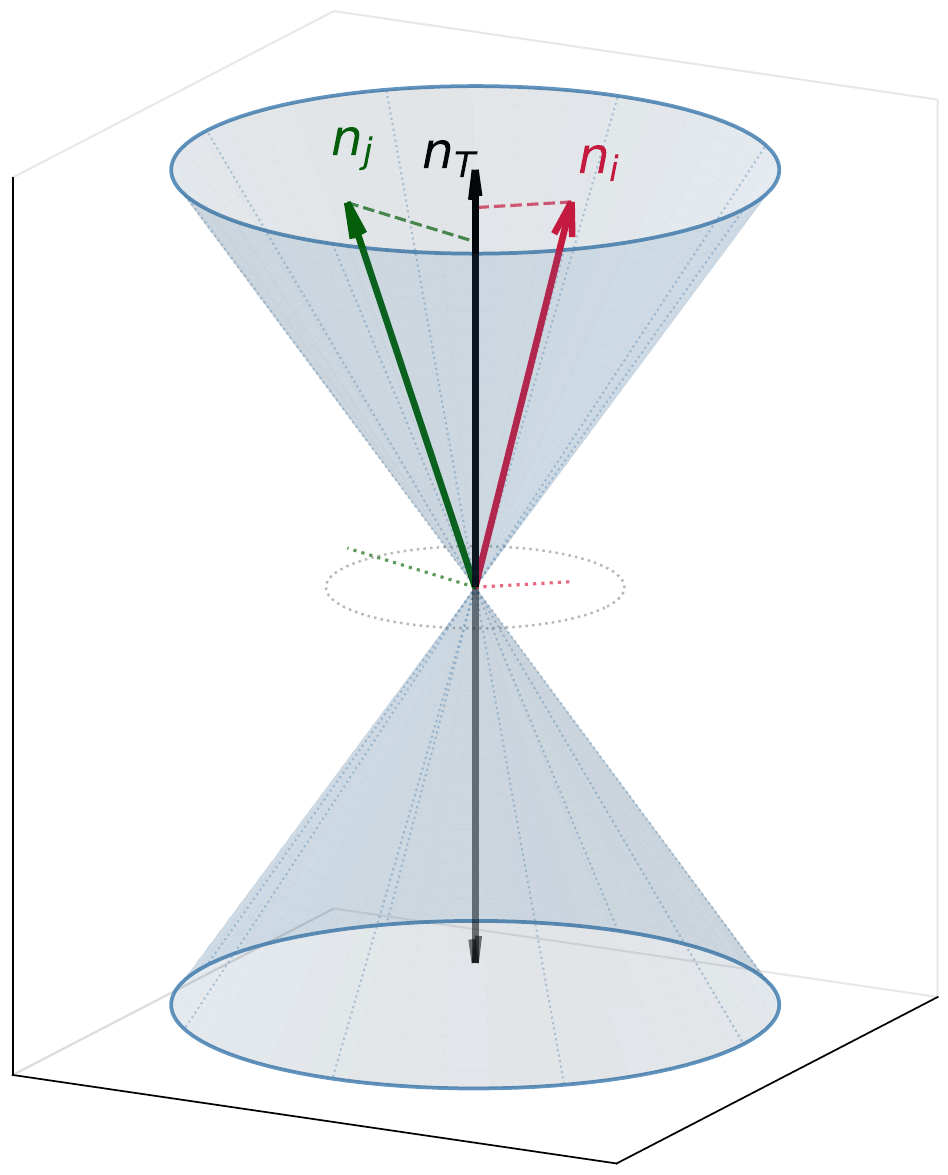}
        \caption{}
        \label{fig:cone_toptop}
    \end{subfigure}
    \caption{Soft emissions inside a thrust cone from two partons along directions $n_i$ and $n_j$ on opposite sides of the cone (\subref{fig:cone_topbottom}) and on the same side (\subref{fig:cone_toptop}). In the latter case an overall $\chi$ dependence arises in the \MS scheme.}
    \label{fig:cone}
\end{figure}

\paragraph{The cone example.}
We close this subsection with the example configuration shown in Figure~\ref{fig:cone}, which is the simplest example we could find, where the extra angle $\chi$ contributes to the result in the \MS scheme. The veto region is the region outside a two-sided cone with half opening angle $\delta$ around the thrust axis $n_T=(1,1,0,0,\vec{0}_{d-4} )$: an emission is vetoed if $\lvert \cos\bar\theta\rvert<\cos\delta\equiv\Delta$, where $\bar\theta$ is the lab-frame polar angle, or, Lorentz-invariantly,
\begin{equation}\label{eq:conedef}
\thetaout(n_q)=\theta\Big(\Delta-\Big\lvert 1-\frac{n_q\cdot n_T}{n_q\cdot v \, n_T\cdot v} \Big\rvert\Big)\, .
\end{equation}
For an arbitrary dipole with lab-frame directions
\begin{equation}\label{eq:conedipoledirs}
  n_i=\left( 1,\cos \bar{\theta}_i,\sin\bar{\theta}_i,0,\vec{0}_{d-4} \right), \quad n_j=\left( 1,\cos \bar{\theta}_j,\sin\bar{\theta}_j\cos\bar{\phi}_{ij},\sin\bar{\theta}_j\sin\bar{\phi}_{ij},\vec{0}_{d-4} \right) ,
\end{equation}
the gap constraint in the COM frame follows from
\begin{equation}
  \label{eq_cos_theta_bar_nq}
\begin{aligned}
   \cos\bar{\theta}&=\frac{1}{2\beta\left( 1-\beta  \sin\theta \cos\phi \right)}\biggl(\cos \bar{\theta}_i (\beta  (\cos\theta +1)-\sin\theta  \cos\phi)\\
   &+\cos \bar{\theta}_j (\beta  (1-\cos\theta )-\sin\theta  \cos\phi)-\sin\theta  \sin\bar{\theta}_i \sin \bar{\theta}_j \cos\chi  \sin \phi  \sin \bar{\phi}_{ij}\biggr)\,, 
\end{aligned}
\end{equation}
where $\beta = \sqrt{1-n_{ij}/2}$.
Note that the $\chi$-dependence of the constraint, entering through \eqref{eq_cos_theta_bar_nq}, is proportional to $\sin\bar\theta_i\sin\bar\theta_j\sin\bar\phi_{ij}$: it drops out if either emitter is (anti-)collinear to the thrust axis or if the two emitters are coplanar with it. Since at LO the two hard partons are back-to-back and aligned with the thrust axis, this implies that in a shower framework the $\chi$-dependence first arises at NLL at order $\mathrm{N}^3\mathrm{LO}$:\footnote{For more complicated constraints or hard functions, $\chi$-dependence can also enter at order $\as$.} one must generate two new hard emitters inside the cones and then emit a soft gluon. Calculating $S_{ij}$, two cases arise: a) $n_i$ and $n_j$ in opposite cones, b) in the same cone (Figure \ref{fig:cone}). We find
\begin{equation}\label{eq:conecases}
  \int_{-1}^{1}\frac{d\!\cos\theta}{\sin^2\!\theta}\int_{-\pi}^{\pi}d\phi\left( \thetaout(\theta,\phi,0)-\thetaout(\theta,\phi,\chi) \right)\begin{cases}
    \;= 0 & \text{for situation a)}\\
    \; \neq 0 & \text{for situation b)}
   \end{cases} \, ,
\end{equation}
where the equality holds for every value of $\chi$: for opposite-side emitters the veto region has global upper and lower rapidity boundaries whose $\phi$-average is $\chi$-independent, so the flat-rapidity integral of the difference vanishes. Thus, even when there are enough directions to generate a $\chi$-dependence, it can still drop out after all integrals are performed --- but not in general, as situation b) shows.

\subsection{The four-dimensional limit}
\label{sec:fourdlimit}

We now revisit the assumption made at the beginning of Section~\ref{sec:schemes}, that the measurement and the emitting dipoles are four-dimensional. A measurement performed in a physical detector determines the veto function only in the physical, four-dimensional spacetime. Its use in $d$ dimensions requires a continuation, i.e.\ a function $\Theta^{(d)}_{\mathrm{out}}$ that agrees with $\thetaout$ on momenta in the physical space.

When does a $d$-dimensional angular integral approach its four-dimensional value? We refer to the directions orthogonal to the four-dimensional physical space --- those with no counterpart in the physical space --- as its \emph{evanescent} components. Consider first an integrand that references only four-dimensional external data, so that it depends on the emission direction solely through the three angles of \eqref{eq:comparam}. The evanescent components of the emission then carry a measure of order $\epsilon$: in \eqref{eq:angularmeasure} the weight $(\sin\chi)^{-1-2\epsilon}$ combines with $\Omega_{d-4}=\mathcal O(\epsilon)$ to localize the emission at $\chi=0$, i.e.\ in the physical space. For a finite integrand, restricting the emission to physical directions is an $\mathcal O(\epsilon)$ approximation, exact for $\epsilon\to0$. This is no longer true if there are explicit $1/\epsilon$ poles multiplying (parts of) the integrand: they promote the $\mathcal O(\epsilon)$ evanescent measure to a finite contribution. For integrals that depend only on four-dimensional external data, poles are thus the only mechanism by which evanescent regions survive the limit $\epsilon\to0$. If the integrand resolves a fifth external direction, this bookkeeping breaks down. The integration over the extra-dimensional directions is no longer trivial, and its weight generates a $1/\epsilon$ term on its own, even for a bounded integrand, as we will see in Section~\ref{sec:msbarfailure}.

These observations bear important consequences for the factorization theorem as a whole. The angular parts $\otimes$ of the convolution in \eqref{eq:fact} formally run over $d$-dimensional hard directions. Suppose that the renormalized ingredients in $\sigma=\sum_m\langle\bH^{\mathrm{ren}}_m\otimes\bS^{\mathrm{ren}}_m\rangle$ are finite pointwise on the $d$-dimensional direction space and not merely after angular integration; whether this is the case depends on the renormalization scheme. Then no pole remains anywhere that could promote the evanescent components of the hard directions, and every convolution may be restricted to physical directions, with an error that vanishes for $\epsilon\to0$: we may make the replacement
\begin{equation}\label{eq:otimes4}
  \bH^{\mathrm{ren}}_m\otimes\bS^{\mathrm{ren}}_m
  \;\;\longrightarrow\;\;
  \bH^{\mathrm{ren}}_m\otimes_2\bS^{\mathrm{ren}}_m\,,
\end{equation}
where $\otimes_2$ denotes the convolution with the angular integral \eqref{eq:dOmega} evaluated with $\e=0$ with all generated hard directions restricted to the physical four-dimensional subspace. The argument proceeds direction by direction, from the outside in: once the inner directions are integrated, the outermost one enters only through scalar products with physical reference vectors; its evanescent components concentrate as above ($\chi\to0$), and the next direction follows, each step protected by the pointwise finiteness. The local schemes, $\XS$, $\LS$ and $\BS$, satisfy the hypothesis by construction: since $\Theta^{\mathrm{RS}}(n_q)=\thetaout(n_q)$ the counterterm carries the same $d$-dimensional constraint as the bare soft-pole density, so the poles cancel pointwise for every emission direction, physical or evanescent.\footnote{At the level of individual multiplicities the collinear singularities must also cancel pointwise; this is achieved by the collinearly subtracted dipoles $\overline W_{ij}$ of \cite{Becher:2021urs}. With unsubtracted dipoles the four-dimensional reduction in \eqref{eq:otimes4} applies only to the multiplicity-summed combination.} For the \MS prescription the replacement \eqref{eq:otimes4} is not legitimate, as we discuss next. 

\subsection{The shortcomings of \texorpdfstring{$\overline{\textbf{MS}}$}{MSbar} for non-global observables}
\label{sec:msbarfailure}

Equation \eqref{eq:DeltaSSchemes} showed that the \MS-renormalized soft function retains the $\chi$-terms: it knows about an additional, unphysical angle. Since the assembled cross section cannot, this dependence must be compensated --- the hard function, that is, the convolution $\otimes$, must know the additional direction as well. To see how, note that the same anomalous dimension renormalizes both sides of the factorization: its in-region part provides the counterterm for the hard functions, with density $\propto -\frac{2}{\epsilon}\,\widetilde W(\epsilon)\,[1-\Theta^{\mathrm{RS}}_{\mathrm{out}}(n_q)]$, differential in the generated direction. For $\Theta^{\mathrm{RS}}(n_q)=\thetaout(\tilde n_q)$ this counterterm knows only four-dimensional boundary data, while the soft-gluon pole of the bare hard function carries the full $d$-dimensional in-region constraint. Their difference,
\begin{equation}\label{eq:HrenMS}
 \bH^{\mathrm{ren},\MS}_{m+1}(n_q)\;=\;\big[\text{pointwise finite}\big]\;+\;\frac{2}{\epsilon}\;4W^q_{ij}\,
 \Big[\Theta^{(d)}_{\mathrm{in}}(n_q)-\Theta_{\mathrm{in}}(\tilde n_q)\Big]\,\bH_m\,,
\end{equation}
vanishes identically for physical $n_q$ but is of order one at evanescent directions. The \MS-renormalized hard function keeps an unsubtracted soft pole at evanescent emission directions, and is finite only after $d$-dimensional integration. Indeed, integrating \eqref{eq:HrenMS} with the $d$-dimensional measure, the $\delta$-term of \eqref{eq:chidist} vanishes but the $\Omega_{d-4}$-suppressed plus-distribution term is promoted by the $2/\epsilon$ to a finite contribution, which is exactly minus the $\chi$-term of $\Delta S^{\MS}_{ij}$ in \eqref{eq:DeltaSSchemes}. One obtains the correct result, but only because the convolution was kept $d$-dimensional: replacing $\otimes\to\otimes_2$ as in \eqref{eq:otimes4} would discard the promoted term and leave the $\chi$-dependence of $\bS^{\mathrm{ren},\MS}$ uncanceled. The four-dimensional reduction \eqref{eq:otimes4} is not available in \MS.

\paragraph{Five directions.} Even for a four-dimensional measurement, the \MS-renormalized hard functions emit partons with five-dimensional components, as required by \eqref{eq:HrenMS}. A secondary soft emission from such a configuration then resolves five linearly independent directions --- the four reference vectors of the measurement together with the extra hard leg. At this stage the renormalized soft function itself is no longer needed (at NLL), but its poles are since they enter the anomalous dimension. In the following we will demonstrate that in the presence of this fifth direction, even the one-loop anomalous dimension starts to depend on the continuation of the angular constraint to $d$ dimensions.

For simplicity we consider a dipole contribution to the bare soft function in which the emitting directions $n_i$ and $n_j$ still lie in the four-dimensional physical space, while the veto region depends on the additional leg $n_5$ generated by $\bH^{(1),\mathrm{ren}}$ with evanescent components --- through a clustering constraint, for example. To compute in this situation we follow the same steps as in the four-dimensional case: we go to the COM frame of the emitting dipole and parametrize the emission as in \eqref{eq:comparam}. The difference is that the fifth reference direction has a non-vanishing component along $\hat n_{d-4}$, so the veto constraint now resolves one additional angle: writing $\hat n_{d-4}=(\cos\chi_2,\,\sin\chi_2\,\hat n_{d-5})$, the constraint becomes a function $\thetaout(\theta,\phi,\chi,\chi_2)$, and the trivial extra-dimensional integration of Section~\ref{sec:schemes} is replaced by the genuine average
\begin{equation}\label{eq:thetabar}
  \bar{\Theta}_{\mathrm{out}}(\theta,\phi,\chi,\epsilon)=
  \frac{\int_0^\pi d\chi_2\,(\sin\chi_2)^{-2-2\epsilon}\;\thetaout(\theta,\phi,\chi,\chi_2)}{\int_0^\pi d\chi_2\,(\sin\chi_2)^{-2-2\epsilon}}\,,
\end{equation}
which then enters the plus-distribution terms in place of $\thetaout(\theta,\phi,\chi)$. This average \emph{does not remain finite} for $\epsilon\to0$: the denominator is of order $\epsilon$, while the numerator stays finite, so the ratio develops a pole.
The pole term of the soft function is thus of the form
\begin{equation}\label{eq:S0p5}
  S_{ij}^{[0]}= \frac{1}{4\pi}\int_{-1}^{1} d\!\cos\theta\int_{-\pi}^{\pi}d\phi\, \frac{2}{\sin^2\!\theta}\left[\, \thetaout\left(\tilde{n}_q\right)\;\;+\;2\int_0^{\frac{\pi}{2}}\frac{d\chi}{\sin\chi}\,T(\theta,\phi,\chi) \right] ,
\end{equation}
where $T(\theta,\phi,\chi)$ is a function that probes the evanescent angle $\chi$, which explicitly depends on the continuation of the veto function into the $d$-dimensional space.
 The subleading term $S^{[1]}_{ij}$ receives analogous modifications. Our discussion shows that even the pole of the bare soft function, and thus the one-loop anomalous dimension in $\MS$, depends on how the observable is continued to $d$ dimensions. 

\paragraph{What an \MS shower would need.}
With these results in hand, the practical requirements for implementing \MS in a parton-shower framework at NLL would be:
\begin{enumerate}
\item $\bH^{(1),\mathrm{ren}}$ with five-dimensional external legs: by \eqref{eq:HrenMS} its emitted parton carries an evanescent component that the subsequent evolution must resolve;  
\item $\bS^{(1),\mathrm{ren}}$ with a five-dimensional internal emission (the $\chi$-integrals of \eqref{eq:DeltaSSchemes});
\item a two-loop anomalous dimension that needs only four-dimensional external data but generates five-dimensional directions $n_q$ and $n_r$. It also contains both the full and the projected veto --- structures $[\Theta^{(d)}_{\mathrm{in}}(n_q)-\Theta_{\mathrm{in}}(\tilde n_q)]$, generated by the scheme-change commutator $[\delta\bGam_\Theta,\bGam^{(1)}]$ (cf.\ Section \ref{sec:schemechange}), where $\delta\bGam_\Theta$ denotes the part of the one-loop scheme change that swaps the full constraint $\thetaout(n_q)$ for its projection $\thetaout(\tilde n_q)$;
\item the one-loop anomalous dimension at five-dimensional \emph{input} directions;
\item at higher logarithmic orders, even more directions will be necessary.
\end{enumerate}

We see that the use of \MS for non-global observable entails significant complications.
The local schemes avoid all of this at the modest price of a non-minimal subtraction: with $\Theta^{\mathrm{RS}}=\thetaout(n_q)$ every renormalized object is pointwise finite, the four-dimensional reduction \eqref{eq:otimes4} applies, and the shower lives entirely in physical space.

It is instructive to contrast this with \emph{global} observables, for which the standard \MS treatment never faces these issues. There the measurement enters the soft function analytically, through weights such as $(f(n_q))^{2\epsilon}=1+2\epsilon\ln f(n_q)+\dots$: the observable dependence carries an extra power of $\epsilon$ relative to the poles, its $d$-dimensional continuation affects only terms one order higher in $\epsilon$, and no evanescent data survives in the $\epsilon^0$ renormalized functions at the orders needed. It is the step-function character of a veto, an $\mathcal O(\epsilon^0)$ constraint with support boundaries, multiplying poles, that makes non-global observables sensitive to the evanescent continuation, and \MS the wrong tool for them.

\section{Anomalous dimension in the \texorpdfstring{\XS}{XSbar} scheme}\label{sec:XSanomdim}

The derivation of the two-loop anomalous dimension in \cite{Becher:2021urs} for fixed angular constraints was based on the \XS scheme. We will first reproduce the result from this reference in a compact notation and then discuss the change into the \LS scheme. In reference \cite{Becher:2021urs}, the change into the \MS scheme was discussed, although an explicit form of the scheme change terms was not given. With our discussion in Section~\ref{sec:msbarfailure} we now understand that the \MS scheme is not suitable for non-global observables, and we will not discuss it further.

The one-loop anomalous dimension for fixed-cone gap-between-jets observables in the \XS scheme is given by
\begin{equation}\label{eq:Gamma1}
  \begin{aligned}
\Rm{m}= & -4 \sum_{i,j=1}^m \bT_{i, L}^{\alpha} \bT_{j, R}^{\tilde{\alpha}} W_{i j}^q \,\thetain\!\left(n_q\right) \\
\Vm{m}= & 2 \sum_{i,j=1}^m\left(\bT_{i, L} \cdot \bT_{j, L}+\bT_{i, R} \cdot \bT_{j, R}\right) \int\!\!\left[d \Omega_q\right] W_{i j}^q \\
& -i \pi \sum_{i,j=1}^m \frac{1}{2}\left[\bT_{i, L} \cdot \bT_{j, L}-\bT_{i, R} \cdot \bT_{j, R}\right] \Pi_{i j}\gamcuspm{0}\,, 
\end{aligned}
\end{equation}
with $\gamcuspm{0}=4$. We sum over all legs $i$ and $j$, rather than restricting the sum to $i\neq j$, as it typically done. The restriction can be dropped for the real part since $W_{i i}^q=0$, and for the imaginary part since the associated color factor vanishes for $i = j$. The color generator $\bT_{i, L}^\alpha$ acts on the $i$-th particle in  the amplitude in the hard function, while $\bT_{i, R}^{\bar{\alpha}}$ acts on the conjugate amplitude. The real emission part $\Rm{m}$ generates a hard function with an additional gluon with open color index $\alpha$ in the amplitude and $\tilde{\alpha}$ in the conjugate amplitude, while these indices are contracted in the virtual part of the anomalous dimension. Introducing a shorthand notation for the color dipoles 
\begin{align}\label{eq:dipoles}
   \bm{D}^{\alpha;\tilde{\alpha}}_{ij} &=  \bT_{i,L}^{\alpha}\, \bT^{\tilde{\alpha}}_{j,R} = \overline{\bm{D}}^{\,\alpha;\tilde{\alpha}}_{ji}\,, &  \bm{D}_{ij} &=  \bT_{i,L}\cdot \bT_{j,L}\,, &  \overline{\bm{D}}_{ij} &= \bT_{i,R}\cdot \bT_{j,R}\,,
\end{align}
where the overline indicates the hermitian conjugate,
we can write the anomalous dimension in the compact form
\begin{equation}\label{eq:oneLoopXS}
\bGam^{(1)} =-4 W_{ij}^q\left(  \bm{D}^{\alpha\tilde{\alpha}}_{ij} -\frac{1}{2} \bm{D}_{ij}- \frac{1}{2}\overline{\bm{D}}_{ij} \right)  - \frac{i \pi}{2} \gamcuspm{0}\, \Pi_{i j} \left(\bm{D}_{ij}- \overline{\bm{D}}_{ij}\right).
\end{equation}
This notation is supplemented with the following set of rules: i.) Repeated leg indices $i,j$ are summed over. ii.) For terms involving a kinematic dependence on a generated momentum $q$, one needs to add in a constraint $\thetain\!\left(n_q\right)$ for the real emission terms, and an integration $\int\left[d \Omega_q\right]$ for the virtual terms. 

At the two-loop level additional color structures arise. They include dipoles for the emission of a pair of particles in the adjoint or fundamental representation
\begin{align}\label{eq:pairdipoles}
\bm{D}_{ij}^{A,\alpha\beta;\tilde{\alpha}\tilde{\beta}} &=\bT_{i,L}^c \bT_{j,R}^d f^{\alpha  \beta  c} f^{ \tilde{\alpha}\tilde{\beta} d}
\,, &        \bm{D}_{ij}^{F,\alpha\beta;\tilde{\alpha}\tilde{\beta}} &=\bT_{i,L}^c \bT_{j,R}^d t^{c,\alpha }{}_{\!\beta } t^{d,\tilde{\beta}}{}_{\!\tilde{\alpha}}\,,
\end{align}    
 which have the property $\overline{\bm{D}}_{ji}^{X,\alpha\beta;\tilde{\alpha}\tilde{\beta}}= \bm{D}_{ij}^{X,\alpha\beta;\tilde{\alpha}\tilde{\beta}}$. In addition, there are tripole structures for double gluon emissions and their counterparts for single emission and virtual corrections
\begin{align}\label{eq:tripoles}
\bm{T}_{ijk}^{\alpha\beta;\tilde{\alpha}\tilde{\beta}} &= i f^{ \tilde{\alpha}\tilde{\beta} c} \left(\bT_{i,L}^{\alpha } \bT_{j,L}^{\beta }\right)_{\!+} \bT_{k,R}^c \, ,&      \overline{\bm{T}}_{ijk}^{\alpha\beta;\tilde{\alpha}\tilde{\beta}}&=-i f^{ \alpha \beta  c} \bT_{k,L}^c \left( \bT_{j,R}^{\tilde{\beta}}\bT_{i,R}^{\tilde{\alpha}}\right)_{\!+}\,,\notag\\
 \bm{T}_{ijk}^{\alpha;\tilde{\alpha}} &=i f^{\tilde{\alpha} b c}  \bT_{i,L}^{\alpha } \left(\bT_{k,R}^c\bT_{j,R}^b \right)_{\!+} \, ,& \overline{\bm{T}}_{ijk}^{\alpha;\tilde{\alpha}} &=-i f^{\alpha b c}  \left(\bT_{j,L}^b \bT_{k,L}^c\right)_{\!+} \bT_{i,R}^{\tilde{\alpha}}\, ,\\
 \bm{T}_{ijk} &= i f^{abc}\left( \bT_{k,R}^c \bT_{j,R}^b\bT_{i,R}^a\right)_{\!+} \, ,&  \overline{\bm{T}}_{ijk} &=-i f^{abc}\left(\bT_{i,L}^a \bT_{j,L}^b \bT_{k,L}^c\right)_{\!+}  \notag\,,  
\end{align}

where $(\bm{T}_i^a\dots )_+$ denotes a symmetrized product of the generators $\bm{T}_i^a$ including an averaging factor. For example
\begin{equation}\label{eq:symprodexample}
\left(\bT_{i,L}^\alpha \bT_{j,L}^\beta \right)_{\!+} = \frac{1}{2}\left( \bT_{i,L}^\alpha \bT_{j,L}^\beta+ \bT_{j,L}^\beta \bT_{i,L}^\alpha \right).
\end{equation}
Since the generators of different legs commute, the ordering is only relevant for generators on the same leg. The final color structure which arises at the two-loop level are contributions involving four legs with color structure
\begin{align}\label{eq:Qstruct}
\bm{Q}_{ijkl}^{\alpha\beta;\tilde{\alpha}\tilde{\beta}} &=\left(\bT_{i,L}^{\alpha } \bT_{j,L}^{\beta }\right)_{\!+}\left(\bT_{k,R}^{\tilde{\alpha}} \bT_{l,R}^{\tilde{\beta}}\right)_{\!+}=\overline{\bm{Q}}_{klij}^{\,\alpha\beta;\tilde{\alpha}\tilde{\beta}}\,.
 \end{align}
 This structure arises for the clustering two-loop anomalous dimension discussed below, but is absent for fixed angular constraints. In our notation, it is always understood that the emitted gluon with color indices $\alpha,\tilde{\alpha}$ has the momentum $q$, while an emitted gluon with color indexes $\beta,\tilde{\beta}$ has the momentum $r$. For example, $\bm{T}_{ijk}^{\alpha\beta;\tilde{\alpha}\tilde{\beta}}$ acts on a soft function containing a Wilson line along direction $n_q$ with color indices $(\alpha,\tilde{\alpha})$, and another Wilson line along direction $n_r$ with color indices  $(\beta,\tilde{\beta})$.

In terms of the above color structures, the anomalous dimension in the $\XS$ scheme for the lepton-collider case can be written in the compact form 
\begin{equation}\label{eq:Gamma2short}
  \begin{aligned}
    \bGam^{(2)}&= K_{i j k; qr} \left( \bm{T}_{ijk}^{\alpha\beta;\tilde{\alpha}\tilde{\beta}}-2 \bm{T}_{ijk}^{\alpha;\tilde{\alpha}}+ \bm{T}_{ijk} \right) +\text{h.c.}\\
    &-2 K^A_{ij;qr} \bm{D}_{ij}^{A,\alpha\beta;\tilde{\alpha}\tilde{\beta}} 
    -2 n_F K^F_{ij;qr}\bm{D}_{ij}^{F,\alpha\beta;\tilde{\alpha}\tilde{\beta}}
    -2 \,n_S K^S_{ij;qr} \bm{D}_{ij}^{S,\alpha\beta;\tilde{\alpha}\tilde{\beta}} \\
    &+2(C_A K^A_{ij;qr}+ n_F T_F K^F_{ij;qr}+ n_S T_S  K^S_{ij;qr}) \bm{D}_{ij}^{\alpha;\tilde{\alpha}} \\
   &+K_{ij;q} \left( 2 \bm{D}^{\alpha;\tilde{\alpha}}_{ij} - \bm{D}_{ij}- \overline{\bm{D}}_{ij} \right)\\
   & +  i \pi I_{ijk;q}\! \left( \bm{T}_{ijk}^{\alpha;\tilde{\alpha}}- \overline{\bm{T}}_{ijk}^{\,\alpha ;\tilde{\alpha}}\right) -i \pi I_{ij}  \left(\bm{D}_{ij}- \overline{\bm{D}}_{ij}\right) ,
  \end{aligned}
\end{equation} 
where the repeated indices $i$, $j$, $k$ are summed over all $m$ legs in \eqref{eq:Gamma2short} and without excluding terms with equal indices.  The hermitian conjugate in \eqref{eq:Gamma2short} only acts on the tripole terms in the first line. The anomalous dimension involves the tripole kinematic functions\footnote{Terms such as $I_{ijj;q}$ must be understood in the sense of a limit $n_k \to n_j$.}
\begin{equation}\label{eq:Gamma2XStripole}
  \begin{aligned}
    K_{i j k; qr}&=8 k_{i j k; qr}\ln\!\left(\frac{n_{k q} \, n_{vr}}{n_{k r} \,n_{vq}}\right) ,\\    
    I_{ijk;q}&= 4\ln\!\left(2 n_{vq}^2 W_{jk}^q\right)\! { \left(W_{i j}^q-W_{i k}^q\right)} ,
    \end{aligned}
\end{equation}
where we introduced the abbreviation 
\begin{equation}\label{eq:kijkqr}
k_{i j k; qr}=W^q_{ik}W^r_{jk}-W^q_{ik}W^r_{jq}-W^q_{ir}W^r_{jk}+W^{qr}_{ij}\,
\end{equation}
for the relevant combination of products of dipoles. Here and below we use the shorthand
\begin{equation}\label{eq:Wqr}
W^{qr}_{ab}=W^{rq}_{ba}=W^q_{ab}W^r_{bq}=W^r_{ab}W^q_{ar}=\frac{n_{ab}}{n_{aq}\,n_{qr}\,n_{br}}
\end{equation}
for the correlated emissions of $q$ and $r$ off the dipole $[ab]$; note that $W^{qr}_{ab}\neq W^{rq}_{ab}$.
The dipole kinematic functions can be written in the form
\begin{equation}\label{eq:Gamma2XSdipole}
  \begin{aligned}    
    K^A_{i j ; q r} &= K_{i j ; q r}^{(a)} - 2 K_{i j ; q r}^{(b)} + K_{i j ; q r}^{(c)}\,, \\
    K^S_{i j ; q r} &=  K_{i j ; q r}^{(c)} \,,\\
    K^F_{i j ; q r} &= K_{i j ; q r}^{(b)} -2  K_{i j ; q r}^{(c)} \, ,\\
     K_{ij;q}&=-\frac{1}{2} W_{ij}^q \left( 4\beta_0 \ln\!\left(2 n_{vq}^2 W_{ij}^q\right) + \gamcuspm{1}\right) ,\\
       I_{ij}&=\frac{1}{2}\gamcuspm{1}\, \Pi_{i j}\,,
\end{aligned}
\end{equation}
with
\begin{equation}
\gamma_1^\text{cusp} = 4 \left(\left(\frac{67}{9}-\frac{\pi ^2}{3}\right) C_A-\frac{20}{9}  n_F T_F -\frac{8}{9}  n_S T_S\right) .
\end{equation}
The individual parts of the dipole coefficients are given by
\begin{equation}\label{eq:Kijfuns}
  \begin{aligned}    
    K_{i j ; q r}^{(a)}&= \frac{2 n_{ij} \left(\frac{n_{ij} n_{qr}}{n_{iq} n_{jr}-n_{ir} n_{jq}}+1\right) \ln\!\left(\frac{n_{iq} n_{jr}}{n_{ir} n_{jq}}\right)}{n_{iq} n_{jr} n_{qr}}+\left( i \leftrightarrow j \right),\\
    K_{i j ; q r}^{(b)}&=\frac{8 n_{i j} \ln\!\left(\frac{n_{i q} n_{j r}}{n_{i r} n_{j q}}\right)}{n_{q r} \left(n_{i q} n_{j r}-n_{i r} n_{j q}\right)}\,,\\
    K_{i j ; q r}^{(c)}&=\frac{4}{n_{q r}^2} \left(\frac{\left(n_{i r} n_{j q}+n_{i q} n_{j r}\right) \ln\!\left(\frac{n_{i q} n_{j r}}{n_{i r} n_{j q}}\right)}{n_{i q} n_{j r}-n_{i r} n_{j q}}-2\right).
  \end{aligned}
\end{equation}
The above decomposition was introduced by Caron-Huot for super Yang-Mills (SYM) theory \cite{Caron-Huot:2015bja}. In $N=4$ SYM, only $K_{i j ; q r}^{(a)}$ is present, as the theory contains $n_S=3$ complex scalar fields and $n_F =2$ Dirac fermion fields ($4$ Weyl fermions) in the adjoint representation. 

We observe that the functions $ K_{ij;q}$, $I_{ijk;q}$ and the tripole coefficient $K_{i j k; qr}$ explicitly depend on the velocity vector $v$. While the $v$-dependence in $I_{ijk;q}$ drops out by color conservation, the remaining $v$-dependence is genuine. Therefore, as stated earlier, the two-loop anomalous dimension in the $\XS$ scheme is frame dependent. A systematic discussion of the transformation properties of all building blocks is given in Appendix~\ref{sec:lorentz}. 

In \cite{Becher:2021urs}, the three-particle color structures were defined without symmetrizing the color structures. The symmetrizers $(\dots)_+$ in \eqref{eq:Gamma2short} can be dropped if the kinematic functions fulfill the identities
\begin{equation}  \label{eq_condition_to_drop_symmetrizers}
   K_{i i j ; q r}+K_{i i j ; r q}=K_{i j i; q r}=K_{j i i; q r}=I_{ijj;qr}=0\, .
\end{equation}
These conditions are fulfilled for the kinematic functions  \eqref{eq:Gamma2XStripole} and \eqref{eq:Gamma2XSdipole} in the \XS scheme and will also hold for the final results in the \LS scheme given in \eqref{eq:KbarFuncs} below. One can therefore drop the symmetrizers in our final results. 

Finally, we stress that even within a given renormalization scheme, there is some ambiguity when defining the kinematic functions \eqref{eq:Gamma2XStripole} and \eqref{eq:Gamma2XSdipole}. As pointed in \cite{Caron-Huot:2015bja}, due to color conservation and the symmetries of the color structures when acting on soft functions, the expression
\begin{equation}\label{eq:ZeroDueToColorConservation}
  \begin{aligned}
  &(E_{ij;qr}+2 F_{ik;qr}-2 F_{jk;rq})\left( \bm{T}_{ijk}^{\alpha\beta;\tilde{\alpha}\tilde{\beta}}-2 \bm{T}_{ijk}^{\alpha;\tilde{\alpha}}+ \bm{T}_{ijk} \right) +\text{h.c.}\\
    &+2 \left( F_{ij;qr}+F_{ji;qr} \right)\left(  C_A\bm{D}_{ij}^{\alpha;\tilde{\alpha}}-\bm{D}_{ij}^{A,\alpha\beta;\tilde{\alpha}\tilde{\beta}} \right) \\
   &-\frac{1}{2} C_A\int\left[d \Omega_r\right]\left(  E_{ij;qr}-E_{ij;rq} \right) \left( 2 \bm{D}^{\alpha;\tilde{\alpha}}_{ij} - \bm{D}_{ij}- \overline{\bm{D}}_{ij} \right)
  \end{aligned}
\end{equation} 
acting on a hard function yields zero for generic $E_{ij;qr}$ and $F_{ij;qr}$, satisfying \mbox{$E_{ij;qr}=-E_{ji;rq}$}. We stress that the kinematic factor $(E_{ij;qr}+2 F_{ik;qr}-2 F_{jk;rq})$ does not satisfy the second, and third condition in \eqref{eq_condition_to_drop_symmetrizers} and thus, the symmetrizers in the tripole color structure cannot be dropped in \eqref{eq:ZeroDueToColorConservation}.

\section{Scheme change}\label{sec:schemechange}

As indicated in \eqref{eq:SRS}, the one-loop anomalous dimension in the general scheme is simply the subtraction multiplied by $(-2\epsilon)$. This is an immediate consequence of the fact that the bare soft function is $\mu$-independent, and 
\begin{equation}\label{eq:couplingrunning}
\frac{d\alpha_s(\mu,\epsilon)}{d\ln\mu} = -2 \epsilon\, \alpha_s(\mu,\epsilon)+\mathcal{O}(\alpha_s^2)\,.
\end{equation}
The one-loop anomalous dimension in the modified dipole schemes is thus the same as in the $\XS$ scheme, but with $W_{ij}^q$ replaced by $\widetilde{W}_{ij}^q(\epsilon)$
\begin{equation}\label{eq:GammaRSoneloop}
\bGam^{\mathrm{RS}(1)}(\epsilon) =-4 \widetilde{W}_{ij}^q(\epsilon) \left(  \bm{D}^{\alpha;\tilde{\alpha}}_{ij} -\frac{1}{2} \bm{D}_{ij}- \frac{1}{2}\overline{\bm{D}}_{ij} \right)  - \frac{i \pi}{2} \gamcuspm{0}\, \Pi_{i j} \left(\bm{D}_{ij}- \overline{\bm{D}}_{ij}\right).
\end{equation}
When working with renormalized quantities, we will take the limit $\epsilon\to 0$ of this result, reproducing the one-loop result \eqref{eq:oneLoopXS} in the \XS scheme, i.e.\ $\bGam^{\mathrm{RS}(1)}(0)=\bGam^{(1)}$. In this sense the one-loop anomalous dimension is scheme independent. The products of the one-loop anomalous dimension in $d=4$ are directly tied to the leading logarithms in the cross section \eqref{eq:LogsSigmaNNLO}. These are physical, and therefore scheme independent. However, the $\epsilon$ terms in the above equation change the renormalization of the one-loop soft  and hard function, as discussed in detail in Section~\ref{sec:soft_function}. Also, the two-loop anomalous dimension in $d=4$ depends on the scheme. We will now derive relations to convert the result in the \XS scheme into the RS schemes. There are two possible ways to extract the conversion terms. One way to perform the conversion would be to write out \eqref{eq:LogsSigmaNNLO} in both schemes and then read off $\bGam^{\mathrm{RS}(2)}$ from the fact that the single-log term at two loops must be scheme independent. A second possibility is to write out the renormalization condition for the soft function
\begin{align}
 \bm{\mathcal{S}}^{\mathrm{RS}(1)}_m &= \bm{\mathcal{S}}^{(1)}_m - \frac{1}{2\epsilon} \bGam^{\mathrm{RS}(1)}(\epsilon)\hat{\otimes}\bm{1} \,, \label{eq:resDivOne} \\
 \bm{\mathcal{S}}^{\mathrm{RS}(2)}_m &=  \bm{\mathcal{S}}^{(2)}_m  
 - \frac{1}{8\epsilon^2} \left[ 
 \bGam^{\mathrm{RS}(1)}(\epsilon)\hat{\otimes} \bGam^{\mathrm{RS}(1)}(\epsilon)\hat{\otimes}\bm{1}+ 2\beta_0 \bGam^{\mathrm{RS}(1)}(\epsilon)\hat{\otimes}\bm{1}  \right] \nonumber \\
& - \frac{1}{4\epsilon} \left[\bGam^{\mathrm{RS}(2)}(\epsilon) \hat{\otimes} \bm{1} +2 (\bGam^{\mathrm{RS}(1)}(\epsilon)+2\beta_0) \hat{\otimes}\, \bm{\mathcal{S}}^{\mathrm{RS}(1)} \right] .    \label{eq:resDivTwo}
\end{align}
Because the renormalized function is finite, and the bare function scheme independent, it follows that the pole terms generated by the anomalous dimensions in \eqref{eq:resDivOne} and \eqref{eq:resDivTwo} are also scheme independent.
To extract the pole terms, let us expand 
\begin{equation}\label{eq:GammaRSexpansion}
\bGam^{\mathrm{RS}(1)}(\epsilon) = \bGam^{(1)} - 2 \epsilon\, \delta\bGam^{\mathrm{RS}(1)} + \dots
\end{equation}
and insert the expansion into \eqref{eq:resDivOne} and \eqref{eq:resDivTwo}. Using the one-loop equation \eqref{eq:resDivOne}, one concludes that
\begin{equation}\label{eq:softRSOne}
\bm{\mathcal{S}}^{\mathrm{RS}(1)} = \bm{\mathcal{S}}^{\XS(1)} +  \delta\bGam^{\mathrm{RS}(1)}\hat{\otimes}\bm{1} \,.
\end{equation}
Comparing the $1/\epsilon$ poles in \eqref{eq:resDivTwo} in the RS and the \XS scheme, we can then read off the conversion between the two schemes after we also convert the one-loop soft function using \eqref{eq:softRSOne}. This leads to the following result for the scheme change in the two-loop anomalous dimension
\begin{equation}\label{eq:DeltaGamma2}
\Delta \bGam^{\mathrm{RS}(2)}= \bGam^{\mathrm{RS}(2)} - \bGam^{(2)} = \left[ \delta\bGam^{\mathrm{RS}(1)}, \bGam^{(1)} \right]-2\beta_0 \,\delta\bGam^{\mathrm{RS}(1)}\,.
\end{equation}
To compute the scheme change explicitly, we start with an arbitrary hard function $\bH_m$, apply the anomalous dimensions and then map the color structures to the ones in the general expression \eqref{eq:Gamma2short}. Even though our result \eqref{eq:Gamma2short} for the two-loop $\XS$ scheme anomalous dimension only holds for lepton colliders, we will still analyze the scheme change terms for the general case, as this exposes interesting structures. For a full derivation of the hadron-collider anomalous dimension, one will need to consider both soft and collinear limit of hard amplitudes. For the one-loop case, such an analysis was performed in \cite{Becher:2023mtx}.

The explicit form of the one-loop anomalous dimension $\bGam^{(1)}$ in the $\XS$ was given in \eqref{eq:Gamma1}. The quantity $\delta\bGam^{\mathrm{RS}(1)}$ has the same form as \eqref{eq:Gamma1} but with $W_{i j}^q$ replaced by $\delta W_{i j}^q$, where
\begin{equation}\label{eq:Wtildeexpansion}
\widetilde{W}_{ij}^q(\epsilon) = W_{i j}^q - 2 \epsilon\, \delta W_{i j}^q + \dots
\end{equation}
Explicitly, we have
  \begin{align}\label{eq:deltaGamma1}
\delta\Rm{m}^\mathrm{RS}= & -4 \sum_{i,j=1}^m \bT_{i, L}^{\alpha} \bT_{j, R}^{\tilde{\alpha}}\, \delta W_{i j}^q \,\thetain\!\left(n_q\right), \nonumber  \\
\delta\Vm{m}^\mathrm{RS}= & 2 \sum_{i,j=1}^m\left(\bT_{i, L} \cdot \bT_{j, L}+\bT_{i, R} \cdot \bT_{j, R}\right) \int\!\!\left[d \Omega_q\right] \delta W_{i j}^q \\
& -i \pi \sum_{i,j=1}^m \frac{1}{2}\left[\bT_{i, L} \cdot \bT_{j, L}-\bT_{i, R} \cdot \bT_{j, R}\right] \delta\Pi_{i j}\gamcuspm{0}\, .\nonumber 
\end{align}
Similarly to the modified dipole, we also introduce a modified imaginary part of the form
\begin{equation}\label{eq:Piexpansion}
\Pi_{i j}(\epsilon) = \Pi_{i j} - 2 \epsilon\, \delta\Pi_{i j} + \dots\,.
\end{equation}

The evaluation of the commutator in \eqref{eq:DeltaGamma2} is now straightforward, but one needs to take into account that the real emission operators $\Rm{m}$ and $\delta\Rm{m}$ add an extra gluon to the hard function. The successive application of the anomalous dimension then involves sums over the $(m+1)$ partons. Consider, as an example, the term 
\begin{equation}\label{eq:extragluonaction}
\begin{aligned}
    \bH_m\, &\Rm{m}\, \delta\Rm{m+1}^\mathrm{RS} = \sum_{k,l=1}^{m+1}\,\sum_{i,j=1}^m \bT_{k}^{\beta} \bT_{i}^{\alpha} \,\bH_m \, \bT_{j}^{\tilde{\alpha}} \,\bT_{l}^{\tilde{\beta}}\, W_{i j}^q\, \delta W_{k l}^r 
     =  \sum_{i,j,k,l=1}^{m} \bT_{k}^{\beta} \bT_{i}^{\alpha} \,\bH_m \, \bT_{j}^{\tilde{\alpha}} \,\bT_{l}^{\tilde{\beta}}\, W_{i j}^q\, \delta W_{k l}^r\\
    & - \sum_{i,j,l=1}^{m}  i f^{ \beta\alpha c} \, \bT_{i}^{c} \,\bH_m \, \bT_{j}^{\tilde{\alpha}} \,\bT_{l}^{\tilde{\beta}}\, W_{i j}^q
    \, \delta W_{q l}^r - \sum_{i,j,k=1}^{m}  \bT_{k}^{\beta} \bT_{i}^{\alpha} \,\bH_m \, \bT_{j}^{c} \,   i f^{c\tilde{\alpha}\tilde{\beta}}\, W_{i j}^q
    \, \delta W_{q k}^r \,.
\end{aligned}
\end{equation}
In the second step, the action of the generator on the extra gluon along direction $q$ with color indices $\alpha$ and $\tilde{\alpha}$, was evaluated explicitly according to the color-space formalism. 

Evaluating all relevant products of anomalous dimensions, the scheme change terms for the double-real ($d$), real-virtual ($r$) and purely virtual terms ($v$) are obtained in the form
\begin{equation}\label{eq:Gamma2shortScheme}
  \begin{aligned}
   \Delta \bGam^{\mathrm{RS}(2)}=& \Delta K^{\bm{d}}_{i j k; qr} \bm{T}_{ijk}^{\alpha\beta;\tilde{\alpha}\tilde{\beta}}-2  \Delta K^{\bm{r}}_{i j k; qr} \bm{T}_{ijk}^{\alpha;\tilde{\alpha}}+ ( \Delta K^{\bm{v}}_{i j k; qr}+\Delta K^{\bm{v}}_{i j k})\, \bm{T}_{ijk} +\text{h.c.}\\
    &-2 \Delta K^{\bm{d}}_{ij;qr} \,\bm{D}_{ij}^{A,\alpha\beta;\tilde{\alpha}\tilde{\beta}}  +2( C_A  \Delta K^{\bm{r}}_{ij;qr}+4 \beta_0 \delta W_{ij}^q) \bm{D}^{\alpha;\tilde{\alpha}}_{ij} -4\beta_0 \delta W_{ij}^q \left(\bm{D}_{ij}+ \overline{\bm{D}}_{ij} \right)\\
   & +  i \pi I^{\bm{r}}_{ijk;q}\! \left( \bm{T}_{ijk}^{\alpha;\tilde{\alpha}}- \bm{\bar{T}}_{ijk}^{\alpha ;\tilde{\alpha}}\right) +  i \pi I^{\bm{v}}_{ijk;q}\! \left( \bm{T}_{ijk}- \bm{\bar{T}}_{ijk}\right) -i \pi C_A I^{\bm{r}}_{ij;q} \bm{D}^{\alpha;\tilde{\alpha}}_{ij} \\
   & + 4 i \pi \beta_0 \delta\Pi_{ij}  \left(\bm{D}_{ij}- \overline{\bm{D}}_{ij}\right) ,
  \end{aligned}
\end{equation}
The tripole coefficient functions  in the first line are
\begin{equation}\label{eq:RStripole}
\begin{aligned}
  \Delta K^{\bm{d}}_{i j k; qr} & = 8\Big[ W_{i k}^q \big(\delta W_{jq}^r-  \delta W_{j  k}^r\big) + W_{jk}^r\big(\delta W_{i k}^q-\delta W_{i r}^q\big) - W_{jq}^r\delta W_{i k}^q +W_{i r}^q \delta W_{j k}^r\Big] ,\\
   \Delta K^{\bm{r}}_{i j k; qr} & = 4 \Big[ W_{i j}^q \big(\delta W_{j k}^r-
   \delta W_{k q}^r \big)+W_{i k}^q \big( \delta W_{j q}^r - \delta W_{jk}^r\big)  +W_{jk}^r \big(\delta W_{i k}^q -\delta W_{i j}^q \big)  \\
   & \hspace{1cm}  + W_{kq}^r \delta W_{i j}^q - W_{jq}^r \delta W_{i k}^q \Big], \\
  \Delta K^{\bm{v}}_{i j k; qr} & =4\Big[W_{i j}^q\big( \delta W_{j k}^r - \delta W_{ik}^r\big) + W_{jk}^q\big(\delta W_{i k}^r -\delta W_{i j}^r \big)\Big] \,,\\
 \Delta K^{\bm{v}}_{i j k} & = 4\pi^2 \Big[\Pi_{i j}\big( \delta \Pi_{i k} - \delta \Pi_{j k})+ \Pi _{j k}\big(\delta\Pi_{i j} -\delta \Pi_{i k}\big) \Big].
\end{aligned}
\end{equation}
The form of the results in \eqref{eq:RStripole} is not unique because one can always add terms which vanish due to the symmetry properties of the color structures and the angular integrations. Indeed, these symmetry properties can be used to write the result in the same general form as the anomalous dimension $\bGam^{(2)}$ in the $\XS$ scheme given in \eqref{eq:Gamma2short}, which has the same coefficient function $K_{i j k; qr}$ multiplying all tripole terms. Doing so makes it manifest that collinear divergences which arise in individual contributions cancel in the sum of real and virtual contributions. 

For purely virtual terms, the color structure is anti-symmetric under the exchange of legs $i$, $j$, $k$ and any symmetric pieces will drop out in the sum over legs. Furthermore, any parts antisymmetric under $q\leftrightarrow r$ will vanish when the angular integral is performed. Similarly, the real-virtual kinematic structure can be anti-symmetrized under $j\leftrightarrow k$. The double-real part can be anti-symmetrized under the simultaneous exchange of $i\leftrightarrow j$ and $q\leftrightarrow r$, see Appendix~\ref{sec:CSsymmetries}. We observe that
\begin{align}
 \Delta K^{\bm{d}}_{[i j k]; \{qr\}} &=  \Delta K^{\bm{v}}_{[i j k]; \{qr\}} \,,&  \Delta K^{\bm{d}}_{i [j k]; qr } &=  \Delta K^{\bm{r}}_{i [j k]; qr}\,,
\label{eq:dKsymmetries}
\end{align}
implying that we can use the real-emission function $\Delta K^{\bm{d}}_{i j k; qr}$ in \eqref{eq:RStripole} also for the real-virtual and double-virtual terms. 

The real-part dipole coefficients in the second line of \eqref{eq:Gamma2shortScheme} are obtained as
\begin{equation}\label{eq:dKdipole}
\begin{aligned}
 \Delta K^{\bm{r}}_{i j; qr} & = 4\left[ W_{i q}^r \delta W_{ij}^q+  W_{j q}^r \delta W_{i j}^q 
    -W_{i j}^q \left(\delta W_{iq}^r+\delta W_{jq}^r\right)\right] ,\\
 \Delta K^{\bm{d}}_{i j; qr} &= \Delta K^{\bm{r}}_{i j; \{qr\}}\,.
\end{aligned}
\end{equation}
Again, this is not of the same form as the anomalous dimension in the \XS scheme in \eqref{eq:Gamma2short}, where the dipole terms involve a single, symmetric function. To match the form of \eqref{eq:Gamma2short}, we split $\Delta K^{\bm{r}}_{i j; qr}$ into a symmetric and anti-symmetric part. The symmetric part is given by $\Delta K^{\bm{d}}_{i j; qr}$ and the anti-symmetric part can be absorbed into the coefficient $K_{ij;q}$, as
\begin{equation}\label{eq:Kijqterm}
\Delta  K_{ij;q}= \int [\rd \Omega_r] \Delta K^{\bm{r}}_{i j; [qr]}\,.
\end{equation}
Since it is antisymmetric in $q$ and $r$, the term does not contribute when multiplying the fully virtual structures $\bm{D}_{ij}$ and $\overline{\bm{D}}_{ij}$.

Finally, we obtain the imaginary terms
\begin{equation}\label{eq:dKim} 
\begin{aligned}
  I^{\bm{r}}_{ij;q} & =8\left[ \left(\delta\Pi_{j q}-\delta\Pi_{iq}\right) W_{ij}^q+\left(\Pi_{i q}-\Pi_{j q}\right) \delta W_{i j}^q \right], \\
   I^{\bm{r}}_{ijk;q} & = 16\left[ W_{i k}^q \left(\delta\Pi_{jk}-\delta\Pi_{j q}\right)+ \delta W_{i j}^q (\Pi_{j k} - \Pi_{kq })  \right],  \\
   I^{\bm{v}}_{ijk}  &= 4 \Big[ \Pi _{i j} \big(\delta W_{j k}^q-\delta W_{i k}^q\big)
   +\Pi_{j k} \big(\delta W_{i k}^q-\delta W_{ij}^q\big) \\
 &  \hspace{1cm} +\big(\delta\Pi_{j k}-\delta\Pi_{i k}\big)
   W_{i j}^q+\left(\delta\Pi_{ik}-\delta\Pi_{i j}\right) W_{j k}^q \Big]\,.
\end{aligned}
\end{equation}
We have derived the general form of the scheme change, but in the present paper, we focus on the lepton-collider case where $\Pi_{i j}=1$ for all $i$ and $j$. In this case, the anti-symmetry of the virtual tripole color structure ensures that there are no contributions from $ I^{\bm{v}}_{ijk}$ and $K^{\bm{v}}_{ijk}$ so these terms can be dropped. More generally, all contributions to the imaginary part from the modified dipoles $\delta W_{i j}^q$ vanish and only terms involving the modified imaginary parts $\delta\Pi_{ij}$ can give rise to non-zero contributions to the imaginary part. This leaves the two functions $I^{\bm{r}}_{ij;q}$ and $I^{\bm{r}}_{ijk;q}$. The coefficient $I^{\bm{r}}_{ij;q}$ can be absorbed into $I^{\bm{r}}_{ijk;q}$ using color conservation which implies the relation
\begin{equation}\label{eq:CADFcolourid}
C_A \bm{D}_{ij}^{\alpha \tilde{\alpha}} F_{ij} = \bm{T}_{ijk}^{\alpha \tilde{\alpha}} (F_{ik} - F_{ij})
\end{equation}
for an arbitrary coefficient function $F_{ij}$ (cf. Appendix \ref{sec:CScc}). To derive this, we use that the coefficient on the right-hand side is antisymmetric under $j\leftrightarrow k$ so that the symmetrizers in the color structure can be dropped, see \eqref{eq_condition_to_drop_symmetrizers}.

\subsection{\texorpdfstring{\LS}{LSbar} scheme}

We now simplify the results further for the \LS scheme for which the one-loop anomalous dimension reads
\begin{equation}\label{eq:GammaLSoneloop}
  \begin{aligned}
\RLSm{m}= & -4 \sum_{(i j)} \bT_{i, L}^a \bT_{j, R}^{\tilde{a}} W_{i j}^q \left( 2 \hat{W}_{i j}^q \right)^{-\epsilon}\thetain\!\left(n_q\right) ,\\
\VLSm{m}= & 2 \sum_{(i j)}\left(\bT_{i, L} \cdot \bT_{j, L}+\bT_{i, R} \cdot \bT_{j, R}\right) \int {\left[d \Omega_q\right]} W_{i j}^q\left( 2 \hat{W}_{i j}^q \right)^{-\epsilon} \\
& -i \pi \sum_{(i j)} \frac{1}{2}\left[\bT_{i, L} \cdot \bT_{j, L}-\bT_{i, R} \cdot \bT_{j, R}\right] \Pi_{i j} \left(2 W_{ij}^v \right)^{-\kappa \,\epsilon} \gamcuspm{0},
\end{aligned}
\end{equation}
For $\kappa=0$, the imaginary part remains unmodified. For the lepton-collider case $\Pi_{i j}=1$, and $\kappa=1$ corresponds to the modification of the  imaginary part introduced in \cite{Caron-Huot:2015bja}. Below, we will see that it has the interesting property of trading the three-particle imaginary part in $\rm{m}$ for a two-particle contribution to $\vm{m}$. For $\kappa\neq 0$, the renormalized one-loop soft function has an imaginary part, even in the lepton-collider case. 

For the \LS scheme in the lepton-collider case, we thus have
\begin{align}\label{eq:modIm}
   \delta W_{ij}^q &= \frac{1}{2} W_{ij}^q \ln(2n_{vq}^2 W_{ij}^q)\,, & \delta \Pi_{ij}^q & = \frac{\kappa}{2} \ln(n_i\cdot n_j)\,.
\end{align}
We have used $\Pi_{i j}=1$ and color-conservation to simplify $\delta\Pi_{ij}^q$. Inserting the explicit forms into the above expressions, we obtain
\begin{equation}
\label{eq_CommutatorKFunctions}
    \begin{aligned}
        \Delta K_{ij;qr}^F&=\Delta K_{ij;qr}^S=0 \,,\\
        \Delta K_{ijk;qr}&=4 \left[\ln (n_{vq}^2 W_{i k}^q)-\ln (n_{vr}^2 W_{j k}^r)\right]\left(W_{ik}^{q} W_{jk}^{r}- W_{ir}^{q} W_{jk}^{r}
- W_{ik}^{q} W_{jq}^{r} \right)  \\
        &\quad -4 W_{i k}^q W_{j q}^r \left(\ln W_{j k}^r-\ln W_{j q}^r\right)+4 W_{i r}^q W_{j k}^r \left(\ln W_{i k}^q-\ln W_{i r}^q\right) ,\\
        \Delta K^A_{ij;qr}&= W_{i j}^r W_{i r}^q \left(\ln W_{i j}^q+\ln W_{i j}^r-\ln W_{i r}^q-\ln W_{j q}^r\right)+\left(q\leftrightarrow r\right) ,\\
        \Delta K_{ij;q}&=C_A \int [\rd \Omega_r] \bigg( W_{i j}^q W_{j q}^r \left[\ln W_{i j}^q-\ln W_{i j}^r+\ln W_{i r}^q-\ln W_{j q}^r+4 \ln \!\left(\frac{n_{vq}}{n_{vr}}\right)\right] \\
        &\hspace{2cm} +\left(i\leftrightarrow j\right)\bigg) +2 \beta_0 W_{i j}^q\ln 2 n_{vq}^2 W_{i j}^q \,, \\
        \Delta I_{ijk;q}&= -4 \,\kappa  \ln\!\left(2 n_{vq}^2 W_{jk}^q\right)\! { \left(W_{i j}^q-W_{i k}^q\right)}
        \, .
    \end{aligned}
\end{equation}
Remarkably, as observed in \cite{Caron-Huot:2015bja}, for $\kappa=1$ the three-particle imaginary part $\Delta I_{ijk;q}$ of the scheme change precisely cancels the  imaginary part $I_{ijk;q}$ present in the $\XS$ scheme. Instead, one is left with the two-particle imaginary part proportional to $\beta_0$ in the last line of \eqref{eq:Gamma2shortScheme} and an imaginary one-loop soft function.

 The above expressions now have the same general form as the result in the \XS scheme given in \eqref{eq:Gamma2short}. However, the kinematic function $\Delta K_{ijk;qr}$ in \eqref{eq_CommutatorKFunctions} does not satisfy the second and third condition in \eqref{eq_condition_to_drop_symmetrizers} and thus, the symmetrizers in the tripole color structure cannot be dropped in \eqref{eq:ZeroDueToColorConservation}. After making a scheme change from the \XS scheme, one can additionally redefine the kinematic functions by adding terms of the form \eqref{eq:ZeroDueToColorConservation} to restore the identities in \eqref{eq_condition_to_drop_symmetrizers}. We will now do this to arrive at our final result in the \LS scheme which involves the functions 
\begin{equation}\label{eq:Kbardefs}
  \begin{aligned}
    \overline{K}_{i j k; qr}&=K_{i j k; qr}+\Delta K_{i j k; qr}+(E_{ij;qr}+2 F_{ik;qr}-2 F_{jk;rq})\,,\\
     \overline{K}^A_{i j; qr}&=K^A_{i j; qr}+\Delta K^A_{i j; qr}+(F_{ij;qr}+F_{ji;qr})\,,\\
     \overline{K}_{i j; q}&=K_{i j; q}+\Delta K_{i j; q}-\frac{1}{2} C_A\int\left[d \Omega_r\right]\left(  E_{ij;qr}-E_{ij;rq} \right)\,,
  \end{aligned}
\end{equation}
with
\begin{equation}\label{eq:EFdefs}
  \begin{aligned}
    E_{ij;qr}&=-4 W_{ij}^q W_{jq}^r \log \left(\frac{n_{iq} n_{jq} n_{rv}^2}{n_{ir} n_{jr} n_{qv}^2}\right) ,\\
    F_{ik;qr}&=-2 W_{ik}^q W_{kq}^r \log \left(\frac{n_{ir} n_{kq}}{n_{ik} n_{qr}}\right) .
  \end{aligned}
\end{equation}
Adding up the individual contributions, the kinematical functions in the \LS scheme read
\begin{equation}\label{eq:KbarFuncs}
  \begin{aligned}
    \overline{K}_{i j k; qr} &=\frac{4}{n_{i q} n_{j r}}\! \Biggl(\!
    \frac{n_{i j} }{n_{q r}}\ln\!\left(\frac{n_{i r} n_{j r} n_{k q}^2}{n_{i q} n_{j q} n_{k r}^2}\right)\!
    +\frac{n_{i k} n_{j k} }{n_{k q} n_{k r}}\ln\!\left(\frac{n_{i k} n_{j r} n_{k q}}{n_{i q} n_{j k} n_{k r}}\right)
    -\frac{n_{i q} n_{j k} }{n_{k q} n_{q r}}\ln\!\left(\frac{n_{j k} n_{q r}}{n_{j q} n_{k r}}\right)\\
    &\hspace{-0.5cm}
    -\frac{n_{i k} n_{j q} }{n_{k q} n_{q r}}\ln\!\left(\frac{n_{i k} n_{j r} n_{k q} n_{q r}}{n_{i q} n_{j q} n_{k r}^2}\right)+\frac{n_{i r} n_{j k} }{n_{k r} n_{q r}}\ln\!\left(\frac{n_{i q} n_{j k} n_{k r} n_{q r}}{n_{i r} n_{j r} n_{k q}^2}\right)+\frac{n_{i k} n_{j r} }{n_{k r} n_{q r}}\ln\!\left(\frac{n_{i k} n_{q r}}{n_{i r} n_{k q}}\right)\Biggr) \,,\\
    \overline{K}_{i j ; q r}&=C_A \overline{K}_{i j ; q r}^{(a)}+\left[n_F T_F-2 C_A\right] K_{i j ; q r}^{(b)}+\left[C_A-2 n_F T_F+n_S T_S\right] K_{i j ; q r}^{(c)}\\
    &=C_A \overline{K}_{i j ; q r}^{A}+n_F T_F K_{i j ; q r}^F+n_S T_S K_{i j ; q r}^S \,,\\
    \overline{K}_{i j ; q r}^{(a)}&=
    \frac{2 n_{i j} \left(2 \ln\!\left(\frac{n_{i j} n_{q r}}{n_{i r} n_{j q}}\right)+\left(\frac{n_{i j} n_{q r}}{n_{i q} n_{j r}-n_{i r} n_{j q}}+1\right) \ln\!\left(\frac{n_{i q} n_{j r}}{n_{i r} n_{j q}}\right)\right)}{n_{i q} n_{j r} n_{q r}}
    +\left( i\leftrightarrow j \right), \\
     \overline{K}_{i j ; q } &= - \frac{1}{2}\gamcuspm{1} W_{i j}^q \,,\\ 
     \overline{I}_{ijk;q}&=I_{ijk;q}+\Delta I_{ijk;q}=  4 (1-\kappa)\ln\!\left(2 n_{vq}^2 W_{jk}^q\right)\! { \left(W_{i j}^q-W_{i k}^q\right)} ,\\
       \overline{I}_{ij}&=I_{ij}+\Delta I_{ij}=\frac{1}{2}\left(\gamcuspm{1}-4 \kappa \beta_0 \ln W_{ij}^v\right).
\end{aligned}
\end{equation}
We point out that the explicit $v$-dependence in $\overline{I}_{ijk;q}$ drops out by color conservation. Thus, the final result for the anomalous dimension in the \LS scheme is independent of the auxiliary vector $v$ and therefore frame independent, as anticipated. 
For the lepton-collider case, the cusp part vanishes by color conservation so that we end up with $\overline{I}_{ij}=0$ for $\kappa=0$, while the choice $\kappa=1$ results in $ \overline{I}_{ijk;q}=0$. The two-loop anomalous dimension $\overline{\bGam}^{(2)}$
in the \LS scheme is obtained by inserting the kinematic functions in \eqref{eq:KbarFuncs} into expression \eqref{eq:Gamma2short}. The above expressions for the anomalous dimension agree with the final result for the anomalous dimension in \cite{Caron-Huot:2015bja}. An alternative form for the three-particle function is
\begin{equation}\label{eq:Kbarnew}
\overline{K}_{ijk;qr}=4\,k_{ijk;qr}\,\ln\!\left(\frac{n_{ir}\,n_{jr}\,n_{kq}^2}{n_{iq}\,n_{jq}\,n_{kr}^2}\right)
+4\,l_{ijk;qr}\,\ln\!\left(\frac{n_{ik}\,n_{qr}}{n_{ir}\,n_{kq}}\right)
-4\,l_{jik;rq}\,\ln\!\left(\frac{n_{jk}\,n_{qr}}{n_{jq}\,n_{kr}}\right),
\end{equation}
with the coefficient $k_{ijk;qr}$ defined \eqref{eq:kijkqr} and the combination
\begin{equation}\label{eq:lijkqrdef}
l_{ijk;qr}= W^q_{ik}\big(W^r_{jk}-W^r_{jq}\big)+W^{qr}_{ik}\,,
\end{equation}
and $W^{qr}_{ab}$ defined in \eqref{eq:Wqr}.
The form \eqref{eq:Kbarnew} makes the relation to the \XS result in \eqref{eq:Gamma2XStripole} manifest.

\section{Leading-color anomalous dimension and shower implementation}\label{sec:largeNc}

The structure of the anomalous dimension simplifies substantially in the large-$N_c$ limit, since the color structure of the amplitude reduces to a product of color dipoles. The leading terms in the limit then arise from applying the anomalous dimension to one of these dipoles, consisting of neighboring legs $i$ and $j$. The purely virtual terms map the dipole onto itself, while a single real gluon emission along the direction $n_q$ splits the dipole $[ij]$ into two neighboring dipoles $[iq]$ and $[qj]$, which we abbreviate as $[iqj]$. To keep track of the dipole structure, the parton shower implementation keeps a list of directions $\left\{\underline{n}\right\} =\{ n_1, n_2, \dots, n_m\}$. The first element corresponds to an outgoing quark, the last one with direction $n_m$ to an outgoing anti-quark, the intermediate elements are gluons. The emission of a real gluon maps this list
\begin{equation}\label{eq:realmapshift}
  \{ n_1, n_2, \dots,n_i, n_j,\dots n_m\} \; \longrightarrow \; \{ n_1, n_2, \dots,n_i, n_q,  n_j,\dots n_m\} \,,
\end{equation}
and the full result for $\boldsymbol{R}_m$ is obtained by summing over all neighboring dipoles $[ij]$. In terms of color structure for real-emission dipoles with a kinematic prefactor $F_{ij}^q$, this corresponds to the replacement
\begin{align}
\sum_{(i j)} \bT_{i, L}^{\alpha} \bT_{j, R}^{\tilde{\alpha}} \,F_{ij}^q  &\to -N_c \sum_{[i j]} F_{ij}^q \,,
\label{eq:LCreal}
\end{align}
where the sum on the right only runs over neighboring dipoles $[ij]$, while the left sum $(ij)$ runs over all pairs of hard partons. Virtual dipoles are evaluated through
\begin{align}
\sum_{(i j)} \left(\bT_{i, L} \cdot \bT_{j, L}+\bT_{i, R} \cdot \bT_{j, R}\right) \,F_{ij}  &\to -2 N_c \sum_{[i j]} F_{ij} \,.
\label{eq:LCvirtual}
\end{align}

For double-real emissions $\bm{d}_m$ in $\bm{\Gamma}^{(2)}$, there are two possibilities of inserting the emissions $q$ and $r$ into an existing dipole $[ij]$. We can either map $[ij]$ to $[iqrj]$ or to $[irqj]$. Since the dipole coefficients $K_{ij;qr}^{(a)}$, $K_{ij;qr}^{(b)}$ and $K_{ij;qr}^{(c)}$ are symmetric under $q\leftrightarrow r$, both insertions give the same result. The only change from the \XS to the \LS scheme, is the replacement   $K^A_{ij;qr} \to \overline K^A_{ij;qr}$ in \eqref{eq:KbarFuncs} in the dipole terms. This quantity can be written as
\begin{equation}\label{eq:KbarAmanifest}
\overline K^A_{ij;qr} = K^A_{ij;qr} +4 W_{ij}^{qr} \ln\left(\frac{n_{ij} n_{qr}}{n_{ir}
   n_{jq}}\right)+4 W_{ij}^{rq} \ln\left(\frac{n_{ij} n_{qr}}{n_{iq}
   n_{jr}}\right)
\end{equation}
to make the scheme difference manifest.

The leading-color limit is more subtle for the terms with tripole color structure $\bm{T}_{ijk}^{\alpha\beta;\tilde{\alpha}\tilde{\beta}}$. Genuine tripoles $i \neq j \neq k$ contributions are suppressed in the large-$N_c$ limit (cf. Appendix~\ref{sec:TripoleCF}), but we get two-leg contributions, specifically
\begin{equation}\label{eq:tripoledouble}
 \sum_{i,j,k} \bm{T}_{ijk}^{\alpha\beta;\tilde{\alpha}\tilde{\beta}} K_{i j k; qr} +\mathrm{h.c.} \; \longrightarrow \;
 \frac{N_c^2}{4} \sum_{[ij]} \left(K_{iij;qr}-K_{iij;rq}-K_{jji;qr}+K_{jji;rq}\right) ,
\end{equation}
where we remapped the $[irqj]$ terms onto $[iqrj]$ so that the end result is a contribution of the form $[iqrj]$, multiplied by an anti-symmetric kinematic function. The remapping is possible since the soft function is symmetric under a swap of the two gluons. At leading color, this corresponds to the simultaneous swap of the direction vectors and color flows. Since the kinematic functions in both the \XS and \LS scheme fulfill the identities \eqref{eq_condition_to_drop_symmetrizers}, we can simplify the above result to 
\begin{equation}\label{eq:tripoledoubleB}
  \sum_{i,j,k} \bm{T}_{ijk}^{\alpha\beta;\tilde{\alpha}\tilde{\beta}} K_{i j k; qr} +\mathrm{h.c.} \; \longrightarrow \; \frac{N_c^2}{2}\sum_{[ij]}\left( K_{iij;qr}-K_{jji;qr} \right).
\end{equation} 
One way to derive this result is to start with a dipole amplitude, evaluate $\bm{T}_{ijk}^{\alpha\beta;\tilde{\alpha}\tilde{\beta}}$ explicitly and take the $N_c \to \infty$ limit. Alternatively one can rewrite the color structure in the color-flow basis as we do in Appendix \ref{sec:colorflow} and then read off the leading term, which is straightforward in this representation. Extracting the leading-color tripole contributions to the real-virtual part $\bm{r}_m$, we find
\begin{equation}\label{eq:tripolesingle}
-2 \sum_{i,j,k} \bm{T}_{ijk}^{\alpha;\tilde{\alpha}} K_{i j k; qr}  \; \longrightarrow \; N_c^2 \sum_{[ij]} \left(K_{iij;qr}+K_{jji;qr}\right),
\end{equation}
where $q$ is the real emission and $r$ the virtual correction. We note the relative sign of the two contributions is opposite to \eqref{eq:tripoledoubleB}.

Written in terms of the combination
\begin{equation}\label{eq:kijqrdef}
k_{ij;qr}= W_{ij}^q W_{ij}^r -W_{ij}^{qr} -W_{ij}^{rq}\,
\end{equation}
the explicit result for the combinations of kinematic functions in \eqref{eq:tripoledoubleB} and \eqref{eq:tripolesingle} in the \XS scheme reads
\begin{equation}
\label{eq:MijqrXS}
\begin{aligned}
   K_{iij;qr}-K_{jji;qr}
     & =-8 k_{ij;qr} \ln\!\left(\frac{n_{iq} n_{jr}}{n_{ir} n_{jq}}\right), \\
   K_{iij;qr}+K_{jji;qr}  & = 
-8 k_{ij;qr}\ln\!\left(\frac{n_{vq}^2 W_{ij}^q }{n_{vr}^2 W_{ij}^r }\right).
\end{aligned}   
\end{equation}
We note that both are antisymmetric under $q\leftrightarrow r$. In the \LS scheme, the combinations are 
\begin{equation}
\label{eq:MijqrLS}
\begin{aligned}
    \overline{K}_{iij;qr}-\overline{K}_{jji;qr}
    & =-8\,k_{ij;qr}\ln\!\left(\frac{n_{iq}n_{jr}}{n_{ir}n_{jq}}\right)
+16\,W^{qr}_{ij}\ln\!\left(\frac{n_{ij}n_{qr}}{n_{ir}n_{jq}}\right)-16\,W^{rq}_{ij}\ln\!\left(\frac{n_{ij}n_{qr}}{n_{iq}n_{jr}}\right) \\
\overline{K}_{iij;qr}+\overline{K}_{jji;qr}  & = 0\,.
\end{aligned}   
\end{equation}
The first term in the result for $\overline{K}_{iij;qr}-\overline{K}_{jji;qr}$ corresponds to the result in the \XS scheme, the other two arise through the scheme change. Interestingly, the real-emission combination in the second line vanishes in the \LS scheme. In this case, the scheme change terms cancel the contribution in the \XS scheme. 

Having listed the individual ingredients, we are now in a position to provide the full anomalous dimension in the large-$N_c$ limit, as it is implemented in the {\sc Marzili} shower~\cite{Becher:2023vrh}. We write the anomalous dimension as a sum of dipole contributions of the form
\begin{align}\label{eq:gamma1}
\bm{\Gamma}^{(1)}_{mn} &=\sum_{[ij]} \left[\bm{V}^{ij}_m \,\delta_{m,n} + \bm{R}^{ij}_m\, \delta_{m,n-1}\right],\\
\bm{\Gamma}^{(2)}_{mn} &=\sum_{[ij]} \left[\bm{v}^{ij}_m \,\delta_{m,n} + \bm{r}^{ij}_m\, \delta_{m,n-1} + \bm{d}^{\spac ij}_m\, \delta_{m,n-2}\right].
\label{eq:Gamma2mnLC}
\end{align}
As discussed above, in the one-loop case, the color structure reduces to a factor of $N_c$ and the anomalous dimension then reads
\begin{align}
\boldsymbol{R}^{ij}_m = &\,  4\spac N_c \spac W_{ij}^q\spac\thetain\!\left(n_q\right), & 
\boldsymbol{V}^{ij}_m = & \,-4\spac N_c  \int [\mathrm{d}^2\Omega_q]  \spac W_{ij}^q\,.
\label{eq:GmLC}
\end{align}
An insertion of the real-emission anomalous dimension maps the dipole $[ij]$ onto $[iqj]$, while the virtual part does not change the dipole structure. 

In terms of the ingredients derived above, the two-loop result for QCD takes the form
\begin{equation}\label{eq:gamma2}
\begin{aligned}
  \bm{d}^{\,ij}_m &= N_c \Big[2(C_A\,K^A_{ij;qr}+n_F T_F\,K^F_{ij;qr})+\tfrac{N_c}{2}\big(K_{iij;qr}-K_{jji;qr}\big)\Big]\thetain(n_q)\thetain(n_r)\,,
 \\
 \bm{r}^{\,ij}_m &= N_c\Bigg[N_c\Dqqq{r}\big(K_{iij;qr}+K_{jji;qr}\big)-2\Dqqq{r}\big(C_A K^A_{ij;qr}+n_FT_FK^F_{ij;qr}\big) \\
 &  \hspace{1.5cm} -2K_{ij;q}\Bigg]\thetain(n_q)\,,\\
 \bm{v}^{\,ij}_m &= 2N_c\Dqqq{q}K_{ij;q}\,.
\end{aligned}
\end{equation}
The double gluon emission terms in $ \bm{d}^{\,ij}_m$ map the dipole $[ij]$ onto $[iqrj]$, while the fermionic part splits the dipole $[ij]\to [i\bar{q}]\, [qj]$, where the anti-quark carries one momentum and the quark the other one. The real-virtual part $\bm{r}^{\,ij}_m$ maps the dipole $[ij]$ onto $[iqj]$. The purely virtual part $\bm{v}^{\,ij}_m$ leaves the dipole structure unchanged. An efficient way to implement the fermionic part into the MC shower is to treat the $\bar{q}q$ emission like a gluon double emission $[iqrj]$, but treat the formal $[qr]$ link as a non-splitting dipole. This can be achieved by setting $\bm{V}_m^{qr}=0$ in the shower algorithm so that the  $[qr]$ dipole is never selected for a new emission, see Section A in the supplemental material to \cite{Becher:2023vrh}.

To obtain the result for the anomalous dimension in the \LS scheme one simply replaces the kinematic functions in \eqref{eq:gamma2} by the functions in this scheme. The anomalous dimension simplifies in this scheme since $\overline{K}_{iij;qr}+\overline{K}_{jji;qr}=0$. In the \LS scheme, the function $ \overline{K}_{i j ; q } = - \frac{1}{2}\gamcuspm{1} W_{i j}^q$ and the two-loop virtual term is proportional to the one-loop anomalous dimension
\begin{equation}\label{eq:vmLC}
\overline{\bm{v}}_m = \frac{\gamcuspm{1}}{\gamcuspm{0}} \bm{V}_m\,.
\end{equation}
Together with the correction from the running coupling, these
cusp terms can be accounted for by modifying the shower evolution variable. At LL the shower evolution time is defined as
\begin{equation}\label{eq:showertime}
t = \int_{\mu_s}^{\mu_h} \frac{d\mu}{\mu} \frac{\alpha_s}{4\pi} = \frac{1}{2\beta_0} \ln \frac{\alpha(\mu_s)}{\alpha(\mu_h)}\,.
\end{equation}
Using the modified soft-emission coupling introduced by Catani, Marchesini and Webber~\cite{Catani:1990rr}
\begin{equation}\label{eq:alphaCKW}
\alpha_s^{\mathrm{CMW}} = \alpha_s\left(1 + \frac{\alpha_s}{4\pi} \frac{\gamcuspm{1}}{\gamcuspm{0}}  \right)
\end{equation}
and including the correction of the running coupling, one obtains a modified evolution time
\begin{equation}\label{eq:ttilde}
\tilde{t} = \int_{\mu_s}^{\mu_h} \frac{d\mu}{\mu} \frac{\alpha_s^{\mathrm{CMW}}(\mu)}{4\pi} = \frac{1}{2\beta_0} \left [\ln \frac{\alpha_s(\mu_s)}{\alpha_s(\mu_h)} + \frac{\alpha_s(\mu_h)-\alpha_s(\mu_s) }{4\pi} \left(\frac{\beta_1}{\beta_0} - \frac{\gamcuspm{1}}{\gamcuspm{0}}\right) \right] ,
\end{equation}
with
\begin{align}
\beta_1 &= \frac{34}{3}C_A^{\spac 2} - \frac{20}{3}C_AT_F \spac n_F-4\spac C_FT_F \spac n_F\,.
\end{align}
If one evolves with $\tilde{t}$, one can thus set $\overline{K}_{i j ; q }=0$ in the two-loop anomalous dimension. This modified evolution time is used in the {\sc Gnole} shower \cite{Banfi:2021xzn}.

\begin{figure}
    \centering
    \includegraphics[scale=0.8]{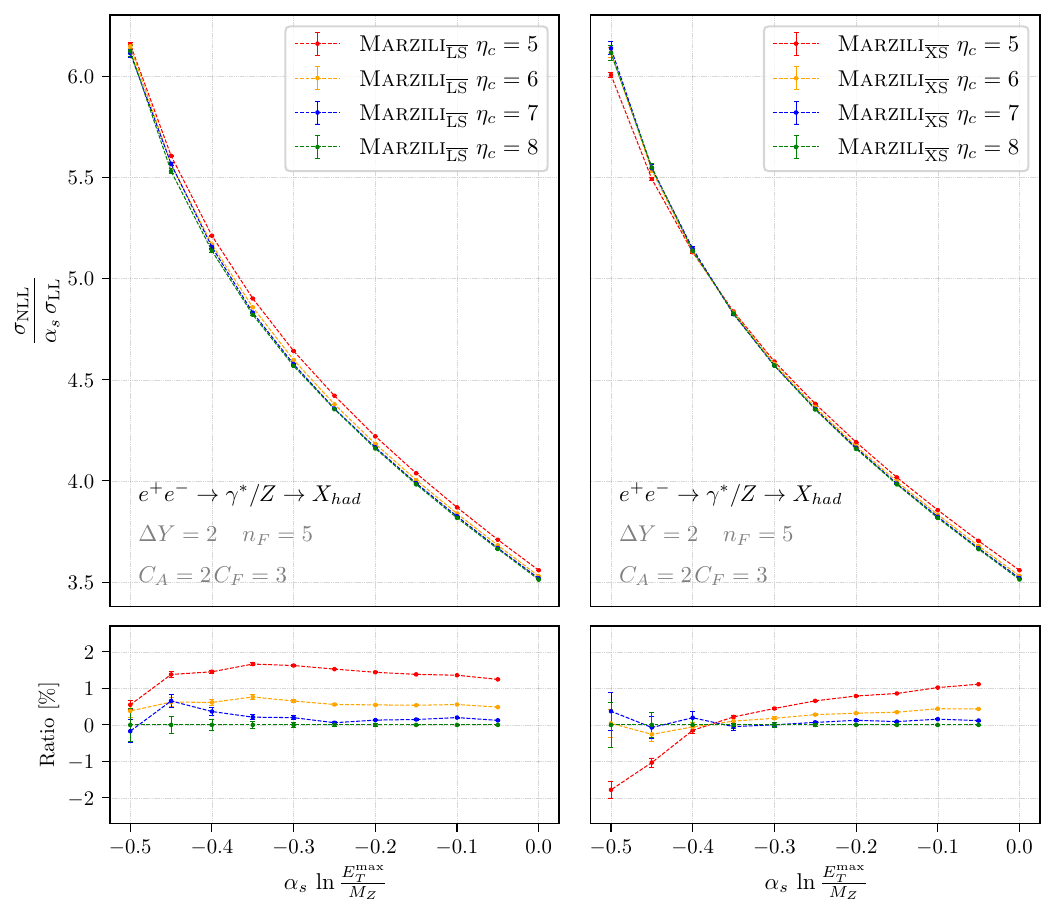}
    \caption{Dependence on the technical cutoff $\eta_c$ using different renormalization schemes.}
    \label{fig:etacut-study}
\end{figure}

The individual terms in the anomalous dimension contain divergences when the parton momenta $q$ and $r$ become collinear to the directions $n_i$ and $n_j$ or to one another. In the Monte Carlo implementation in {\sc Marzili}, the singularities in the individual pieces are regularized with a cutoff parameter $\eta_{c}$, which determines the rapidity cutoff in the dipole center-of-mass frame to
\begin{equation}\label{eq:etac}
y_{\max}= \ln  \!\left(\sqrt{1-\frac{M^2}{4}}+e^{\eta_c} \frac{M}{2}\right) ,
\end{equation}
with $M^2=2n_{ij}$. For a back-to-back dipole $y_{\max}=\eta_{c}$. In the two-loop anomalous dimension \eqref{eq:gamma2}, collinear singularities only arise when $n_q$ and $n_r$ become collinear to one another, which corresponds to an unresolved
splitting of a soft parent gluon, $g\to gg$ and $g\to q\bar q$, respectively. In this case {\sc Marzili} imposes a lab-frame cut
$\tan(\theta_{qr}/2)>e^{-\eta_c}$, where $\theta_{qr}$ is the opening
angle between $n_q$ and $n_r$ \cite{Schalch:2024ylx}.

Collinear safety guarantees that the limit $\eta_c \to \infty$ of the cross section exists. 
The anomalous dimensions \eqref{eq:gamma1} and \eqref{eq:gamma2} make collinear safety manifest since the same kinematic functions enter real and virtual terms with opposite signs, so that the singular regions cancel in the sum. The only exception are the terms $K_{iij;qr}-K_{jji;qr}$ and $K_{iij;qr}+K_{jji;qr}$ induced by the tripole functions $K_{ijk;qr}$. Both combinations are separately collinear finite, since they both are anti-symmetric under $q \leftrightarrow  r$ and finite in the limits where $q$ or $r$ become collinear to $n_i$ and $n_j$. To ensure an efficient cancellation of the collinear divergences, we evaluate the antisymmetric pieces by randomly generating two directions $n_q$ and $n_r$ and computing the difference between two showers for $[iqrj]$ and $[irqj]$. These tripole pieces are the only leading-color terms that distinguish the two color orderings; the symmetric terms weight them equally. In the absence of subsequent evolution, the two orderings are related by exchanging the emitted gluon momenta, and their difference therefore vanishes. Consequently, these terms were not visible in the fixed-order check performed in Ref.~\cite{Becher:2021urs}, where the two-loop insertion was not followed by further evolution. Subsequent emissions, however, resolve different daughter dipoles in the two flows and therefore generate different radiation patterns, making the color-flow assignment essential for the all-order evolution. Although the daughter spins are summed over, the kernels are not averaged
over the splitting azimuth: their full angular dependence therefore retains
the correlation between the parent-gluon polarization and the splitting plane, a physical four-dimensional effect.
\begin{figure}
    \centering
    \begin{subfigure}[b]{1.0\textwidth}
        \centering
        \includegraphics[width=0.95\textwidth]{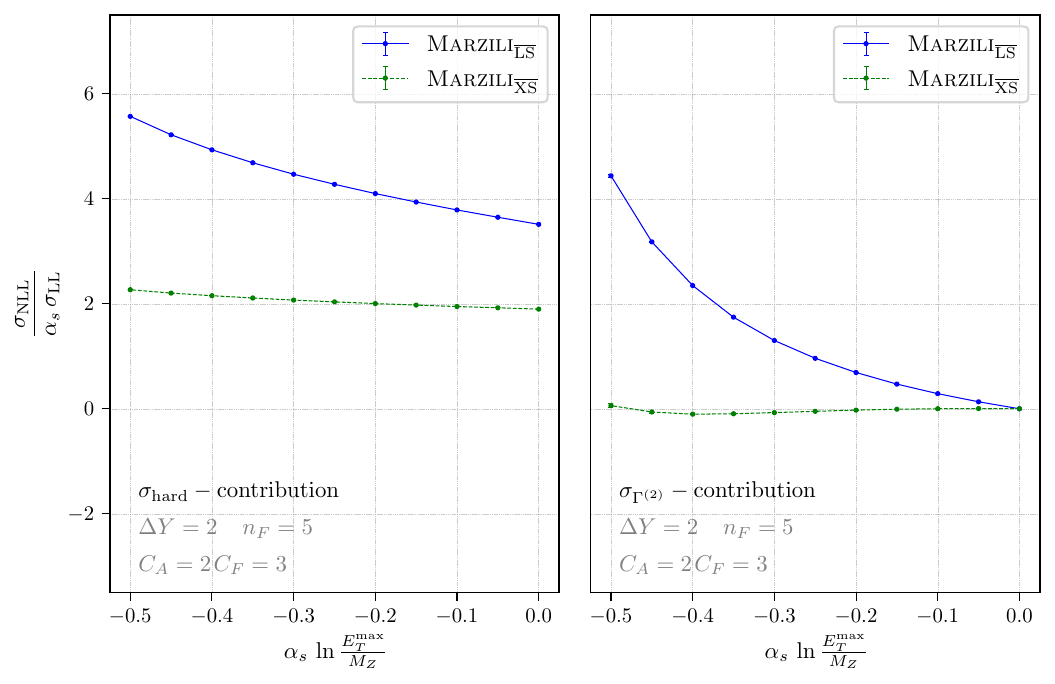}
        \caption{Hard and two-loop contribution.}
        \label{fig:top}
    \end{subfigure}
    \textcolor{white}{space}
    \begin{subfigure}[b]{1.0\textwidth}
        \centering
        \includegraphics[width=0.95\textwidth]{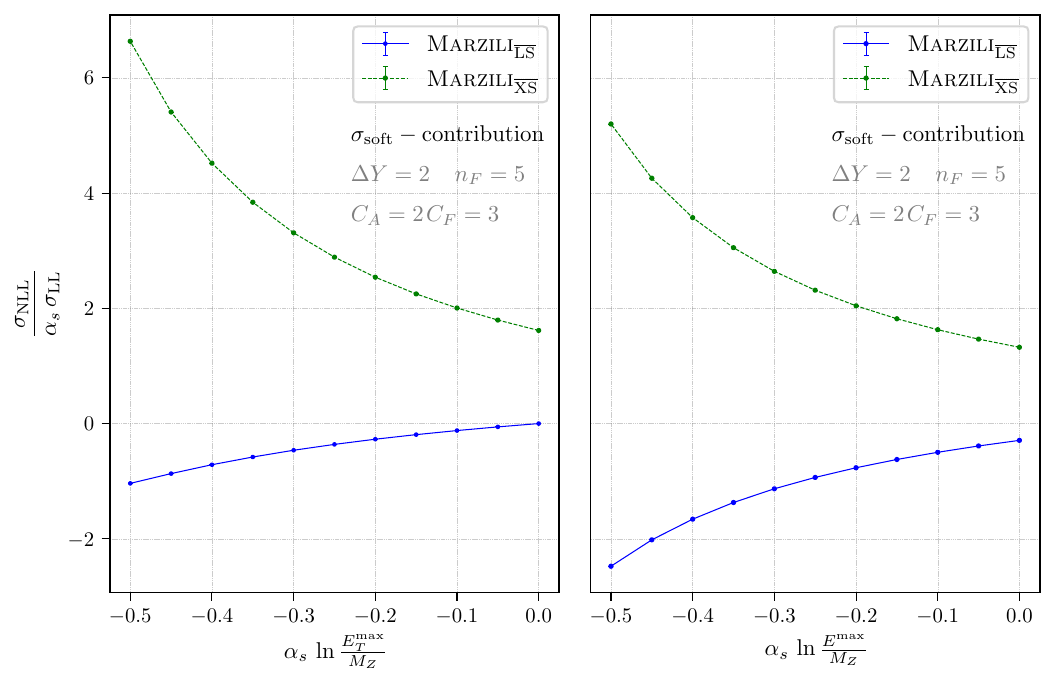}
        \caption{Soft contribution.}
        \label{fig:bottom}
    \end{subfigure}
    
    \caption{Different contributions to the transverse energy $E_T^{\mathrm{max}}$ and the energy $E^{\mathrm{max}}$ flow in a rapidity slice. We use a technical cutoff $\eta_c=8$.}
    \label{fig:stacked}
\end{figure}

\begin{figure}
    \centering
    \includegraphics[scale=0.8]{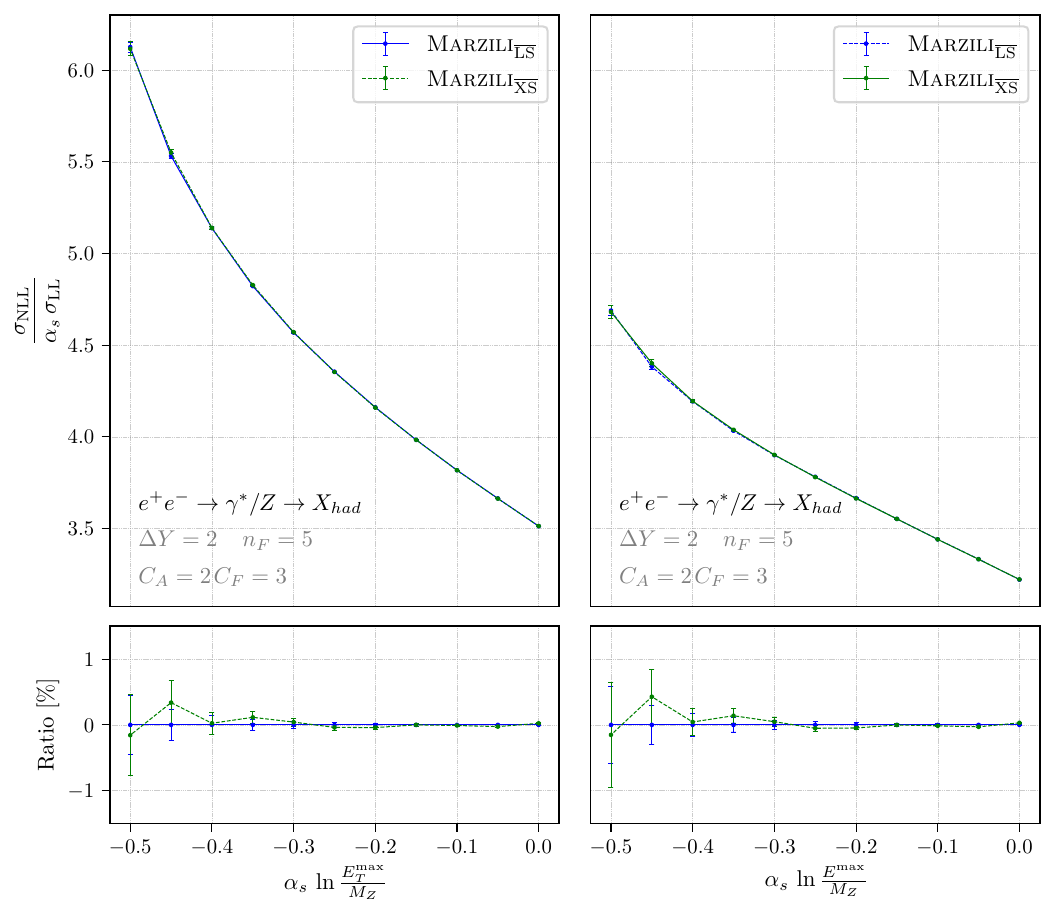}
    \caption{Comparison between different renormalization schemes for both the transverse energy $E_T^{\mathrm{max}}$ and the energy $E^{\mathrm{max}}$ flow in a rapidity slice. We use a technical cutoff $\eta_c=8$.}
    \label{fig:total}
\end{figure}

As we have discussed in detail in Section \ref{sec:soft_function}, the standard \MS scheme would require the introduction of an additional $(d-4)$-dimensional angle, which was not taken into account in \cite{Becher:2023vrh}. For this reason, the $M_{ij;qr}$ function used in this reference is not appropriate. We have now implemented the anomalous dimensions both for the \XS and the \LS scheme in {\sc Marzili} and have computed the gap fraction for $e^+ e^- \to 2\, \text{jets}$ with cones around the thrust axis. The cone size is chosen such that the region outside the cones covers the rapidity $\Delta Y = 2$ relative to the thrust axis, and we impose that the total momentum transverse to the thrust axis is below $E_T^{\mathrm{max}}$ in the region outside the cones. To verify cutoff independence we have performed runs with successively lower cutoffs, i.e.,\ larger values of $\eta_c$, see Figure \ref{fig:etacut-study}. The plots in this figure show the NLL correction $\sigma_{\mathrm{NLL}}$, normalized to $\alpha_s(M_Z) \,\sigma_{\mathrm{LL}}$, as \cite{FerrarioRavasio:2023kyg}. We observe that for cutoffs $\eta_{c}>6$, the obtained results for the NLL correction agree within about half a percent. Raising $\eta_c$ increases computer time since it exponentially slows down the evolution in the shower time $t \approx \alpha_s \ln(M_Z/E_{T}^{\mathrm{max}})$ so that many more emissions are needed for a given value  of $E_T^{\mathrm{max}}$. Very large values of $\eta_c$ can also lead to numerical precision loss due to the strong cancellations between real and virtual terms.

Working with two different renormalization schemes provides a highly non-trivial consistency check on our computation, since the individual contributions depend on the scheme choice, while their sum is physical and therefore scheme independent. The scheme dependence is illustrated in Figure \ref{fig:stacked}. In the upper part of the figure, we show the contributions from the NLO terms in the hard function, as well as the contribution from the two-loop anomalous dimension. Individually, these contributions strongly differ between the two schemes, but the difference must vanish in the sum with the soft correction shown in the lower plots in Figure \ref{fig:stacked}. For the soft function, we show results both for a veto on transverse energy $E_{T}$ and the energy $E$. Only the soft function NLO correction is sensitive to the exact definition of the veto, the other parts are only sensitive to the choice of the soft scale $\mu_s$. In Figure \ref{fig:total}, we add up the different ingredients, including the running-coupling contribution, which is common to both schemes and is therefore not shown separately in Figure~\ref{fig:stacked}, and observe that the total NLL corrections agree between the two schemes, for both $E_{T}$ and $E$. This agreement is highly non-trivial: although the hard, two-loop-anomalous-dimension, and soft contributions differ, their sum agrees within numerical accuracy for two distinct veto measurements. This provides a stringent validation of the scheme-change relations and their implementation within the parton shower framework {\sc Marzili}.

\section{Energy dependence and clustering effects}\label{sec:clustering}

A factorization theorem which accounts for clustering effects in non-global jet observables was obtained in \cite{Becher:2023znt}. An important insight in \cite{Becher:2023znt} was that to LL accuracy, clustering effects necessitate only a small modification of the one-loop anomalous dimension used for fixed-cone cross sections introduced in \cite{Becher:2015hka}. In fact, the virtual part $\Vclm{m}$ is the same as for the fixed-cone case, and the real part, generating a particle with momentum $q$, only attaches a delta function $\delta(x_{q})$, where $x_{q}$ is the energy fraction of the emitted particle. Explicitly, we have
\begin{equation}\label{eq:Rclm}
\Vclm{m} = \Vm{m}\, , \qquad  \Rclm{m} = \delta^{({m+1})}(x_q)  \Rm{m}\,.
\end{equation}
The delta function ensures that the one-loop anomalous dimension produces strongly-ordered soft gluons. The energy ordering of the gluons needs to be considered when evaluating the gap constraint on the jets. In \cite{Becher:2023znt}, the clustering sequence for a set of strongly-ordered emissions was analyzed for different  jet finding algorithms. When performing the energy-fraction integrations after multiple emissions, it is important to integrate the softer energies first. To keep track of the ordering, we have added a superscript to the delta function in \eqref{eq:Rclm}, where higher numbers indicate softer particles. In a product
\begin{equation}
    \delta^{({m+1})}(x_q)\, \delta^{({m+2})}(x_r)
\end{equation}
we thus need to integrate first over $x_r$, then over $x_q$. In our previous paper \cite{Becher:2023znt}, we worked with energy-ordered hard functions and ratios $z_{m+2} = x_{m+2}/ x_{m+1}$ which avoids this issue but is inconvenient in the context of NLL evolution.

\begin{figure}[t]
  \centering
  \begin{minipage}{0.48\textwidth}
    \centering
    \includegraphics[width=\linewidth]{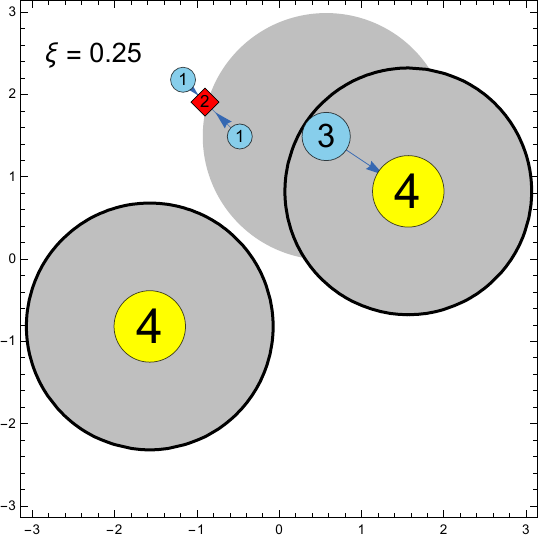}
  \end{minipage}
  \hfill
  \begin{minipage}{0.48\textwidth}
    \centering
    \includegraphics[width=\linewidth]{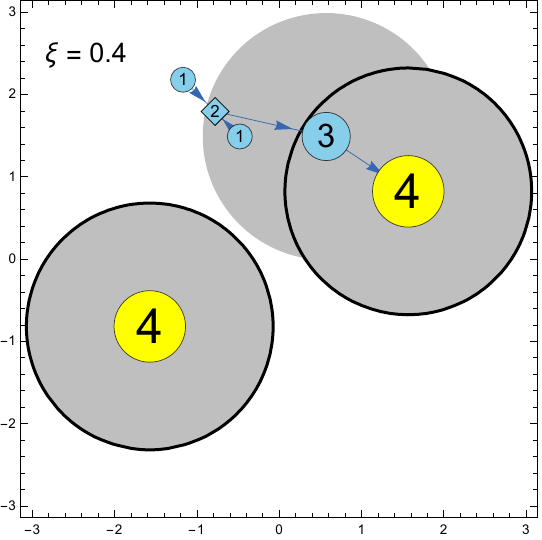}
  \end{minipage}
  \caption{We show two events that can be generated in the shower through the term $\Hfun{2}\otimes_x\RLSclm{2}\hat{\otimes}_x\dLSclm{3}$ in the NLL treatment of dijet production at lepton colliders. Here the veto energy is put on all jets different from the two hardest jets and the clustering is performed with the $k_T$ algorithm using the distance $d_{ij}^2=\min\left( E_i^2,E_j^2 \right)\left( \Delta\eta^2+\Delta \phi^2 \right)$. The figure shows the direction of the partons in the $\eta-\phi$ plane as circles---note that all partons have the same directions in both events. Pseudojet directions that do not align with any parton are displayed as diamonds. The size of the circles and diamonds indicates the energy scale at which the emissions happen---partons live at three well-separated energy scales corresponding to $\Hfun{2}$, $\RLSclm{2}$ and $\dLSclm{3}$ for this particular situation. The two hardest jets are colored yellow and any additional jet, leading to a vetoed event, is colored red. The arrows indicate where a parton ends up after a clustering step and the numbers indicate the step in the $k_T$ clustering when the respective pseudojet gets clustered with another pseudojet or becomes a jet. \textbf{Left:} The partons generated in $\dLSclm{2}$ cluster into a jet that gets vetoed. \textbf{Right:} The clustered pseudojet is also clustered with the parton generated in $\RLSclm{2}$ and thus this event is not vetoed. \label{fig:xidepktclustering}}
\end{figure}

Throughout this section, the anomalous dimensions are functions of the energy variables of the generated partons, and the corresponding integration measures are part of the convolution $\otimes_x$ defined in \eqref{eq:otimesz}, not of the anomalous dimensions themselves. At NLL, following the logic in \cite{Becher:2023znt}, the double-virtual anomalous dimension $\vm{m}$ is again the same as for the fixed-cone case,  and the real-virtual anomalous dimension $\rclm{m}$ can be obtained by multiplying the fixed-cone anomalous dimension $\rm{m}$ by $\delta^{(m+1)}(x_q)$. However, at NLL, two soft particles with momenta $q$ and $r$ can be emitted at comparable energies $E_q$ and $E_r$. The contribution to the anomalous dimension where both of these emissions are real is contained in the contribution $\dclm{m}$ to the two-loop anomalous dimension. For general clustering algorithms, the veto region depends non-trivially upon the energy fraction 
\begin{equation}
  \label{eq_xi_definition}
  \xi = \frac{E_q}{E_q+E_r}=\frac{x_{q}}{x_{q}+x_{r}}\, 
\end{equation}
of the two emissions. This means that the double-real emission clustering anomalous dimension, $\dLSclm{m}$, needs to retain the dependence on $\xi$ to evaluate general veto constraints. For example, there are phase space contributions where the emissions $q$ and $r$ with fixed directions $n_q$ and $n_r$ always cluster together in the $k_T$ clustering algorithm. The direction of the clustered pseudojet depends on the value of $\xi$, and for some values of $\xi$, the pseudojet may become a jet while it might be clustered into another jet for other values of $\xi$. This situation is illustrated in Figure~\ref{fig:xidepktclustering}. In the fixed-cone anomalous dimension $\dm{m}$ calculated in \cite{Becher:2021urs}, the integral over the energy fraction $\xi$ was already performed; thus, we cannot obtain $\dLSclm{m}$ from $\dm{m}$. In the next subsection, we derive $\dLSclm{m}$ following the same steps as \cite{Becher:2021urs} but keeping the result differential in $\xi$.

\subsection{Derivation of the Two-Loop Clustering Anomalous Dimension}
\label{sec:GammaClustLSScheme}
The double-real emission anomalous dimensions for clustering cross sections $\dclm{m}$ can be derived from the double-soft IR poles of $\Hfun{m+2}$, where two emissions with momenta $q$ and $r$ become soft simultaneously. 
Instead of the individual energy fractions $x_{q}$ and $x_{r}$, we characterize the energies of the pair by the fraction $\xi$ defined in \eqref{eq_xi_definition} and by
\begin{equation}
  \label{eq_z_definition}
  x=\frac{2\left(E_q+E_r\right)}{Q}=x_{q}+x_{r}\, ,
\end{equation}
so that $x_{q}=x\,\xi$, $x_{r}=x\,\barxi$ and $\rd x_{q}\rd x_{r}=x\,\rd x\,\rd\xi$. In the following, $\Hfun{m+2}\,\rd x\,\rd\xi$ denotes the hard function differential in $x$ and $\xi$.
In the limit where $E_q$ and $E_r$  become soft simultaneously, we  get poles that have to be reproduced by
\footnote{$\Hfun{m+2}$ also has poles multiplying $\Hfun{m+1}$, but those are related to the limit where only the last emission is soft.}
\begin{equation}\label{eq:Hmp2factorized}
  \begin{aligned}
      \Hfun{m+2}\,\rd x\,\rd\xi &\approx \left(\frac{\alpha_s}{4 \pi}\right)^2\Bigl[\frac{1}{8 \epsilon^2} \Hfun{m}  \frac{1}{2}\left( \Rclm{m} (q) \Rclm{m+1}(r) +(\;(q,\alpha,\tilde{\alpha})\leftrightarrow (r,\beta,\tilde{\beta})\;)\right)\\
       &\quad\quad-\frac{1}{4 \epsilon} \Hfun{m}  \dclm{m}\Bigr]\rd x\,\rd\xi\, .
  \end{aligned}
\end{equation}
Here we explicitly symmetrized the iterated one-loop anomalous dimension to make it easier to compare to the left-hand side of the equation, which is naturally symmetric in the two gluons. In the first term in  \eqref{eq:Hmp2factorized}, the one-loop anomalous dimensions impose the energy ordering $x_r \ll x_q$  through $\delta^{(m+1)}(x_q) \delta^{(m+2)}(x_r)$; the ordering is reversed in the second term.

If the two soft emissions are both gluons, we can approximate the matrix element as
\begin{equation}\label{eq:Mggamp}
  \left|\mathcal{M}_{m+2}^{g g}(\{\underline{p}, q, r\})\right\rangle=\left[\frac{1}{2}\!\left\{\boldsymbol{J}^{\mu, \alpha}(q), \boldsymbol{J}^{\nu, \beta}(r)\right\}+g_s^2 i f^{\alpha \beta c} \sum_i \bT_i^c J_i^{\mu \nu}(q, r)\right]\! \varepsilon_{q,\mu}^{* } \varepsilon_{r,\nu}^{* }\left|\mathcal{M}_m(\{\underline{p}\})\right\rangle ,
\end{equation}
where
\begin{equation}\label{eq:currentdef}
 \begin{aligned}
  \boldsymbol{J}^{\mu, \alpha}(q)&=g_s\sum_{i=1}^m \bT_i^\alpha J_i^\mu(q)\, , \quad  J_i^\mu(q)=\frac{ n_i^\mu}{n_i \cdot q}\,,\\
   J_i^{\mu \nu}(q, r)&=\frac{1}{2 n_i \cdot(q+r)}\left[\frac{n_i^\mu n_i^\nu}{n_i \cdot q}-\frac{n_i^\mu n_i^\nu}{n_i \cdot r}+\frac{g^{\mu \nu} n_i \cdot(r-q)+2 n_i^\mu q^\nu-2 n_i^\nu r^\mu}{q \cdot r}\right] .
 \end{aligned}
\end{equation}
In the following, we will also use the notation
\begin{equation}\label{eq:currenttilde}
  \tilde{J}_i^{\mu}(q)=E_q J_i^\mu(q)\, ,\quad \tilde{J}_i^{\mu \nu}(q, r)=E_q E_r J_i^{\mu \nu}(q, r)\, .
\end{equation}
These rescaled currents depend only on the directions $n_q$, $n_r$ and, in the case of the two-particle current, on the energy fraction $\xi$.

Using the soft approximation for the amplitude and working in the lab frame with $v^\mu = (1, \vec{0})$, we obtain
\begin{equation}
  \label{eq:Hmp2Poles1}
  \begin{aligned}
   & \Hfun{m+2} \,\rd x\,\rd\xi\approx  \frac{1}{2}
 \rd x \rd \xi  g_s^4 \Hfun{m} \int \frac{d E_q E_q^{d-5}}{\tilde{c}^\epsilon(2 \pi)^2} \int \frac{d E_r E_r^{d-5}}{\tilde{c}^\epsilon(2 \pi)^2} 
 \delta\!\left( x-\frac{2\left(E_q +E_r\right) }{Q} \right)\\
&\Biggl[ \tilde{J}_i^\mu(q) \tilde{J}_{j}^\nu(r) \tilde{J}_{k,\mu}(q) \tilde{J}_{l,\nu}(r) \bm{Q}_{ijkl}^{\alpha\beta;\tilde{\alpha}\tilde{\beta}}  \\
&-\tJ_k^{\mu\nu}(q,r)\tJ_{i,\mu}(q) \tJ_{j,\nu}(r) \left(\bm{T}_{ijk}^{\alpha\beta;\tilde{\alpha}\tilde{\beta}}+\overline{\bm{T}}_{ijk}^{\alpha\beta;\tilde{\alpha}\tilde{\beta}}\right)\\
&+ \tJ_i^{\mu\nu}(q,r)\tJ_{j,{\mu\nu}}(q,r) \bm{D}_{ij}^{A,\alpha\beta;\tilde{\alpha}\tilde{\beta}}\Biggr]\thetain\!\left(n_q,n_r,\xi \right) \theta(\Lambda-E_q-E_r)\delta\!\left( \xi-\frac{E_q}{E_{q}+E_r} \right) ,
  \end{aligned}
\end{equation}
 where we use a UV cutoff $\Lambda$ to isolate the IR divergences as in  \cite{Becher:2021urs} and a summation over repeated indices is understood. In the double-soft limit, we can furthermore approximate \mbox{$\delta\!\left( x-\frac{2\left(E_q+E_r\right)}{Q} \right)\approx \delta(x)$}. The energy integrals can now be performed, and we obtain
\begin{equation}
  \label{eq:Energ Integration}
  \begin{aligned}
   & g_s^4\int \frac{d E_q E_q^{d-5}}{\tilde{c}^\epsilon(2 \pi)^2} \int \frac{d E_r E_r^{d-5}}{\tilde{c}^\epsilon(2 \pi)^2 }\delta\left( \xi-\frac{E_q }{E_{\mathrm{tot}}} \right) \theta(\Lambda-E_{\mathrm{tot}}) =
    &-\left( \frac{\alpha_s}{4\pi} \right)^2  \frac{4}{\epsilon }  \left(\frac{\mu }{2 \Lambda }\right)^{4 \epsilon } (\xi \barxi)^{-1-2 \epsilon }\, ,
    \end{aligned}
\end{equation}
with $E_{\mathrm{tot}}=E_q +E_r$.
To expand the integrand in $\epsilon$, we define the distributional expansion
\begin{equation}
  \label{eq:xiBarxiExpansion}
  (\xi \barxi)^{-1-2\epsilon}\equiv\frac{-1}{2\epsilon}\delta(\xi\barxi)+\frac{1}{(\xi \barxi)_+}-2\epsilon \left(\frac{\ln(\xi\barxi)}{\xi\barxi}\right)_+ +O(\epsilon^2)\,,
\end{equation}
with $\delta(\xi\barxi)=\delta(\xi)+\delta(\barxi)$. For a trivial test function, this expression integrates to 
\begin{equation}
  \label{eq:xiBarxiIntegral}
    \int_0^1 d\xi  (\xi \barxi)^{-1-2\epsilon}=\frac{\Gamma(-2\epsilon)^2}{\Gamma(-4\epsilon)}=-\frac{1}{\epsilon}+\frac{2\pi^2}{3}\epsilon +O(\epsilon^2)\, .
\end{equation}
The action of the plus-distributions on the right-hand side of \eqref{eq:xiBarxiExpansion} on a non-trivial test function $f(\xi)$ is obtained by expanding the subtracted expression
\begin{equation}\label{eq:xidist}
    \int_0^1 \! d\xi \left[ (\xi \barxi)^{-1-2\epsilon} f(\xi)  -  \xi^{-1-2\epsilon} f(0) -   \barxi^{-1-2\epsilon} f(1)\right ]
\end{equation} 
in $\epsilon$. For the first term, this gives
\begin{equation}\label{eq:xibarxiplus}
  \int_0^1 \! d\xi  \frac{1}{(\xi \barxi)_+} f(\xi) =   \int_0^1 \! d\xi \left[\frac{1}{\xi \barxi} f(\xi)  -  \frac{1}{\xi} f(0) -   \frac{1}{\barxi} f(1)\right ] .
\end{equation} 
 Even though we only need the poles of \eqref{eq:Hmp2Poles1}, we still need the expansion \eqref{eq:xiBarxiExpansion} up to order $\epsilon$, because \eqref{eq:Hmp2Poles1} contains hidden poles that are generated in the angular integrals when $q$ and $r$ become collinear. 

To proceed further, we now evaluate the Lorentz contractions of the eikonal currents in \eqref{eq:Hmp2Poles1}. These contractions define a set of functions which encode the kinematical dependence of the anomalous dimension. The simplest structure is the term involving four different legs
\begin{equation}
\label{eq:CurrentContractions4}
\tJ_i^\mu(q) \tJ_{j}^\nu(r) \tJ_{k,\mu}(q) \tJ_{l,\nu}(r) = W_{ik}^q W_{jl}^r\,.
\end{equation}
Next, we consider the three-leg terms

\begin{align}\label{eq:CurrentContractions3}
\tJ_k^{\mu\nu}(q,r)\tJ_{i,\mu}(q) \tJ_{j,\nu}(r)=\,&\frac{\xi  \barxi \left(n_{k r}-n_{k q}\right) }{\xi  n_{k q}+\barxi n_{k r}} k_{i j k; qr}\nonumber \\
& \; +\xi  W_{i k}^q W_{j q}^r-\barxi W_{i r}^q W_{j k}^r+\frac{1}{2} (\barxi-\xi ) W_{i k}^q W_{j k}^r+\frac{1}{2} (\barxi-\xi ) W_{i j}^q W_{j q}^r  \, ,
\end{align}
where the function $k_{i j k; qr}$ was defined in  \eqref{eq:kijkqr}. We note that the final term in this result is independent of $n_k$. Due to color conservation, it gives a vanishing contribution to \eqref{eq:Hmp2Poles1} when the sum over $k$ is performed and can therefore be dropped. The most complicated kinematic function arises in the two-leg terms generated by contracting the non-abelian part of the two-emission current with itself. We obtain
\begin{align}\label{eq:CurrentContractions2}
&\tJ_i^{\mu\nu}(q,r)\tJ_{j,{\mu\nu}}(q,r)-\frac{1}{2}\tJ_i^{\mu\nu}(q,r)\tJ_{i,{\mu\nu}}(q,r)-\frac{1}{2}\tJ_j^{\mu\nu}(q,r)\tJ_{j,{\mu\nu}}(q,r)=\nonumber\\
&\hspace{1cm}\frac{\xi  \barxi n_{i j}}{\left(\xi  n_{i q}+\barxi n_{i r}\right) \left(\xi  n_{j q}+\barxi n_{j r}\right)}\Biggl(-\frac{1}{2} n_{i j} \left(\frac{1}{n_{i q} n_{j r}}+\frac{1}{n_{i r} n_{j q}}\right)+\frac{\frac{n_{i r} n_{j q}}{n_{i q} n_{j r}}+\frac{n_{i q} n_{j r}}{n_{i r} n_{j q}}+6}{2 n_{q r}}\nonumber\\
&\hspace{3cm}-\frac{\xi  \barxi (1-\epsilon ) \left(n_{i r} n_{j q}-n_{i q} n_{j r}\right){}^2}{n_{i j} n_{q r}^2 \left(\xi  n_{i q}+\barxi n_{i r}\right) \left(\xi  n_{j q}+\barxi n_{j r}\right)}\Biggr)+\frac{1}{4} W_{i j}^q W_{i j}^r-\frac{1}{2}\left(W_{i j}^{q r}+W_{i j}^{r q}\right) .
\end{align}
In evaluating this contribution, we have subtracted two single-leg terms which vanish by color conservation. This subtraction has the dual benefit of eliminating power divergences in the collinear limit $n_q\to n_r$ as well as removing the unphysical term $i=j$.

We now subtract the poles in \eqref{eq:Hmp2Poles1} that are generated by iterating the
 one-loop anomalous dimension. They read 
 \begin{equation}
  \label{eq:Hmp2RtimesRpoles}
  \begin{aligned}
    &\frac{1}{8 \epsilon^2} \Hfun{m}  \frac{1}{2}\left( \Rclm{m} (q) \Rclm{m+1}(r) +(q\leftrightarrow r)\right)\rd x\,\rd\xi=\frac{1}{\epsilon^2}\rd x\rd\xi \left( \delta(\xi)+\delta(\barxi) \right)\delta^{(m+1)}(x)\,\Hfun{m}\\
&\Biggl[\left(  W_{i k}^q W_{j l}^r\right)\bm{Q}_{ijkl}^{\alpha\beta;\tilde{\alpha}\tilde{\beta}}  \\
&- \left( \xi W_{i k}^q W_{j q}^r-\bar\xi W_{i r}^q W_{j k}^r+\frac{1}{2} (\bar\xi-\xi) W_{i k}^q W_{j k}^r\right) \left(\bm{T}_{ijk}^{\alpha\beta;\tilde{\alpha}\tilde{\beta}}+\overline{\bm{T}}_{ijk}^{\alpha\beta;\tilde{\alpha}\tilde{\beta}}\right)\\
&+ \frac{1}{2}\left( \frac{1}{2} W_{i j}^q W_{i j}^r-W_{i j}^{q r}-W_{i j}^{r q} \right)\bm{D}_{ij}^{A,\alpha\beta;\tilde{\alpha}\tilde{\beta}}\Biggr]\thetain\!\left(n_q,n_r,\xi \right) .
  \end{aligned}
 \end{equation}
  Due to the overall delta functions, we get $\xi=0$ or $\xi=1$ in each term, but we have inserted prefactors of $\xi$ and $\bar{\xi}$ in the third line to make it manifest, that this expression reproduces the strongly-ordered limits ($\xi \to 0$ or $\barxi\to 0$) of the general expression \eqref{eq:Hmp2Poles1}, after inserting the explicit expression \eqref{eq:CurrentContractions4}, \eqref{eq:CurrentContractions3} and \eqref{eq:CurrentContractions2} for the contracted currents. This means that all terms in \eqref{eq:Hmp2Poles1} multiplying $\delta(\xi)+\delta(\barxi)$ are canceled by the iterated real anomalous dimension. The remaining terms are part of the anomalous dimension $\dclm{m}$ for double-real emissions. 
To encode their kinematical dependence, we now introduce $\xi$-dependent kinematic functions, obtained by multiplying the products of currents with the plus distribution from \eqref{eq:xidist}. Specifically, we define
\begin{equation}
\label{eq:dKclustDefs}
  \begin{aligned}
   K_{i j k l; qr}(\xi)&=8\frac{1}{(\xi \barxi)_+} W_{i k}^q W_{j l}^r \,,\\
       K_{i j k; qr}(\xi)&=-8\frac{1}{(\xi \barxi)_+} \tJ_k^{\mu\nu}\tJ_{i,\mu} \tJ_{j,\nu} \,,\\
    &=2 \left( \delta (\xi ) +\delta (\barxi ) \right) k_{i j k ; q r}\ln\!\left(\frac{n_{k q} }{n_{k r} }\right)-4  (\delta (\xi )-\delta (\barxi)) W_{i r}^q W_{j k}^r \ln\!\left(\frac{n_{k q} }{n_{k r} }\right)\\
    &\quad +4  \left(\frac{n_{k r}}{\xi  \left(\xi  n_{k q}+\barxi n_{k r}\right)}\right)_+\!\!\! \left(2 W_{i r}^q W_{j k}^r-W_{i k}^q W_{j k}^r-W_{i j}^{q r}\right)-\left((i,q,\xi)\leftrightarrow (j,r,\barxi)\right) ,\\
     K^A_{i j ; qr}(\xi)&=-4\frac{1}{(\xi \barxi)_+} \left( \tJ_i^{\mu\nu} \tJ_{j,\mu\nu}- \frac{1}{2}\tJ_i^{\mu\nu} \tJ_{i,\mu\nu}- \frac{1}{2}\tJ_j^{\mu\nu} \tJ_{j,\mu\nu}\right)\Biggr \rvert_{\epsilon=0}\\
     &=\frac{-4 n_{i j}}{\left(\xi  n_{i q}+\barxi n_{i r}\right) \left(\xi  n_{j q}+\barxi n_{j r}\right)}\Biggl(-\frac{1}{2} n_{i j} \left(\frac{1}{n_{i q} n_{j r}}+\frac{1}{n_{i r} n_{j q}}\right)+\frac{\frac{n_{i r} n_{j q}}{n_{i q} n_{j r}}+\frac{n_{i q} n_{j r}}{n_{i r} n_{j q}}+6}{2 n_{q r}}\\
&\quad-\frac{\xi  \barxi \left(n_{i r} n_{j q}-n_{i q} n_{j r}\right){}^2}{n_{i j} n_{q r}^2 \left(\xi  n_{i q}+\barxi n_{i r}\right) \left(\xi  n_{j q}+\barxi n_{j r}\right)}\Biggr)-\frac{1}{(\xi \barxi)_+}\left(  W_{i j}^q W_{i j}^r-2 W_{i j}^{q r}-2 W_{i j}^{r q} \right) .
  \end{aligned}
\end{equation}
 In $K_{ijk;qr}(\xi)$ we dropped the $n_k$-independent term $\frac12(\barxi-\xi)W^q_{ij}W^r_{jq}$ of \eqref{eq:CurrentContractions3}, which vanishes after the sum over $k$ by color conservation. These definitions hold in the lab frame, where $v=(1,0,0,0)$, so that the $\xi$-integral of $K_{ijk;qr}(\xi)$ reproduces $K_{ijk;qr}$ of \eqref{eq:Gamma2XStripole} with $n_{vq}=n_{vr}=1$. We will explain at the end of the section how this result can be generalized to a general frame. These functions satisfy the equations
\begin{equation}\label{eq:xisumrules}
  \begin{aligned}
    \int \rd    \xi K_{i j k l; qr}(\xi) &=0 \,,\\
    \int \rd \xi  K_{i j k; qr}(\xi) &=K_{i j k; qr} \,,\\
    \int \rd \xi  K^A_{i j ; qr}(\xi) &=K^A_{i j ; qr}\, .
  \end{aligned}
\end{equation}
Setting $\mu=2\Lambda$ we find
  \begin{align}\label{eq:Hmp2Final}
    &\Hfun{m+2}\,\rd x\,\rd\xi\approx -\frac{ \delta^{(m+1)}(x)\rd x \rd \xi}{4\epsilon }\Hfun{m} \left(\frac{\alpha_s}{4 \pi}\right)^{2}\Biggl[K_{i j k l; qr}(\xi)\bm{Q}_{ijkl}^{\alpha\beta;\tilde{\alpha}\tilde{\beta}} \nonumber  \\
&+ K_{i j k; qr}(\xi)\left(\bm{T}_{ijk}^{\alpha\beta;\tilde{\alpha}\tilde{\beta}}+\overline{\bm{T}}_{ijk}^{\alpha\beta;\tilde{\alpha}\tilde{\beta}}\right)  \nonumber\\
&+\left( -2 K^A_{i j ; qr}(\xi) \right)  \bm{D}_{ij}^{A,\alpha\beta;\tilde{\alpha}\tilde{\beta}} \Biggr]\thetain\!\left(n_q,n_r,\xi \right)\\
&+\frac{\rd x\,\rd\xi}{8 \epsilon^2} \Hfun{m}  \frac{1}{2}\left( \Rclm{m} (q) \Rclm{m+1}(r) +(q\leftrightarrow r)\right)  \nonumber\\
&+\delta^{(m+1)}(x)\rd x \rd \xi \frac{2}{\epsilon}\Hfun{m} \left(\frac{67}{9}+\frac{c_R}{6}-\frac{2 \pi^2}{3}\right) \bm{D}_{ij}^{A,\alpha\beta;\tilde{\alpha}\tilde{\beta}} W_{i j}^q \thetain\!\left(n_q\right) \delta\left(n_q-n_r\right) ,  \nonumber
\end{align}
where $c_R$ tracks the number of gluon polarizations and has to be set to $c_R=1$ for conventional dimensional regularization. The terms in the last line in \eqref{eq:Hmp2Final} are generated from terms in \eqref{eq:Hmp2Poles1} which have finite energy integrations but suffer from collinear singularities in the angular integrals over $n_q$ and $n_r$. To capture these poles, $\Hfun{m+2}$ has to be expanded as a distribution in the directions $n_q$ and $n_r$. The additional poles produced are proportional to $\delta(n_q-n_r)$, which is defined such that
\begin{equation}\label{eq:collineardelta}
  \Dqqq{r}f(n_r)\delta(n_q-n_r)=f(n_q)
\end{equation} 
for any smooth test function $f(n_r)$. 
Since the clustering algorithm used to define the veto, as well as the veto energy $\Eout$, are collinear-safe, all $\xi$ dependence drops out in the collinear limit. Thus, the $\xi$ integral can be performed once and for all in the terms multiplying $\delta(n_q-n_r)$. The extraction of these collinear terms is therefore the same as for the case of fixed angular constraints discussed in detail in \cite{Becher:2021urs}. 

From the result \eqref{eq:Hmp2Final} we can immediately read off the gluon-gluon emission contribution to the anomalous dimension
\begin{equation}
  \label{eq:dclAm}
\begin{aligned}
   \dclAm{m}&= \delta^{(m+1)}(x)\Biggl[ K_{i j k l; qr}(\xi)\bm{Q}_{ijkl}^{\alpha\beta;\tilde{\alpha}\tilde{\beta}}  \\
  &\quad+ K_{i j k; qr}(\xi)\left(\bm{T}_{ijk}^{\alpha\beta;\tilde{\alpha}\tilde{\beta}}+\overline{\bm{T}}_{ijk}^{\alpha\beta;\tilde{\alpha}\tilde{\beta}}\right)\\
  &\quad+\left( -2K^A_{i j ; qr}(\xi) \right) \bm{D}_{ij}^{A,\alpha\beta;\tilde{\alpha}\tilde{\beta}} \Biggr]\thetain\!\left(n_q,n_r,\xi \right)\\
  &\quad-8\delta^{(m+1)}(x)\left(\frac{67}{9}+\frac{c_R}{6}-\frac{2 \pi^2}{3}\right) \bm{D}_{ij}^{A,\alpha\beta;\tilde{\alpha}\tilde{\beta}} W_{i j}^q \thetain\!\left(n_q\right) \delta\left(n_q-n_r\right) .
\end{aligned}
\end{equation}

We now turn to the emission of a soft $q\bar{q}$ pair, for which the approximate matrix element reads
\begin{equation}\label{eq:Mqqbamp}
  \left|\mathcal{M}_{m+2}^{q\bar{q} }(\{\underline{p}, q, r\})\right\rangle=g_s \bar{u}(q)\, \gamma^\mu \,(\bm{t}^c)^{\alpha}_{\phantom{\alpha}\beta}\,  v(r) \frac{1}{2 q \cdot r} \boldsymbol{J}_{\mu, c}(q+r)\left|\mathcal{M}_m(\{\underline{p}\})\right\rangle\, ,
\end{equation}
where $\alpha$ and $\beta$ are the color indices of the outgoing quark and anti-quark. In analogy to the gluon case, we define
\begin{equation}\label{eq:KFxi}
\begin{aligned}
  K^F_{i j ; qr}(\xi)&=-8\frac{1}{\xi \barxi} \frac{1}{\left( n_q\cdot n_r \right)^2}\left( -g_{\mu \nu} q \cdot r+q_\mu r_\nu+r_\mu q_\nu \right)\left( J_i^{\mu} J_{j}^\nu-\frac{1}{2}J_i^{\mu} J_{i}^\nu -\frac{1}{2}J_j^{\mu} J_{j}^\nu\right)\\
     &=\frac{8}{n_{q r}^2 \left(\xi  n_{i q}+\barxi n_{i r}\right) \left(\xi  n_{j q}+\barxi n_{j r}\right)}\Biggl(n_{i j} n_{q r}-n_{i r} n_{j q}-n_{i q} n_{j r}\\
     &\qquad+ \frac{n_{i q} n_{i r} \left(\xi  n_{j q}+\barxi n_{j r}\right)}{\xi  n_{i q}+\barxi n_{i r}}+\frac{n_{j q} n_{j r} \left(\xi  n_{i q}+\barxi n_{i r}\right)}{\xi  n_{j q}+\barxi n_{j r}}\Biggr)\, ,
\end{aligned}
\end{equation}
where the currents in the first line are evaluated with the total momentum $q+r$. Treating poles appearing due to the collinear singularity $q\to r$ as in the gluon case, we find 
\begin{equation}
  \label{eq:dclFm}
      \dclFm{m}= \delta^{(m+1)}(x) \Biggl[-2 n_F K^F_{i j ; qr}(\xi)\,\thetain\!\left(n_q,n_r,\xi \right)+\frac{208}{9}\, n_F W_{i j}^q\, \thetain\!\left(n_q\right)\delta(n_q-n_r)\Biggr] \bm{D}_{ij}^{F,\alpha\beta;\tilde{\alpha}\tilde{\beta}}\, .
\end{equation}
Finally, we consider the emission of two scalar particles which we assume to be in the adjoint representation for concreteness. The matrix element reads

\begin{equation}\label{eq:Mssamp}
  \left|\mathcal{M}_{m+2}^{s s}(\{\underline{p}, q, r\})\right\rangle=i g_s f^{\alpha \beta c} \frac{r^\mu-q^\mu}{2 q \cdot r} \boldsymbol{J}_{\mu, c}(q+r)\left|\mathcal{M}_m(\{\underline{p}\})\right\rangle\, .
\end{equation}
We then define
\begin{equation}\label{eq:KSxi}
\begin{aligned}
  K^S_{i j ; qr}(\xi)&=-2\frac{1}{\xi \barxi}\frac{1}{ n_{qr}^2} 
  \left(q_\mu-r_\mu \right) \left(q_\nu-r_\nu \right) 
  \left( J_i^{\mu} J_{j}^\nu-\frac{1}{2}J_i^{\mu} J_{i}^\nu -\frac{1}{2}J_j^{\mu} J_{j}^\nu\right)\\
     &= \frac{4 \xi  \barxi  \left(n_{i r} n_{j q}-n_{i q} n_{j r}\right)^2}{n_{q r}^2 \left(\xi  n_{i q}+\barxi n_{i r}\right)^2 \left(\xi  n_{j q}+\barxi n_{j r}\right)^2}\, .
\end{aligned}
\end{equation}
Treating the poles appearing due to the collinear singularity $q\to r$ as in the gluon case, we find
\begin{equation}
  \label{eq:dclSm}
    \dclSm{m}= \delta^{(m+1)}(x)\Biggl[-2 n_S K^S_{ {i j}; qr}(\xi)\,\thetain\!\left(n_q,n_r,\xi \right)+\frac{40}{9}\, n_S W_{ij}^q\, \thetain\!\left(n_q\right)\delta(n_q-n_r)\Biggr] \bm{D}_{ij}^{A,\alpha\beta;\tilde{\alpha}\tilde{\beta}}\, ,
\end{equation}
where $n_S$ is the number of complex scalar fields.

To finalize the anomalous dimension, we note that we can use color coherence to move the terms multiplying $\delta(n_q-n_r)$ to the real-virtual anomalous dimension. In fact, the terms multiplying $\delta(n_q-n_r)$ in \eqref{eq:dclAm}, \eqref{eq:dclFm}, and \eqref{eq:dclSm} are the same ones that were also found for the soft anomalous dimension in \cite{Becher:2021urs}, i.e., we move over the same terms as in \cite{Becher:2021urs}, and we find the final result 
\begin{equation}
  \label{eq:Gam2XSClust}
  \begin{aligned}
    \dclm{m}&=\delta^{(m+1)}(x) \Biggl[K_{i j k l; qr}(\xi)\bm{Q}_{ijkl}^{\alpha\beta;\tilde{\alpha}\tilde{\beta}}\\
   &\quad+K_{i j k; qr}(\xi) \left(\bm{T}_{ijk}^{\alpha\beta;\tilde{\alpha}\tilde{\beta}}+\overline{\bm{T}}_{ijk}^{\alpha\beta;\tilde{\alpha}\tilde{\beta}}\right)\\
   &\quad-2\left(  K_{ {i j}; qr}^A(\xi)+n_S K_{ {i j}; qr}^S(\xi)\right)\bm{D}_{ij}^{A,\alpha\beta;\tilde{\alpha}\tilde{\beta}}-2 n_F K_{i j; qr}^F(\xi) \bm{D}_{ij}^{F,\alpha\beta;\tilde{\alpha}\tilde{\beta}}\Biggr] \,,\\   \rclm{m}&=\rm{m}\delta^{(m+1)}(x_q) \,,\\
    \vclm{m}&=\vm{m}\, .
  \end{aligned}
\end{equation}
This is the result in the lab frame; the calculation can be repeated in a Lorentz-invariant way in the usual way by introducing an energy direction vector $v$, see Appendix~\ref{sec:lorentz} for the transformation properties of the relevant building blocks. This would change the energy integrals in \eqref{eq:Energ Integration} effectively leading to the replacement
\begin{equation}\label{eq:framefactor}
  (\xi \barxi)^{-1-2\epsilon}\to   (\xi \barxi)^{-1-2\epsilon}\left(\frac{n_{vq} n_{vr}}{\left( \xi n_{vq} + \barxi n_{vr} \right)^2}\right)^{-2\epsilon}\, .
\end{equation}
The terms in \eqref{eq:Hmp2RtimesRpoles} however do not come with this extra factor. This leads to a mismatch in the pieces multiplying $\delta(\xi)+\delta(\barxi)$ in \eqref{eq:Hmp2Poles1} and \eqref{eq:Hmp2RtimesRpoles}, introducing a $v-$dependent contribution in the $\frac{1}{\epsilon}$ pole and thus in the anomalous dimension:
\begin{equation}\label{eq:clustlogshift}
  \dclm{m}\to \dclm{m} + \frac{1}{2}\left( \Rclm{m} (q)\hat{\otimes}_x \Rclm{m+1}(r) +(q\leftrightarrow r)\right)\ln\!\left(  \frac{n_{vq} n_{vr}}{\left( \xi n_{vq} + \barxi n_{vr} \right)^2}\right) .
\end{equation}
The total effect of this change can also be summarized in the replacement rules 
\begin{equation}\label{eq:framerules}
  \begin{aligned}
   \left( \frac{1}{ \xi \barxi }\right)_+ &\to \left( \frac{1}{\xi \barxi }\right)_++\ln\!\left(  \frac{n_{vq} }{n_{vr} }\right) \left(\delta(\xi)-\delta(\barxi)\right) ,\\
    \ln\!\left(\frac{n_{k q} }{n_{k r} }\right)&\to \ln\!\left(\frac{n_{k q} n_{vr}}{n_{k r}n_{vq} }\right) ,
  \end{aligned}
\end{equation}
that have to be applied to the lab-frame result \eqref{eq:Gam2XSClust}.

We can get a frame-independent  anomalous dimension that does not depend on $v$ by applying the scheme change of Section \ref{sec:schemechange} to the above results. The clustering anomalous dimensions in the \LS scheme are unchanged at one-loop level
\begin{equation}
  \label{eq:Gam1LSClust}
  \begin{aligned}   \RLSclm{m}&=\RLSm{m}\delta^{(m+1)}(x_q) \, ,\\
    \VLSclm{m}&=\VLSm{m}\,,
  \end{aligned}
\end{equation}
while the two-loop result takes the form
\begin{equation}
  \label{eq:Gam2LSClust}
  \begin{aligned}
   \dLSclm{m} &=\delta^{(m+1)}(x)\Biggl[\overline{K}_{i j k l; qr}(\xi)\bm{Q}_{ijkl}^{\alpha\beta;\tilde{\alpha}\tilde{\beta}}
   +\overline{K}_{i j k; qr}(\xi) \left(\bm{T}_{ijk}^{\alpha\beta;\tilde{\alpha}\tilde{\beta}}+\overline{\bm{T}}_{ijk}^{\alpha\beta;\tilde{\alpha}\tilde{\beta}}\right)\\
   &\quad -2\left(  \overline{K}_{ {i j}; qr}^A(\xi)+n_S K_{ {i j}; qr}^S(\xi)\right)\bm{D}_{ij}^{A,\alpha\beta;\tilde{\alpha}\tilde{\beta}}-2 n_F K_{i j; qr}^F(\xi) \bm{D}_{ij}^{F,\alpha\beta;\tilde{\alpha}\tilde{\beta}}\Biggr] \,,\\
   \rLSclm{m}&=\rLSm{m}\delta^{(m+1)}(x_q) \,,\\
    \vLSclm{m}&=\vLSm{m}\, ,
  \end{aligned}
\end{equation}
with coefficient functions 
\begin{equation}\label{eq:KbarijklClust}
  \begin{aligned}
\overline{K}_{i j k l; qr}(\xi)&=4W_{i k}^q W_{j l}^r \left(\left( \delta \left(\barxi\right)-\delta(\xi) \right) \ln\!\left(\frac{n_{i k} n_{j r} n_{l r}}{n_{i q} n_{j l} n_{k q}}\right)+2 \left(\frac{1}{\xi \barxi}\right)_+\right),
  \end{aligned}
\end{equation}
  \begin{equation}\label{eq:KbarijkClust}
  \begin{aligned}
    \overline{K}_{i j k; qr}(\xi) &=  \Biggl[\delta(\xi\barxi)\, k_{i j k; q r} \log \left(\frac{n_{{ik}} n_{{jr}} n_{{kq}}}{n_{{iq}} n_{{jk}} n_{{kr}}}\right)+\delta(\xi\barxi) \Biggr(W_{i j}^q W_{j q}^r \log \left(\frac{n_{{ir}} n_{{jk}} n_{{kq}}}{n_{{ik}} n_{{jq}} n_{{kr}}}\right)\\
    &\hspace{2.5cm} -2 W_{i r}^q W_{j k}^r \log \left(\frac{n_{{ir}} n_{{kq}}}{n_{{ik}} n_{{qr}}}\right)+2 W_{j k}^q W_{j q}^r \log \left(\frac{n_{{jq}} n_{{kr}}}{n_{{jk}} n_{{qr}}}\right)\Biggr)\\
    &\quad+4 \left(\frac{n_{{kr}}}{\xi \left( \bar{\xi } n_{{kr}}+\xi n_{{kq}}\right)}\right)_+ \left(W_{i r}^q W_{j k}^r+W_{i k}^q W_{j q}^r-W_{i k}^q W_{j k}^r-W_{i j}^q W_{j q}^r\right)\\
    &\quad-4 W_{i k}^q W_{j q}^r \left(\frac{1}{2} \left(\delta (\xi )-\delta (\bar{\xi })\right) \log \left(\frac{n_{{iq}} n_{{jq}} n_{{kq}}}{n_{{ik}} n_{{jr}} n_{{qr}}}\right)+\left(\frac{1}{\xi  \bar{\xi }}\right)_+\right)\Biggr] \\
    &\quad -\left[(i,q,\xi)\leftrightarrow (j,r,\barxi)\right] ,
  \end{aligned}
\end{equation}
 \begin{equation}\label{eq:KbarAClust}
  \begin{aligned}
    \overline{K}_{i j; qr}^{A}(\xi) &=2  \delta(\xi\barxi) \left(W_{i j}^q W_{i q}^r \ln\!\left(\frac{n_{i j} n_{q r}}{n_{i q} n_{j r}}\right)+W_{i j}^r W_{i r}^q \ln\!\left(\frac{n_{i j} n_{q r}}{n_{i r} n_{j q}}\right)\right)\\
&\quad+\frac{2 }{\left(\xi  n_{i q}+\barxi n_{i r}\right) \left(\xi  n_{j q}+\barxi n_{j r}\right)}\Biggl(\frac{2 \xi  \barxi \left(n_{i r} n_{j q}-n_{i q} n_{j r}\right){}^2}{n_{q r}^2 \left(\xi  n_{i q}+\barxi n_{i r}\right) \left(\xi  n_{j q}+\barxi n_{j r}\right)} \\
&\hspace{1cm} \quad+n_{i j}^2 \left(\frac{1}{n_{i q} n_{j r}}+\frac{1}{n_{i r} n_{j q}}\right) -\frac{n_{i j} \left(\frac{n_{i r} n_{j q}}{n_{i q} n_{j r}}+\frac{n_{i q} n_{j r}}{n_{i r} n_{j q}}+6\right)}{n_{q r}}\Biggr)\\
&\quad+\left(2 W_{i j}^{q r}+2 W_{i j}^{r q}-W_{i j}^q W_{i j}^r\right) \left(\frac{1}{2} (\delta (\xi )-\delta (\barxi)) \ln\!\left(\frac{n_{i q} n_{j q}}{n_{i r} n_{j r}}\right)+\left(\frac{1}{\xi  \barxi}\right)_+\right) ,
  \end{aligned}
\end{equation}
which are entirely built from the frame-independent building blocks listed in Appendix~\ref{sec:lorentz}.

\subsection{Leading-color clustering anomalous dimension}
\label{sec:GammaClustLC}
In the large-$N_c$ limit, the clustering anomalous dimension can be organized in the same way as the fixed-cone result of Section~\ref{sec:largeNc}. The dipole and tripole color structures reduce exactly as in \eqref{eq:tripoledoubleB}, with the kinematic functions replaced by their $\xi$-dependent counterparts of \eqref{eq:dKclustDefs}. For the tripole, the identity $K_{iij;rq}(\barxi)=-K_{iij;qr}(\xi)$, which follows from the antisymmetry of \eqref{eq:dKclustDefs} under $(i,q,\xi)\leftrightarrow(j,r,\barxi)$, plays the role of \eqref{eq_condition_to_drop_symmetrizers} and yields the same combination $K_{iij;qr}(\xi)-K_{jji;qr}(\xi)$ as in \eqref{eq:tripoledoubleB}. The dipole functions $K^A_{ij;qr}(\xi)$ and $K^F_{ij;qr}(\xi)$ are symmetric under this exchange.

The only new ingredient is the quadrupole structure $\bm{Q}_{ijkl}^{\alpha\beta;\tilde{\alpha}\tilde{\beta}}$, which is absent for fixed angular constraints because its coefficient integrates to zero, see \eqref{eq:xisumrules}, but contributes at leading color in the clustering case. Its action in the color-flow basis is worked out in \eqref{eq_Q_LC}. As usual there are two ways to insert the gluon pair into an existing dipole $[ij]$: $[ij]\to[iqrj]$ or $[ij]\to[irqj]$. The new feature of the quadrupole correlations is that the gluons can also be inserted into two \emph{different} dipoles, $[ij]\,[kl]\to[iqj]\,[krl]$. For the kinematic function $K_{ijkl;qr}(\xi)=8\,W_{ik}^qW_{jl}^r/(\xi\barxi)_+$ the combinations in \eqref{eq_Q_LC} collapse to products of dipole functions, and we find
\begin{equation}\label{eq:quadrupoleLC}
\sum_{i,j,k,l} K_{ijkl;qr}(\xi)\,\bm{Q}_{ijkl}^{\alpha\beta;\tilde{\alpha}\tilde{\beta}} \to \frac{N_c^2}{(\xi\barxi)_+} \bigg[\sum_{[ij]} W_{ij}^q W_{ij}^r\,\Big([iqrj]+[irqj]\Big)
+\!\! \sum_{[ij]\neq[kl]} \!\!\! 8 W_{ij}^q W_{kl}^r\,[iqj]\,[krl] \bigg] ,
\end{equation}
where we have indicated the color flow generated by each term, and the last sum runs over ordered pairs of distinct dipoles, with the gluon $q$ being inserted into $[ij]$ and $r$ into $[kl]$. Distinct dipoles may share a leg (adjacent dipoles included). Remapping $[irqj]$ onto $[iqrj]$ as in Section~\ref{sec:largeNc}, which now involves the simultaneous exchange $q\leftrightarrow r$ and $\xi\leftrightarrow\barxi$, doubles the color-connected term since $1/(\xi\barxi)_+$ is symmetric. 

The leading-color clustering anomalous dimension therefore contains, in addition to the single-dipole terms of \eqref{eq:Gamma2mnLC}, a contribution acting on two dipoles at once,
\begin{equation}\label{eq:gammaClustLC}
\bm{\Gamma}^{cl(2)}_{mn} =\sum_{[ij]} \left[\bm{v}^{ij}_m \,\delta_{m,n} + \bm{r}^{ij}_m\,\delta^{(m+1)}(x)\, \delta_{m,n-1} + \bm{d}^{\spac cl,ij}_m\, \delta_{m,n-2}\right] + \sum_{[ij]\neq[kl]} \bm{d}^{\spac cl,[ij][kl]}_m\, \delta_{m,n-2}\,,
\end{equation}
with $\bm{v}^{ij}_m$ and $\bm{r}^{ij}_m$ as in \eqref{eq:gamma2}. In the \XS scheme, the double-real terms read
\begin{equation}\label{eq:dclLC}
\begin{aligned}
  \bm{d}^{\,cl,ij}_m &= \delta^{(m+1)}(x)\,N_c \Big[2\big(C_A K^A_{ij;qr}(\xi)+n_F\,T_F\,K^F_{ij;qr}(\xi)\big) +\tfrac{N_c}{2}\big(K_{iij;qr}(\xi)-K_{jji;qr}(\xi)\big)\\
  &\hspace{1.5cm}+\frac{2N_c}{(\xi\barxi)_+}\,W_{ij}^q W_{ij}^r\Big]\thetain(n_q,n_r,\xi)\,,\\
  \bm{d}^{\,cl,[ij][kl]}_m &= \delta^{(m+1)}(x)\,\frac{8N_c^2}{(\xi\barxi)_+}\, W_{ij}^q W_{kl}^r\;\thetain(n_q,n_r,\xi)\,.
\end{aligned}
\end{equation}
The first line maps the dipole $[ij]$ onto $[iqrj]$ (the fermionic term onto $[i\bar q]\,[qj]$), exactly as in \eqref{eq:gamma2}, while the second line maps the pair of dipoles $[ij]\,[kl]$ onto $[iqj]\,[krl]$. The single-dipole term integrates to the fixed-cone result \eqref{eq:gamma2}, $\int d\xi\,\bm{d}^{\,cl,ij}_m=\bm{d}^{\,ij}_m\,\delta^{(m+1)}(x)$, by virtue of the sum rules \eqref{eq:xisumrules}, the quadrupole contribution integrates to zero. The two-dipole term is a pure clustering effect: it describes two independent emissions into different dipoles, whose energy sharing is correlated by the clustering constraint $\thetain(n_q,n_r,\xi)$, and it vanishes identically whenever the constraint is $\xi$-independent. 

In the \LS scheme the difference to the \XS result is supported at the endpoints $\xi=0,1$, where it is given by the scheme-change commutator $\delta\bm{R}_m\bm{R}_{m+1}-\bm{R}_m\delta\bm{R}_{m+1}$ of \eqref{eq:DeltaGamma2} evaluated for the respective strongly-ordered configuration. At leading color the commutator is elementary, $\delta\bm R^{ij}_m=4N_c\,\delta W^q_{ij}\, \delta^{(m+1)}(x_q)$ with $\delta W^q_{ij}=\frac12W^q_{ij}\ln(2\hat{W}^q_{ij})$, and we find
\begin{equation}\label{eq:dclLCLS}
\begin{aligned}
  \overline{\bm{d}}^{\,cl,ij}_m &= \bm{d}^{\,cl,ij}_m + \delta^{(m+1)}(x)\, 8N_c^2\, W^{qr}_{ij}\left[\delta(\barxi)\ln\!\left(\frac{W^q_{ij}}{W^r_{jq}}\right)+\delta(\xi)\ln\!\left(\frac{W^r_{ij}}{W^q_{ir}}\right)\right]\thetain(n_q,n_r,\xi)\,,\\
  \overline{\bm{d}}^{\,cl,[ij][kl]}_m &= \delta^{(m+1)}(x)\,N_c^2\, W_{ij}^q W_{kl}^r\left[\frac{8}{(\xi\barxi)_+}+4\left(\delta(\barxi)-\delta(\xi)\right)\ln\!\left(\frac{W^q_{ij}}{W^r_{kl}}\right)\right]\thetain(n_q,n_r,\xi)\,,
\end{aligned}
\end{equation}
where $W^{qr}_{ij}=W^q_{ij}W^r_{jq}=W^r_{ij}W^q_{ir}$ according to \eqref{eq:Wqr}, and $\bm{d}^{\,cl,ij}_m$ is the \XS result \eqref{eq:dclLC} in the frame $v=(1,\vec 0)$. 
The logarithms in \eqref{eq:dclLCLS} are ratios of the one-loop dipole functions of the two emissions; in a general frame they become $\ln(n_{vq}^2 W^q_{ij}/(n_{vr}^2 W^r_{jq}))$ etc., and the $v$-dependence cancels against that of the \XS terms.

The same expressions can also be obtained by applying the color-flow rules directly to the \LS functions $\overline{K}_{ijkl;qr}(\xi)$, $\overline{K}_{ijk;qr}(\xi)$ and $\overline{K}^A_{ij;qr}(\xi)$ given in \eqref{eq:KbarijklClust}, \eqref{eq:KbarijkClust}, and \eqref{eq:KbarAClust}. 
The real-virtual and double-virtual terms are the fixed-cone \LS results, $\overline{\bm{r}}^{\,ij}_m\delta^{(m+1)}(x_q)$ and $\overline{\bm{v}}^{\,ij}_m$, for which $\overline{K}_{iij;qr}+\overline{K}_{jji;qr}=0$ and $\overline{K}_{ij;q}=-\frac12\gamcuspm{1}W^q_{ij}$ as discussed in Section~\ref{sec:largeNc}. 

In practice, the last term in \eqref{eq:gammaClustLC} is implemented by selecting a pair of distinct dipoles, inserting $q$ into the first and $r$ into the second, and assigning the weight $\bm d_m^{\,cl,[ij][kl]}$. This is not a new correlated double-soft matrix element: its kernel factorizes as $W_{ij}^q W_{kl}^r$ for two independent emissions and becomes non-trivial through the $\xi$-dependent clustering constraint. A logarithmically accurate shower, such as {\sc PanScales}, must generate this structure through its real dipole kernels and associated Sudakov evolution \cite{Dasgupta:2020fwr,FerrarioRavasio:2023kyg}. In the present RG-evolution framework, where an anomalous-dimension insertion normally acts on a single dipole, the two-dipole term must instead be included explicitly. One interesting application of our framework will be generating NLL benchmarks to test logarithmically accurate parton-showers with observables sensitive to clustering effects. 

\section{Conclusion}

In perturbative computations in dimensional regularization, minimal subtraction schemes, in particular the \MS scheme, are standard. This scheme is also commonly adopted in low-energy effective theories such as SCET to perform resummations based on RG evolution. In this paper, we have shown that \MS is problematic in the context of non-global observables: in minimal subtraction, the non-global angular constraints are subtracted in the physical, four-dimensional subspace rather than in the $d$-dimensional space used to regularize the theory. The resulting mismatch between the angular constraints leads to a dependence on unphysical, evanescent angles, which already arises in the one-loop soft function relevant for NLL resummation. Since the coupled RG equations are solved iteratively in a parton-shower framework, the dependence on evanescent angles would therefore have to be retained at every step of the evolution, forcing the shower to handle higher-dimensional kinematics and distribution-valued angular kernels.

In this work, we have shown that these complications can be avoided by using local subtraction schemes in which the non-global constraint is imposed in the full $d$-dimensional phase space. Such schemes subtract the poles everywhere, even in the evanescent regions of phase space, and avoid the difficulties of $\MS$: they only involve physical angles in the renormalized factorization theorem, so that the evolution can be formulated in the four-dimensional physical space. The simplest local scheme subtracts the entire angular integral in $d$ dimension. This \XS scheme was used in the original derivation of the two-loop anomalous dimension but has the disadvantage that the result is frame dependent. We discussed a set of alternative renormalization schemes for NLL resummation, based on modifying the soft dipole emitter and have shown in detail how to perform the change between different schemes at the two-loop level. By performing the subtraction with a suitably modified dipole, one can cancel the frame-dependent terms and obtain a frame-invariant two-loop result for the anomalous dimension. The simplest possibility is the \LS scheme introduced by Caron-Huot. In the process, we have shed light on the color and Lorentz structure of the two-loop anomalous dimension; in the appendices its most important properties are summarized. 

To test these results at the level of a physical prediction, we derived the leading-color form of the two-loop anomalous dimension and implemented both the \XS and \LS schemes in the {\sc Marzili} parton-shower framework. The separate contributions from the one-loop hard and soft functions and from the two-loop anomalous dimension differ between the two schemes. Once all ingredients are combined, the total NLL corrections agree within numerical accuracy, imposing both a veto on the energy and transverse-energy of the radiation within a rapidity slice. This agreement provides a stringent check on the internal consistency and numerical stability of the implementation. 

We have then derived the two-loop anomalous dimension for clustering algorithms, treating the relative energies of partons emitted at commensurate angles in a fully differential way. This anomalous dimension retains the full color structure, so that ours is the only framework that would allow the systematic inclusion of subleading color effects for clustering cross sections at NLL accuracy. We have provided the clustering anomalous dimension also in the \LS scheme, written in a form that makes its frame independence manifest, and  have derived its leading-color limit. We highlight that even at leading color, the clustering anomalous dimension contains new quadrupole color structures, which require the insertion of gluons into two different dipoles and are absent in the fixed-cone case. We also give the representation of the full anomalous dimension in the color-flow basis, which can form the basis for including sub-leading color effects in a parton shower. 

In future work, the clustering anomalous dimension can serve as a key ingredient for the NLL resummation of non-global observables whose definition involves a non-trivial clustering sequence. Potential applications include jet-mass distributions defined with sequential-recombination algorithms and rapidity-gap vetoes in vector-boson-fusion measurements, where the interplay of jet clustering and fiducial cuts generate non-global sensitivity. Until a full-color NLL shower becomes available, the resulting predictions can be supplemented with estimates of subleading-color effects derived from fixed-order expansions to high loop orders.

\begin{appendix}

\subsection*{Acknowledgments}
We thank Pier Francesco Monni, Simon Plätzer, Gavin Salam, and Rudi Rahn for useful discussions. The authors would like to thank the Mainz Institute of Theoretical Physics (MITP) for hospitality and support during the workshop \emph{Factorization and Evolution in Full Color}. This research has received funding from the Swiss National Science Foundation (SNSF) under grant 200021\_219377.

\pagebreak
 
\section{Symmetry properties of the color structures}
\label{sec:CSsymmetries}

In this appendix we collect the symmetry properties of the color structures defined in \eqref{eq:dipoles}, \eqref{eq:pairdipoles}, \eqref{eq:tripoles} and \eqref{eq:Qstruct}, and their consequences for the kinematic functions that multiply them in the anomalous dimension \eqref{eq:Gamma2short}. Two kinds of statements arise. First, the symmetrized structures obey exact operator identities, valid for arbitrary, also coincident, leg indices. Second, when a structure is summed against a kinematic function and applied to a soft function, further equivalences follow from relabeling the summed leg indices, from relabeling the virtual momenta --- which are integration variables --- and from the symmetry of the soft function under the exchange of the two real gluons, i.e.\ the simultaneous exchange $(q,\alpha,\tilde\alpha)\leftrightarrow(r,\beta,\tilde\beta)$ of their momenta and color indices. Color conservation is not used for the relations below, the associated identities will be presented in Appendix \ref{sec:CScc}.

\paragraph{Operator identities.}
Since generators on different legs commute and the products $(\dots)_+$ are symmetrized, one finds
\begin{gather}
\bm D_{ij}=\bm D_{ji}\,,\qquad \overline{\bm D}_{ij}=\overline{\bm D}_{ji}\,,\notag\\
\bm T_{ijk}=-\bm T_{jik}=-\bm T_{ikj}\,,\qquad
\bm T^{\alpha;\tilde\alpha}_{ijk}=-\bm T^{\alpha;\tilde\alpha}_{ikj}\,,\qquad
\bm T^{\beta\alpha;\tilde\beta\tilde\alpha}_{jik}=-\bm T^{\alpha\beta;\tilde\alpha\tilde\beta}_{ijk}\,,\label{eq:CSids}\\
\bm D^{X,\beta\alpha;\tilde\beta\tilde\alpha}_{ij}=\bm D^{X,\alpha\beta;\tilde\alpha\tilde\beta}_{ij}\quad(X=A,S)\,,\qquad
\bm Q^{\beta\alpha;\tilde\beta\tilde\alpha}_{jilk}=\bm Q^{\alpha\beta;\tilde\alpha\tilde\beta}_{ijkl}\,,\notag
\end{gather}
together with the conjugation relations given in the main text; the one-emission dipole $\bm D^{\alpha;\tilde\alpha}_{ij}$ has no symmetry of its own. The total antisymmetry of the virtual tripole and the $j\leftrightarrow k$ antisymmetry of $\bm T^{\alpha;\tilde\alpha}_{ijk}$ imply
\begin{equation}\label{eq:CSvanishing}
\bm T_{ijk}=0\ \ \text{whenever two legs coincide}\,,\qquad \bm T^{\alpha;\tilde\alpha}_{ijj}=0\,.
\end{equation}
Both statements hold only for the symmetrized structures: without the $(\dots)_+$ they are violated by commutator terms, which is the origin of the conditions \eqref{eq_condition_to_drop_symmetrizers}. For the fermion-pair dipole $\bm D^{F}$ the analogue of the $X=A,S$ relation in \eqref{eq:CSids} involves charge conjugation, which simultaneously exchanges the quark and antiquark Wilson lines of the soft function; its net effect below is the same as for $\bm D^{A}$.

\paragraph{Effective symmetries of the kinematic functions.}
Let $\simeq$ denote equality after summation over legs, angular integration, and application to a soft function, without using color conservation. Combining \eqref{eq:CSids} with the relabeling freedoms one finds, sector by sector:
\begin{itemize}
\item \emph{Double-real.} By the third relation in \eqref{eq:CSids} and the gluon-exchange symmetry of the soft function, only the part of $K_{ijk;qr}$ that is odd under the simultaneous exchange $(i\leftrightarrow j,\,q\leftrightarrow r)$ is relevant; the joint-even part drops identically. Likewise, we need only the $q\leftrightarrow r$-even parts of $K^{A,F,S}_{ij;qr}$ (the legs $i,j$ keep fixed roles), and only the part of a quadrupole coefficient $K_{ijkl;qr}$ even under the joint exchange $(i\leftrightarrow j,\,k\leftrightarrow l,\,q\leftrightarrow r)$.
\item \emph{Real-virtual.} Only the $j\leftrightarrow k$-odd parts of $K_{ijk;qr}$ and $I_{ijk;q}$ contribute. The $\bm D^{\alpha;\tilde\alpha}_{ij}$ does not have any symmetries, while only the $(i\leftrightarrow j)$-even part of $I_{ij}$ multiplying $\bm D_{ij}-\overline{\bm D}_{ij}$ contributes.
\item \emph{Double-virtual.} Both momenta are integrated, so the coefficient can be replaced by its $q\leftrightarrow r$-average, and only its totally leg-antisymmetric projection acts.
\end{itemize}
The three tripole sectors share no common nontrivial symmetry: the double-virtual structure possesses (the analogues of) both the joint $(i\leftrightarrow j,\,q\leftrightarrow r)$ antisymmetry of the double-real and the $j\leftrightarrow k$ antisymmetry of the real-virtual structure, while the latter two share only the identity. A single kinematic function multiplying all three sectors, as in \eqref{eq:Gamma2short}, is therefore constrained by no common symmetry; each sector projects it differently.

\paragraph{Residual freedom in the coefficient functions.}
It follows that the decomposition \eqref{eq:Gamma2short} does not determine the kinematic functions uniquely even before color conservation is invoked: the shifts
\begin{enumerate}
\item $K_{ijk;qr}\to K_{ijk;qr}+\delta K_{ijk;qr}$ with $\delta K$ even under the joint exchange $(i\leftrightarrow j,\,q\leftrightarrow r)$ and even under $j\leftrightarrow k$; the shift $\delta K$ then drops from the double-real and real-virtual sectors by the rules above, and its totally antisymmetric leg projection vanishes because of the $j\leftrightarrow k$ symmetry, so the double-virtual sector is blind to it as well;
\item $K^{X}_{ij;qr}\to K^{X}_{ij;qr}+\delta K^{X}_{ij;qr}$ ($X=A,F,S$) with $\delta K^{X}$ odd under $q\leftrightarrow r$, accompanied by $K_{ij;q}\to K_{ij;q}-\int[d\Omega_r]\,\big(C_A\,\delta K^{A}+n_FT_F\,\delta K^{F}+n_ST_S\,\delta K^{S}\big)_{ij;qr}$; the pair-emission terms drop by the double-real rule, the compensation cancels the shift of the $\bm D^{\alpha;\tilde\alpha}_{ij}$ coefficient, and the induced purely virtual contribution vanishes as the double angular integral of a $q\leftrightarrow r$-odd function, cf.\ \eqref{eq:Kijqterm};
\item $I_{ijk;q}\to I_{ijk;q}+\delta I_{ijk;q}$ with $\delta I$ even under $j\leftrightarrow k$;
\item $I_{ij}\to I_{ij}+\delta I_{ij}$ with $\delta I_{ij}$ odd under $i\leftrightarrow j$;
\end{enumerate}
all leave the action of $\bGam^{(2)}$ on any soft function unchanged. The identities \eqref{eq_condition_to_drop_symmetrizers} should be viewed in this light: they fix part of this freedom. The first identity and the combination $K_{iji;qr}+K_{jii;rq}$ constrain components that never contribute: at coincident legs the double-real rule lets only the $q\leftrightarrow r$-odd part of $K_{iij;qr}$ and the joint-odd combination $K_{iji;qr}-K_{jii;rq}$ contribute. The last identity in \eqref{eq_condition_to_drop_symmetrizers} constrains $I_{ijj;qr}$, the coefficient of the identically vanishing structure \eqref{eq:CSvanishing}. The only content of \eqref{eq_condition_to_drop_symmetrizers} that symmetry alone cannot arrange is the vanishing of the non-zero, joint-odd combination at coincident legs; removing such a combination requires the color-conservation identities \eqref{eq:ZeroDueToColorConservation}, as in the scheme change of Section~\ref{sec:schemechange}.

\section{Color-conservation identities of the color structures}
\label{sec:CScc}

On color-singlet hard functions the total color charge vanishes, 
\begin{align}
\sum_i\bT^\alpha_{i,L}\,\bH &= \sum_i\bT^\alpha_{i}\,\bH = 0\,, & \sum_i\bT^{\tilde{a}}_{i,R}\,\bH &= \sum_i\bH \,\bT^{\tilde{\alpha}}_{i} =0 \, .
\end{align}
For each color structure of Section~\ref{sec:XSanomdim} and each of its leg slots we list here the sum over one index when acting on a singlet hard function. Commuting the total charge through the symmetrized products generates commutator terms; these either cancel or produce structures with fewer legs. All single-slot sums of the two-leg dipoles vanish,
\begin{equation}\label{eq:ccVanish2}
 \sum_i \bm D_{ij}=\sum_i\overline{\bm D}_{ij}=\sum_i \bm D^{\alpha;\tilde\alpha}_{ij}
 =\sum_i \bm D^{X,\alpha\beta;\tilde\alpha\tilde\beta}_{ij}=0\,,
\end{equation}
for any $X=A,F,S$, and likewise for the sums over $j$. The same is true for the virtual tripoles
\begin{align}\label{eq:ccTripoleV}
 &\sum_i \bm T_{ijk}=\sum_j \bm T_{ijk}=\sum_k \bm T_{ijk}=0\,,
\end{align}
while there are non-vanishing dipole terms for some of the real-virtual and double emission terms:
\begin{align}\label{eq:ccTripoleDR}
 &\sum_i \bm T^{\alpha;\tilde\alpha}_{ijk}=0\,,\qquad
 \sum_j \bm T^{\alpha;\tilde\alpha}_{ijk}=+\frac{C_A}{2}\,\bm D^{\alpha;\tilde\alpha}_{ik}\,,\qquad
 \sum_k \bm T^{\alpha;\tilde\alpha}_{ijk}=-\frac{C_A}{2}\,\bm D^{\alpha;\tilde\alpha}_{ij}\,,\\
 &\sum_k \bm T^{\alpha\beta;\tilde\alpha\tilde\beta}_{ijk}=0\,,\qquad
 \sum_i \bm T^{\alpha\beta;\tilde\alpha\tilde\beta}_{ijk}=-\frac{1}{2}\,\bm D^{A,\alpha\beta;\tilde\alpha\tilde\beta}_{jk}\,,\qquad
 \sum_j \bm T^{\alpha\beta;\tilde\alpha\tilde\beta}_{ijk}=+\frac{1}{2}\,\bm D^{A,\alpha\beta;\tilde\alpha\tilde\beta}_{ik}\,,\notag
\end{align}
with the conjugate relations
\begin{align}\label{eq:ccTripoleBar}
 &\sum_i \overline{\bm T}^{\alpha;\tilde\alpha}_{ijk}=0\,,\qquad
 \sum_j \overline{\bm T}^{\alpha;\tilde\alpha}_{ijk}=+\frac{C_A}{2}\,\bm D^{\alpha;\tilde\alpha}_{ki}\,,\qquad
 \sum_k \overline{\bm T}^{\alpha;\tilde\alpha}_{ijk}=-\frac{C_A}{2}\,\bm D^{\alpha;\tilde\alpha}_{ji}\,,\notag\\
 &\sum_k \overline{\bm T}^{\alpha\beta;\tilde\alpha\tilde\beta}_{ijk}=0\,,\qquad
 \sum_i \overline{\bm T}^{\alpha\beta;\tilde\alpha\tilde\beta}_{ijk}=-\frac{1}{2}\,\bm D^{A,\alpha\beta;\tilde\alpha\tilde\beta}_{kj}\,,\qquad
 \sum_j \overline{\bm T}^{\alpha\beta;\tilde\alpha\tilde\beta}_{ijk}=+\frac{1}{2}\,\bm D^{A,\alpha\beta;\tilde\alpha\tilde\beta}_{ki}\,.
\end{align}
The vanishing of all three sums of the purely virtual tripole terms in \eqref{eq:ccTripoleV} is special: for $\bm T_{ijk}$ the two commutator terms generated by moving the total charge through the symmetrized product cancel against each other, by the same contractions of structure constants that produce the $C_A$-terms above. The quadrupole reduces to tripoles,
\begin{equation}\label{eq:ccQ}
 \sum_i \bm Q^{\alpha\beta;\tilde\alpha\tilde\beta}_{ijkl}=-\frac{1}{2}\,\overline{\bm T}^{\alpha\beta;\tilde\alpha\tilde\beta}_{klj}\,,\qquad
 \sum_k \bm Q^{\alpha\beta;\tilde\alpha\tilde\beta}_{ijkl}=-\frac{1}{2}\,\bm T^{\alpha\beta;\tilde\alpha\tilde\beta}_{ijl}\,,
\end{equation}
and the sums over $j$ and $l$ follow from the exchange relation $\bm Q^{\beta\alpha;\tilde\beta\tilde\alpha}_{jilk}=\bm Q^{\alpha\beta;\tilde\alpha\tilde\beta}_{ijkl}$ of Appendix~\ref{sec:CSsymmetries}.

The identities above make the derivation of \eqref{eq:ZeroDueToColorConservation} transparent. Consider first the $E$-term: since $E_{ij;qr}$ does not reference the third leg, the sums over $k$ apply directly, and by \eqref{eq:ccTripoleV}, \eqref{eq:ccTripoleDR},\eqref{eq:ccTripoleBar} the double-real and double-virtual parts of $E_{ij;qr}\big(\bm T^{\alpha\beta;\tilde\alpha\tilde\beta}_{ijk}-2\bm T^{\alpha;\tilde\alpha}_{ijk}+\bm T_{ijk}\big)+\text{h.c.}$ vanish identically. The real-virtual part leaves
\begin{equation}\label{eq:ccEresidue}
 -2\int[d\Omega_r]\,E_{ij;qr}\Big(\sum_k\bm T^{\alpha;\tilde\alpha}_{ijk}+\sum_k\overline{\bm T}^{\alpha;\tilde\alpha}_{ijk}\Big)
 = C_A\!\int[d\Omega_r]\big(E_{ij;qr}-E_{ij;rq}\big)\,\bm D^{\alpha;\tilde\alpha}_{ij}\,,
\end{equation}
where the constraint $E_{ij;qr}=-E_{ji;rq}$ was used to combine the conjugate contribution with the direct one after a relabeling of $\bm D^{\alpha;\tilde\alpha}_{ji}$ relabeled. The residual term is odd under $q\leftrightarrow r$ and canceled by the $2\bm D^{\alpha;\tilde\alpha}_{ij}$ part of the last line of \eqref{eq:ZeroDueToColorConservation}, while the accompanying $-\bm D_{ij}-\overline{\bm D}_{ij}$ pieces vanish upon the remaining angular integration, because their coefficient is the double integral of a function odd under $q\leftrightarrow r$; they are included so that the compensation has exactly the form of the $K_{ij;q}$-term of \eqref{eq:Gamma2short}. For the $F$-terms the sums over the slot that $F$ does not reference leave, by \eqref{eq:ccTripoleDR} and \eqref{eq:ccTripoleBar}, the real-virtual remainder $-2C_A(F_{ij;qr}+F_{ji;qr})\bm D^{\alpha;\tilde\alpha}_{ij}$ and the double-real remainder $(F_{ij;qr}+F_{ij;rq}+F_{ji;qr}+F_{ji;rq})\bm D^{A,\alpha\beta;\tilde\alpha\tilde\beta}_{ij}$, while the double-virtual part vanishes identically. The second line of \eqref{eq:ZeroDueToColorConservation} cancels the real-virtual residue exactly and the double-real residue up to the combination $\big(F_{ij;rq}+F_{ji;rq}-F_{ij;qr}-F_{ji;qr}\big)\bm D^{A}_{ij}$, which is odd under $q\leftrightarrow r$ and therefore drops against the exchange-symmetric soft function by Appendix~\ref{sec:CSsymmetries}. Thus \eqref{eq:ZeroDueToColorConservation} vanishes on any color-singlet hard function; its double-real part does so only in conjunction with the gluon-exchange symmetry of the soft function.

\section{Color structures in the color-flow basis}\label{sec:colorflow}
It is useful to write down the action of the color structures appearing in the one- and two-loop anomalous dimensions on a basis of the color space of the hard function. The color-flow basis is particularly useful because it allows one to read off the terms that dominate in a large-$N_c$ expansion and to systematically organize the corrections in $\frac{1}{N_c}$. 

To define the color-flow states, we replace all adjoint color indices appearing in the amplitude and conjugate amplitude with a pair of a fundamental and anti-fundamental color indices. This conversion involves a normalization factor of $1/\sqrt{T_F}$, where $T_F$ fixes the normalization of the generators through $\mathrm{tr}\left(t^{\alpha}t^{\beta}\right)=T_F\,\delta^{\alpha\beta}$. Here, and in the remainder of this appendix, we set $T_F=\tfrac{1}{2}$. For example, if $\alpha_m$ is the adjoint color index of the $m$-th gluon in the amplitude, we can express it in terms of a pair of a fundamental and anti-fundamental indices $(\iota_m,\overline{\iota}_m)$ using
\begin{equation}\label{eq_adjoint_to_fund}
  \Big\langle\,\begin{matrix} \iota_m\\
  \overline{\iota}_m
  \end{matrix}\, \Big |\,\mathcal{M}\Big\rangle= \sqrt{2}\left( t^{\alpha_m} \right)^{\iota_m}{}_{\overline{\iota}_m}\braket{\alpha_m|\mathcal{M}}\, ,
\end{equation}
where repeated indices are summed over.

The notation in \eqref{eq_adjoint_to_fund} with upper fundamental and lower anti-fundamental indices is convenient, because they always arise in pairs. An amplitude with $N_q$ outgoing quark lines and $N_g$ outgoing gluon lines has $N=N_q+N_g$ fundamental color indices $\iota_1,\dots, \iota_{N}$ and corresponding anti-fundamental color indices $\overline{\iota}_1,\dots, \overline{\iota}_{N}$. A color-flow state of the amplitude can now be defined through a permutation $\sigma\in S_N$, where $S_N$ is the group of permutations of $N$ elements. The color-flow state $\ket{\sigma}$ is the amplitude
\begin{equation}\label{eq:cfbraket}
 \Big\langle\,\begin{matrix} \iota_1 & \dots & \iota_N \\
  \overline{\iota}_1& \dots & \overline{\iota}_N
  \end{matrix}\, \Big |\,\sigma\Big\rangle=\delta^{\iota_1}{}_{\overline{\iota}_{\sigma(1)}}\dots \delta^{\iota_N}{}_{\overline{\iota}_{\sigma(N)}}\, ,
\end{equation}
where $\sigma(i)$ refers to the index obtained by acting on $i$ with the permutation $\sigma$. While we refer to the set of all $\ket{\sigma}$ as the color-flow basis, this is actually a misnomer, and the set is in general overcomplete for two reasons. First, if there are gluons in the amplitude, we have to account for the fact that gluons live in the adjoint representation of $\mathrm{SU}(N_c)$: The amplitudes built from linear sums of color-flow states are more general than the right-hand side of \eqref{eq_adjoint_to_fund}, i.e., we can write down amplitudes, where the produced gluons live in the adjoint representation of $\mathrm{U}(N_c)$ rather than $\mathrm{SU}(N_c)$. When re-expressing an amplitude written with (anti-)fundamental indices through adjoint indices, we use the equation
\begin{equation}
   \label{eq_fund_to_adjoint}
  \braket{\alpha_m|\mathcal{M}}=  \sqrt{2}\left( t^{\alpha_m} \right)^{\overline{\iota}_m}{}_{\iota_m}  \Big\langle\,\begin{matrix} \iota_m\\
  \overline{\iota}_m
  \end{matrix}\, \Big |\,\mathcal{M}\Big\rangle\, .
\end{equation}
If the indices are in a color singlet configuration
\begin{equation}\label{eq:cfsinglet}
  \Big\langle\,\begin{matrix} \iota_m\\
  \overline{\iota}_m
  \end{matrix}\, \Big |\,\mathcal{M}\Big\rangle\propto \delta^{\iota_m}{}_{\overline{\iota}_m}\, ,
\end{equation}
then the right-hand side of \eqref{eq_fund_to_adjoint} yields zero. In other words, the representation of a physical amplitude $  \braket{\alpha_m|\mathcal{M}}$ in the color-flow representation is not unique because we can always add terms proportional to $\delta^{\iota_m}{}_{\overline{\iota}_m}$ without changing the physical amplitude. We can get rid of the ambiguity by multiplying a color-flow amplitude by a projector 
\begin{equation}\label{eq:cfprojector}
  \left( P_m \right)^{\iota_m}{}_{\iota_m^\prime \overline{\iota}_m}{}^{\overline{\iota}_m^\prime}=\delta^{\iota_m}{}_{\iota_m^\prime}\delta^{\overline{\iota}_m^\prime}{}_{\overline{\iota}_m}-\frac{1}{N_c}\delta^{\iota_m}{}_{\overline{\iota}_m}\delta^{\overline{\iota}_m^\prime}{}_{\iota_m^\prime}\, ,
\end{equation}
i.e.,
\begin{equation}\label{eq:cfprojaction}
   \braket{(\iota_m,\overline{\iota}_m)|P_m|\mathcal{M}}=  \left( P_m \right)^{\iota_m}{}_{\iota_m^\prime \overline{\iota}_m}{}^{\overline{\iota}_m^\prime}  \braket{(\iota_m^\prime,\overline{\iota}_m^\prime)|\mathcal{M}}\,.
\end{equation}
To get rid of all ambiguities, we have to multiply the amplitude by a projector for each gluon. After getting rid of all $\mathrm{U}(1)$ gluons, the color-flow basis is still overcomplete for specific integer values of $N_c$. To understand this, we note that the scalar product matrix
\begin{equation}
  \label{eq_SP_matrix}
  \braket{\tau|\sigma}=N_c^{N-\#(\sigma,\tau)}\, ,
\end{equation}
with $\#(\sigma,\tau)$ being the number of transpositions required to transform $\sigma$ into $\tau$, does not have full rank for all values of $N$ and $N_c$. For example, for a color singlet decaying into a $q \bar{q}\, q \bar{q}\, q \bar{q}$ system, i.e., $N=3$, and for $N_c=2$, the rank of the scalar product matrix would be $5$ instead of $N!=6$, meaning that we can rewrite one of the six color-flow states through the other five without changing physical measurements.

For our purposes, we do not need to worry about the color-flow basis being overcomplete. In the following we go through all color structures entering the one-loop and two-loop anomalous dimensions. 
\subsection{Dipole color structures}
The dipole structure $\bm{D}_{mn}$ appearing in purely virtual anomalous dimensions acts on color flows as
\begin{equation}
  \label{eq_dipole_v_cf}
\begin{aligned}
    &\sum_{m,n}F_{mn}\bm{D}_{mn}\ket{\sigma}=\frac{N_c}{2}\Biggl(\sum_i^N \left(F_{i i}+F_{\sigma(i) \sigma(i)}-F_{i \sigma(i)}- F_{\sigma(i) i } \right)\ket{\sigma} \\
    & +\frac{1}{N_c}\sum_{(i j)}^N\left( F_{ij}+F_{\sigma(i)\sigma(j)} -F_{i\sigma(j)}-F_{\sigma(i)j} \right)\ket{\sigma \chi_{ij}}+\sum_{i,j}^N\frac{1}{N_c^2}\left( F_{i \overline{j}}+F_{\overline{i} j} -F_{ij}-F_{\overline{i}\overline{j}} \right)\ket{\sigma}\Biggr)\, ,
\end{aligned}
\end{equation}
where, here, and in the remainder of this appendix, the indices on the left-hand side, $m$ and $n$, are leg indices running over all external particles, and $F_{mn}$ is any kinematic function depending on the legs $m$ and $n$. The indices $i$ and $j$ on the other hand run from $1$ to $N$, enumerating the fundamental color indices $\iota_i$ present in the amplitude $\ket{\sigma}$ from $\iota_1$ to $\iota_N$. Parentheses around summation indices, here $(ij)$, indicate that we only sum terms where all indices are different. In subscripts of the kinematic function $F$ we use the following shorthand notation: $i$ refers to the leg carrying the index $\iota_i$, $\bar{i}$ refers to the leg carrying the index $\overline{\iota}_i$, and $\sigma(i)$ refers to the leg carrying the index $\overline{\iota}_{\sigma(i)}$. Note that while $i$ and $\bar{i}$ denote different legs if $\iota_i$ is carried by a quark, they refer to the same leg if $\iota_i$ is carried by a gluon. Finally, $\sigma\chi_{ij}$ refers to the permutation where one first acts with the transposition $\chi_{ij}$ that swaps $i$ and $j$ and then applies $\sigma$,\footnote{Composing flow permutations with the operations $\chi$ from the right corresponds to acting with the color-flow operators of Ref.~\cite{Martinez:2018ffw} from the left: our $\ket{\sigma\chi_{ij}}$ equals $\mathbf{s}_{ij}\ket{\sigma}$, with $\mathbf{s}_{ij}$ the operator exchanging $\sigma(i)$ and $\sigma(j)$, and the state $\ket{\tilde{\sigma}\chi_{iq}}$ appearing below equals $\mathbf{t}_{i}\ket{\sigma}$, with $\mathbf{t}_{i}$ the operator inserting the color line of the extra gluon $q$ on leg $i$.} i.e., 
\begin{equation}\label{eq:cfsigmaij}
 \sigma\chi_{ij}(k)=\begin{cases}
    \sigma(k)\,, &\text{ if }k\notin \{i,j\}\, ,\\
    \sigma(j)\,, &\text{ if }k=i \, , \\
    \sigma(i)\,, & \text{ if }k=j\, .
  \end{cases}
\end{equation}
The action of the hermitian conjugate $\overline{\bm{D}}_{mn}$ of $\bm{D}_{mn}$ on a color-flow state is obtained as
\begin{equation}
  \label{eq_Db_action}
  \bra{\sigma}\sum_{m,n}F_{mn}\overline{\bm{D}}_{mn}=\left( \sum_{m,n}F_{mn}^*\bm{D}_{mn} \ket{\sigma}\right)^\dagger\, ,
\end{equation}
where $*$ refers to complex conjugation and $\ket{\sigma}^\dagger=\bra{\sigma}$. The leading-color action of $\bm{D}_{mn} $ reads

\begin{equation}
  \label{eq_dipole_v_cf_LC}
    \sum_{m,n}F_{mn}\bm{D}_{mn}\ket{\sigma}=\frac{N_c}{2} \sum_{i}^N\left(F_{i i}+F_{\sigma(i)\sigma(i)}-F_{i \sigma(i)}- F_{\sigma(i) i } \right)\ket{\sigma}+\dots\,,
\end{equation}
where the dots indicate terms subleading in color. Note that only color-connected leg pairs contribute. We point out that in the two-loop anomalous dimensions the color structure $\bm{D}_{mn}$ comes with an additional factor of $C_A$, enhancing its $N_c$ power counting.

The dipole structure $\bm{D}^{\alpha\tilde{\alpha}}_{mn}$ appearing in single-emission anomalous dimensions should be understood as shorthand notation for the operator 
\begin{equation}
  \label{eq_Dmnalphaalphatild_operator}
  \sum_{\alpha,\tilde{\alpha}}\bm{D}^{\alpha\tilde{\alpha}}_{mn}\otimes\ket{\alpha}\!\!\bra{\tilde{\alpha}}
\end{equation}
that maps from the color space of the hard function to the larger color space with an extra gluon carrying color indices $\alpha$ in the amplitude, and $\tilde{\alpha}$ in the conjugate amplitude. The fundamental and anti-fundamental index of the added gluon are $\iota_{q}$ and $\overline{\iota}_{q}$, where we will use the shorthand notation $q=N+1$ throughout this appendix. 

The operator $\bm{D}^{\alpha\tilde{\alpha}}_{mn}$ is built from the operator $\bT_{m,L}^\alpha$, which we understand as shorthand notation for the operator
\begin{equation}
  \label{eq_Tialpha_as_an_operator}
  \sum_{\alpha}\bT_{m,L}^\alpha\otimes\ket{\alpha}\, ,
\end{equation}
and its hermitian conjugate action on the conjugate amplitude from the right.
The action of the operator $\bT_{m,L}^\alpha$ on an arbitrary color-flow state $\ket{\sigma}$ is given by
\begin{equation}
    \label{eq_TialphaAction}
  \sum_m F_m \bT_{m,L}^\alpha\ket{\sigma}=\frac{1}{\sqrt{2}}\sum_{i}^N \left( F_i-F_{\sigma(i)} \right)\left( \ket{ \tilde{\sigma}\chi_{iq} }-\frac{1}{N_c}\ket{\tilde{\sigma}}  \right) ,
\end{equation}
where we used the notation from \eqref{eq_dipole_v_cf}, and additionally $\tilde{\sigma}$ refers to the permutation 
\begin{equation}
  \label{eq_sigtilde_1}
  \tilde{\sigma}(i)=\sigma(i) \text{ for }i\leq N\, ,\quad \tilde{\sigma}(q)=q\, ,
\end{equation}
i.e., $\ket{\tilde{\sigma}}$ is the color-flow state where the extra gluon is color-connected to itself, i.e., it is a $\mathrm{U}(1)$ gluon. 
It is worth noting that the action of $\bT_{m,L}^\alpha$ inserts $\mathrm{U}(1)$ gluon states $\ket{\tilde\sigma}$ with a matrix element that is independent of the color flow we start with. To see this we can rewrite \eqref{eq_TialphaAction} as 
\begin{equation}
    \label{eq_TialphaAction_B}
  \sum_m F_m \bT_{m,L}^\alpha\ket{\sigma}=\frac{1}{\sqrt{2}}\sum_{i}^N \left( F_i-F_{\sigma(i)} \right)\ket{ \tilde{\sigma}\chi_{iq} }  - \frac{1}{\sqrt{2}N_c}\sum_i^N(F_i-F_{\overline{i}})\ket{\tilde{\sigma}}\, ,
\end{equation}
where we relabeled the summation index in the second term.
Using \eqref{eq_TialphaAction}, we find that the action of the dipole operator on a generic color-flow state is given by
\begin{equation}\label{eq:cfdipoleoff}
\begin{aligned}   \sum_{m,n}F_{mn}\bm{D}^{\alpha\tilde{\alpha}}_{mn}\ket{\sigma}\!\!\bra{\tau}&=\frac{1}{2}\sum_{i,j}^N\left( \ket{\tilde{\sigma}\chi_{iq}}-\frac{1}{N_c}\ket{\tilde{\sigma}}  \right)\left(  \bra{\tilde{\tau}\chi_{jq}}-\frac{1}{N_c}\bra{\tilde{\tau}} \right)\\
    &\times \left( F_{ij}+F_{\sigma(i)\tau(j)}- F_{i \tau(j)}- F_{\sigma(i) j}\right) ,
\end{aligned}
\end{equation}
 where we used the notation $\tilde{\tau}$ analogously to \eqref{eq_sigtilde_1}.

For the leading-color contribution we only need to look at states that have the same color-flow in the amplitude and conjugate amplitude---other configurations will be suppressed by the scalar product matrix introduced in equation \eqref{eq_SP_matrix}. The leading-color action is thus given by
\begin{equation}\label{eq:cfdipolediag}
\begin{aligned}  \sum_{m,n}F_{mn}\bm{D}^{\alpha\tilde{\alpha}}_{mn}\ket{\sigma}\!\!\bra{\sigma}&=\frac{1}{2}\sum_{i}^N \ket{\tilde{\sigma}\chi_{iq}} \bra{\tilde{\sigma}\chi_{iq}}\\
    &\times \left( F_{ii}+F_{\sigma(i)\sigma(i)}- F_{i\sigma(i)}- F_{\sigma(i)i}\right)+\dots\, .
\end{aligned}
\end{equation}
We point out that in the two-loop anomalous dimensions the color structure $\bm{D}^{\alpha\tilde{\alpha}}_{mn}$ comes with an additional factor of $C_A$, enhancing its $N_c$ power counting.

The double-gluon-emission dipole structure $\bm{D}_{mn}^{A,\alpha\beta;\tilde{\alpha}\tilde{\beta}}$ has to be understood as shorthand notation for the operator
\begin{equation}
  \label{eq_double_emission_dipole_operator}
  \sum_{\alpha, \tilde{\alpha}}\sum_{\beta, \tilde{\beta}}\bm{D}_{mn}^{A,\alpha\beta;\tilde{\alpha}\tilde{\beta}}\otimes \ket{\alpha}\!\!\bra{\tilde{\alpha}}\otimes \ket{\beta}\!\!\bra{\tilde{\beta}}
\end{equation}
that maps from the color space of the hard function with $N$ fundamental color lines to the larger space with two extra gluons. The fundamental color index of the gluon with adjoint index $\alpha$($\beta$) is $\iota_{q}$($\iota_{r}$). We will use the shorthand notation $r=N+2$ throughout this appendix. The operator $\bm{D}_{mn}^{A,\alpha\beta;\tilde{\alpha}\tilde{\beta}}$ is built by acting with the operator, $i f^{\alpha \beta c}\bT_{m,L}^c$ on the amplitude, and with its hermitian conjugate on the conjugate amplitude. Here we interpret $i f^{\alpha \beta c}\bT_{m,L}^c$ as shorthand notation for the operator
\begin{equation}\label{eq:cfifTop}
  \sum_{\alpha, \beta} i f^{\alpha \beta c}\bT_{m,L}^c \otimes \ket{\alpha}\otimes \ket{\beta}\, .
\end{equation} 
Then the action of $i f^{\alpha \beta c}\bT_{m,L}^c$ on a color-flow state $\ket{\sigma}$ is given by
\begin{equation}
  \label{eq_ifalphabetacTmc}
  \sum_{m}F_m i f^{\alpha \beta c}\bT_{m,L}^c\ket{\sigma}=\frac{1}{2}\sum_{i}^N \left( F_i-F_{\sigma(i)} \right)\left( \ket{\tilde{\sigma}\chi_{iqr}}- \ket{\tilde{\sigma}\chi_{irq}}  \right) ,
\end{equation}
where now $\tilde{\sigma}$ refers to the permutation
\begin{equation}
  \label{eq_sigtilde_2}
  \tilde{\sigma}(i)=\sigma(i)\, , \text{ for }i\leq N\, ,\quad \tilde{\sigma}(q)=q\, ,\quad \tilde{\sigma}(r)=r\, ,
\end{equation}
and 
$\tilde{\sigma}\chi_{iqr}$ is the permutation, where one first acts with the cycle $i\to q \to r\to i$ and then with $\tilde{\sigma}$, i.e.,
\begin{equation}\label{eq:cfsigmatilde}
  \tilde{\sigma}\chi_{iqr}(i)=q\, ,\quad   
  \tilde{\sigma}\chi_{iqr}(q)=r\, ,\quad   
  \tilde{\sigma}\chi_{iqr}(r)=\sigma(i)\, .
\end{equation}

Using \eqref{eq_ifalphabetacTmc}, the action of the operator $\bm{D}_{mn}^{A,\alpha\beta;\tilde{\alpha}\tilde{\beta}}$ on a generic color-flow state is found to be
\begin{equation}\label{eq:cfDAoff}
  \begin{aligned}
    \sum_{mn}\bm{D}_{mn}^{A,\alpha\beta;\tilde{\alpha}\tilde{\beta}}\ket{\sigma}\!\!\bra{\tau} &=\frac{1}{4}\sum_{i,j}^N \left( F_{ij}+F_{\sigma(i)\tau(j)}-F_{i\tau(j)} -F_{\sigma(i)j}\right)\\
    &\quad \Biggl( \ket{\tilde{\sigma}\chi_{iqr}}\!\!\bra{\tilde{\tau}\chi_{jqr}} + \ket{\tilde{\sigma}\chi_{irq}}\!\!\bra{\tilde{\tau}\chi_{jrq}} \\
    &\qquad- \ket{\tilde{\sigma}\chi_{iqr}}\!\!\bra{\tilde{\tau}\chi_{jrq}} - \ket{\tilde{\sigma}\chi_{irq}}\!\!\bra{\tilde{\tau}\chi_{jqr}}\Biggr)\, .
  \end{aligned}
\end{equation}

The leading-color action of $\bm{D}_{ij}^{A,\alpha\beta;\tilde{\alpha}\tilde{\beta}}$ is given by
\begin{equation}
\label{eq_DA_LC}
  \begin{aligned}  \sum_{mn}\bm{D}_{mn}^{A,\alpha\beta;\tilde{\alpha}\tilde{\beta}}\ket{\sigma}\!\!\bra{\sigma} &=\frac{1}{4}\sum_{i}^N \left( F_{ii}+F_{\sigma(i)\sigma(i)}-F_{i\sigma(i)} -F_{\sigma(i)i}\right)\\
    &\quad \left( \ket{\tilde{\sigma}\chi_{iqr}}\!\!\bra{\tilde{\sigma}\chi_{iqr}} + \ket{\tilde{\sigma}\chi_{irq}}\!\!\bra{\tilde{\sigma}\chi_{irq}}  \right)+\dots\, .
  \end{aligned}
\end{equation}

The interpretation of \eqref{eq_DA_LC} is that we can choose any of the color-connected dipoles $[i\sigma(i)]$ and insert a color-connected gluon pair into the dipole. There are two ways of doing this: we can either end up with the color connected dipoles $[iqr\sigma(i)]$, which corresponds to the first term in the second line of \eqref{eq_DA_LC}, or we can end up with the color-connected dipoles $[irq\sigma(i)]$, which corresponds to the second term in the second line of \eqref{eq_DA_LC}.

Finally, we analyze the dipole color structure $\bm{D}_{ij}^{F,\alpha\beta;\tilde{\alpha}\tilde{\beta}} $ for the emission of a $q\bar{q}$ pair. In the notation of this appendix, $\bm{D}_{ij}^{F,\alpha\beta;\tilde{\alpha}\tilde{\beta}} $ must be understood as shorthand notation for the operator

\begin{equation}\label{eq:cfgluonproj}  \sum_{\iota_{q}\tilde{\iota}_{q}}\sum_{\overline{\iota}_{q}\overline{\tilde{\iota}}_{q}}\bm{D}_{ij}^{F,\iota_{q}\overline{\iota}_{q};\tilde{\iota}_{q}\overline{\tilde{\iota}}_{q}} \otimes \Big | \,\begin{matrix} \iota_{q}\\
  \overline{\iota}_{q}
  \end{matrix}\, \Big\rangle\!  \Big\langle \,\begin{matrix} \tilde{\iota}_{q}\\
  \overline{\tilde{\iota}}_{q}
  \end{matrix}\, \Big |
\end{equation}

that maps from the color space of the hard function to the larger color space containing an additional quark with fundamental color index $\iota_{q}$ and an antiquark with anti-fundamental color index $\overline{\iota}_{q}$. The operator $\bm{D}_{ij}^{F,\alpha\beta;\tilde{\alpha}\tilde{\beta}} $ is constructed by acting with the operator $\bT_{m,L}^c t^{c,\alpha }{}_{\!\beta }$ on the left, and with its hermitian conjugate on the right. The operator $\bT_{m,L}^c t^{c,\alpha }{}_{\!\beta }$ is shorthand for 
\begin{equation}\label{eq:cfquarkproj}
  \sum_{\iota_q,\overline{\iota}_q} \bT_{m,L}^c t^{c,\iota_q }{}_{\!\overline{\iota}_q} 
  \Big | \,\begin{matrix} \iota_{q}\\
  \overline{\iota}_{q}
  \end{matrix}\, \Big\rangle\, .
\end{equation}
Its action on a color-flow state $\ket{\sigma}$ is given by
\begin{equation}
  \label{eq_Tctc_action}
  \sum_m F_m \bT_{m,L}^c t^{c,\iota_q }{}_{\!\overline{\iota}_q}\ket{\sigma}=\frac{1}{2}\sum_{i}^N \left( F_i-F_{\sigma(i)} \right)\left( \ket{\tilde{\sigma}\chi_{iq}}-\frac{1}{N_c}\ket{\tilde{\sigma}}  \right) ,
\end{equation}
where $\tilde{\sigma}$ is defined as in \eqref{eq_sigtilde_1}. We point out that \eqref{eq_Tctc_action} and \eqref{eq_TialphaAction} are identical up to a prefactor of $\sqrt{T_F}=1/\sqrt{2}$.

Using \eqref{eq_Tctc_action}, we find that the action of $\bm{D}_{ij}^{F,\alpha\beta;\tilde{\alpha}\tilde{\beta}} $ on an arbitrary color-flow state reads
\begin{equation}\label{eq:cfDFoff}
  \begin{aligned}
    \sum_{m,n} F_{mn}\bm{D}_{mn}^{F,\alpha\beta;\tilde{\alpha}\tilde{\beta}} \ket{\sigma}\!\!\bra{\tau}&=\frac{1}{4}\sum_{i,j}^N\left( F_{ij}+F_{\sigma(i)\tau(j)}-F_{i\tau(j)} -F_{\sigma(i)j}\right)\\
    &\hspace{1.5cm}\left( \ket{\tilde{\sigma}\chi_{iq}}-\frac{1}{N_c}\ket{\tilde{\sigma}} \right)\left( \bra{\tilde{\tau}\chi_{jq}}-\frac{1}{N_c}\bra{\tilde{\tau}} \right) .
  \end{aligned}
\end{equation}
The leading-color contribution of the fermionic term $\bm{D}_{mn}^{F,\alpha\beta;\tilde{\alpha}\tilde{\beta}}$ is suppressed by one power of $\frac{1}{N_c}$ compared to the other leading-color contributions we have listed above. This is because the emitted $q\bar{q}$ pair only introduces one new color line, at variance with the gluon case, which introduces two. This means that we only gain one additional power of $N_c$ through the scalar product matrix \eqref{eq_SP_matrix} rather than two. However, $n_F T_F N_c$ is comparable in size to $N_c^2$ in QCD, thus it makes sense to count $n_F$ as order $N_c$. In this case, $\bm{D}_{ij}^{F,\alpha\beta;\tilde{\alpha}\tilde{\beta}} $ contributes at leading color as
\begin{equation}\label{eq:cfDFdiag}
    \sum_{m,n}\bm{D}_{mn}^{F,\alpha\beta;\tilde{\alpha}\tilde{\beta}}\ket{\sigma}\!\!\bra{\sigma} =\frac{1}{4}\sum_{i}^N\left( F_{ii}+F_{\sigma(i)\sigma(i)}-F_{i\sigma(i)} -F_{\sigma(i)i}\right) \ket{\tilde{\sigma}\chi_{iq}} \!\! \bra{\tilde{\sigma}\chi_{iq}} +\dots\, .
\end{equation}

\subsection{Tripole color structures}
\label{sec:TripoleCF}

The tripole color structure $\bm{T}_{ijk}$ appearing in purely virtual contributions to the two-loop anomalous dimensions acts on an arbitrary color-flow state as
\begin{equation}
  \label{eq_Tlmn_v_CF}
  \begin{aligned}    \sum_{l,m,n}F_{lmn}\bm{T}_{lmn}\bra{\tau} &=\frac{N_c}{4}\sum_{(ij)}^N F^{T_v}_{\tau;ij}\left( \bra{\tau\chi_{ij}}-\frac{1}{N_c}\bra{\tau} \right)\\
    &\quad+\frac{1}{4}\sum_{(ijk)}^N F^{T_v}_{\tau;ijk}\left( \bra{\tau\chi_{ikj}}-  \bra{\tau\chi_{ijk}}\right) ,
  \end{aligned}
\end{equation}
where 
\begin{equation}\label{eq:cfFTv}
  \begin{aligned}
    F^{T_v}_{\tau;ij}&=-F_{\tau (i) j i}+F_{\tau (i)
   \tau (j) i}+F_{\tau (i) i
   j}-F_{i \tau (i) j}+F_{i \tau(j) j}-F_{\tau (i) \tau (j)
   j}\\
   &\quad +F_{i j \tau(i)}-F_{i \tau(j) \tau (i)}-F_{\tau (i) i \tau(j)}
   -F_{i j \tau(j)}+F_{\tau (i) j \tau(j)}+F_{i \tau (i) \tau (j)}\\
     F^{T_v}_{\tau;ijk}&=-F_{\tau (i) j k}-F_{i \tau (j)
   k}+F_{\tau (i) \tau (j)
   k}-F_{i j \tau (k)}+F_{\tau(i) j \tau (k)}\\
   &\quad +F_{i \tau
   (j) \tau (k)}-F_{\tau (i)
   \tau (j) \tau (k)}+F_{i j k}\, ,
  \end{aligned}
\end{equation}
and we use the same notation as above. Additionally, $\tau\chi_{kml}$ refers to the permutation obtained by first acting with the cycle $(k\to m \to l\to k)$ and then with $\tau$, i.e., 
\begin{equation}\label{eq:cftaukml}
   \tau\chi_{kml}(i)=\begin{cases}
    \tau(i)\, , &\text{for }i \notin \{k,l,m\}\\
    \tau(l)\, ,& \text{for }i = k\\
     \tau(m)\, ,& \text{for }i = l\\
      \tau(k)\, ,&\text{for }i = m\, .
  \end{cases}
\end{equation}
The leading-color contribution from $\bm{T}_{ijk}$ is suppressed compared to the dipole contribution  $\bm{D}_{mn}$ because there is no contribution proportional to $N_c$ and $\bra{\tau}$ simultaneously.
The action of $\bm{T}_{ijk}$ is obtained along the lines of \eqref{eq_Db_action}.

The tripole operator appearing in the single-emission two-loop anomalous dimension $ \bm{T}_{lmn}^{\alpha;\tilde{\alpha}}$ is built by acting with the operator $\bT_{l,L}^\alpha$ from the left, and with the hermitian conjugate of $i f^{\alpha b c}\left( \bT_{m,L}^b \bT_{n,L}^c \right)_+\!$ from the right. The operator $i f^{\alpha b c}\left( \bT_{m,L}^b \bT_{n,L}^c \right)_+\!$ is to be understood as shorthand notation along the lines of \eqref{eq_Tialpha_as_an_operator}. Its action on a color-flow state $\ket{\sigma}$ is given by
\begin{equation}
  \label{eq_ifalphabcsymTmbTnc}
\begin{aligned}
  &\sum_{m,n} F_{mn}i f^{\alpha b c}\left( \bT_{m,L}^b \bT_{n,L}^c \right)_+\!\ket{\sigma}=\frac{N_c}{2\sqrt{2}}\sum_i^N \left( F_{i\sigma(i)}-F_{\sigma(i)i} \right)\left( \ket{\tilde{\sigma}\chi_{iq}}-\frac{1}{N_c}\ket{\tilde{\sigma}} \right)\\
  &\qquad +\frac{1}{2\sqrt{2}}\sum_{(ij)}^N \biggl(F_{ij}+ F_{\sigma(i)\sigma(j)}-F_{\sigma(i)j}-F_{i\sigma(j)} \biggr)\left( \ket{\tilde{\sigma}\chi_{ijq}} - \ket{\tilde{\sigma}\chi_{jiq}}\right) .
\end{aligned}
\end{equation}
We point out that the prefactor of $\ket{\tilde\sigma}$ on the right-hand side is independent of the color flow we start with, as was the case for the operator $\bT_{m,L}^\alpha$ in \eqref{eq_TialphaAction}.
Using the result \eqref{eq_ifalphabcsymTmbTnc} and \eqref{eq_TialphaAction}, we find that the action of $\bm{T}_{ijk}^{\alpha;\tilde{\alpha}}$ on an arbitrary color-flow state can be written as
\begin{equation}\label{eq:cfTroff}
  \begin{aligned}   \sum_{lmn}F_{lmn}\bm{T}_{lmn}^{\alpha;\tilde{\alpha}}\ket{\sigma}\!\!\bra{\tau}&=-\frac{N_c}{4}\sum_{i,j}^N\left(\ket{\tilde{\sigma}\chi_{iq}} -\frac{1}{N_c}\ket{\tilde{\sigma}} \right)\left( \bra{\tilde{\tau}\chi_{jq}}-\frac{1}{N_c}\bra{\tilde{\tau}} \right)F^{T_r}_{\sigma \tau;ij}\\
    &\quad+\frac{1}{4}\sum_{i}^N\sum_{\left( jk \right)}^N \left( \ket{\tilde{\sigma}\chi_{iq}} -\frac{1}{N_c}\ket{\tilde{\sigma}} \right)\left( \bra{\tilde{\tau}\chi_{jqk}}-\bra{\tilde{\tau}\chi_{jkq}} \right)F^{T_r}_{\sigma \tau;ijk}\, ,
  \end{aligned}
\end{equation}
where
\begin{equation}\label{eq:cfFTr}
  \begin{aligned}
    F^{T_r}_{\sigma \tau;ij}&=F_{\sigma (i) \tau (j)
   j}-F_{\sigma (i) j \tau
   (j)}-F_{i \tau (j) j}+F_{i j
   \tau (j)}\,,\\
F^{T_r}_{\sigma \tau;ijk}&=F_{\sigma (i) \tau (j)
   k}+F_{\sigma (i) j \tau
   (k)}-F_{\sigma (i) \tau (j)
   \tau (k)}-F_{\sigma (i) j
   k}-F_{i \tau (j) k}-F_{i j
   \tau (k)}+F_{i \tau (j)
   \tau (k)}+F_{i j k}\, ,
  \end{aligned}
\end{equation}
and $\bm{T}_{klm}^{\alpha;\tilde{\alpha}}$ should be understood as shorthand notation analogous to \eqref{eq_Dmnalphaalphatild_operator}. The action of $\bm{T}_{ijk}^{\alpha;\tilde{\alpha}}$ can be obtained along the lines of \eqref{eq_Db_action}.

The leading color contribution from $ \bm{T}_{ijk}^{\alpha;\tilde{\alpha}}$ is non-vanishing and given by
\begin{equation}\label{eq:cfTrdiag}
    \sum_{klm}F_{klm}\bm{T}_{klm}^{\alpha;\tilde{\alpha}}\ket{\sigma}\!\!\bra{\sigma}=-\frac{N_c}{4}\sum_{i}^N \ket{\tilde{\sigma}\chi_{iq}}   \bra{\tilde{\sigma}\chi_{iq}}F^{T_r}_{\sigma \sigma;ii} +\dots\, ,
\end{equation}
where the kinematic structure reads
\begin{equation}\label{eq:cfFTrdiag}
  F^{T_r}_{\sigma \sigma;ii}=-F_{i \sigma (i) i}+F_{\sigma
   (i) \sigma (i) i}+F_{i i
   \sigma (i)}-F_{\sigma (i) i \sigma (i)}\, .
\end{equation}

The double emission tripole operator $\bm{T}_{lmn}^{\alpha\beta;\tilde{\alpha}\tilde{\beta}}$ should be understood as shorthand notation 
analogously to \eqref{eq_double_emission_dipole_operator}. Following the definition in \eqref{eq:tripoles}, the operator $\bm{T}_{lmn}^{\alpha\beta;\tilde{\alpha}\tilde{\beta}}$ is built by acting with the operator $(T_{l,L}^\alpha T_{m,L}^\beta)_+$ from the left, and with $i f^{\tilde{\alpha}\tilde{\beta} c}\,T_{n,R}^c$, which multiplies the conjugate amplitude, from the right.
The action of $(\bm{T}_{l,L}^\alpha \bm{T}_{m,L}^\beta)_+$ on a color-flow state $\ket{\sigma}$ is given by
\begin{equation}
  \label{eq_TialphaTjbetaSymAction}
\begin{aligned}
    &\sum_{l,m}F_{lm}(\bm{T}_{l,L}^\alpha \bm{T}_{m,L}^\beta)_+\ket{\sigma}=\frac{1}{4}\sum_i^N \!\left( F_{\sigma(i)i}-F_{i\sigma(i)} \right)\left( \ket{\tilde{\sigma}\chi_{iqr}} - \ket{\tilde{\sigma}\chi_{irq}} \right)\\
    &+\frac{1}{4}\sum_i^N \!\left( F_{ii}+F_{\sigma(i)\sigma(i)}-F_{i\sigma(i)}-F_{\sigma(i)i} \right)\!\left(\! \ket{\tilde{\sigma}\chi_{iqr}} +\ket{\tilde{\sigma}\chi_{irq}}-\frac{2}{N_c}\left( \ket{\tilde{\sigma}\chi_{iq}}+\ket{\tilde{\sigma}\chi_{ir}} \right) +\frac{2}{N_c^2}\ket{\tilde{\sigma}} \!\right)\\
    &+\frac{1}{2}\sum_{(ij)}^N \!\left( F_{ij}+F_{\sigma(i)\sigma(j)}-F_{i\sigma(j)}-F_{\sigma(i)j} \right)\!\left( \!\ket{\tilde{\sigma}\chi_{iq}\chi_{jr}}-\frac{1}{N_c}\left( \ket{\tilde{\sigma}\chi_{iq}}+\ket{\tilde{\sigma}\chi_{jr}} \right) +\frac{1}{N_c^2}\ket{\tilde{\sigma}} \!\right) .
\end{aligned}
\end{equation}
Using \eqref{eq_ifalphabetacTmc} and \eqref{eq_TialphaTjbetaSymAction}, the action of $\bm{T}_{ijk}^{\alpha\beta;\tilde{\alpha}\tilde{\beta}}$ on an arbitrary color-flow state can be written as

\begin{equation}\label{eq:cfTdoff}
  \begin{aligned}
    \sum_{l,m,n}F_{lmn}\bm{T}_{lmn}^{\alpha\beta;\tilde{\alpha}\tilde{\beta}} &\ket{\sigma}\!\!\bra{\tau}=\frac{1}{8}\sum_{k}^N \Biggl(\\
    &\sum_{i}^N F^{T_d;o}_{\sigma \tau;ik}\left( \ket{\tilde{\sigma}\chi_{iqr}}+\ket{\tilde{\sigma}\chi_{irq}}-\frac{2}{N_c}\left( \ket{\tilde{\sigma}\chi_{iq}}+ \ket{\tilde{\sigma}\chi_{ir}} \right)+\frac{2}{N_c^2}\ket{\tilde{\sigma}} \right)\\
    &+\sum_{i}^N F^{T_d;e}_{\sigma \tau;ik}\left( \ket{\tilde{\sigma}\chi_{iqr}}-\ket{\tilde{\sigma}\chi_{irq}} \right)\\
     &+\sum_{(ij)}^N F^{T_d;o}_{\sigma \tau;ijk}\left( 2\ket{\tilde{\sigma}\chi_{iq}\chi_{jr}}-\frac{2}{N_c}\left( \ket{\tilde{\sigma}\chi_{iq}}+ \ket{\tilde{\sigma}\chi_{jr}} \right)+\frac{2}{N_c^2}\ket{\tilde{\sigma}} \right)\Biggr)\\
    &\times \left( \bra{\tilde{\tau}\chi_{kqr}}-\bra{\tilde{\tau}\chi_{krq}} \right) ,
  \end{aligned}
\end{equation}
where the permutation 
$ \tilde{\sigma}\chi_{iq}\chi_{jr}$ is the permutation where one first acts with the transposition $\chi_{jr}$, then with $\chi_{iq}$ and then with $\tilde{\sigma}$. The kinematic functions are given by
\begin{equation}\label{eq:cfFTd}
  \begin{aligned}
    F^{T_d;o}_{\sigma \tau;ik}&=F^{T_d;o}_{\sigma \tau;iik}\,,\\
    F^{T_d;o}_{\sigma \tau;ijk}&
=
F_{\sigma(i)\sigma(j)\tau(k)}
+ F_{\sigma(i)jk}
+ F_{i\sigma(j)k}
+ F_{ij\tau(k)}
- F_{\sigma(i)j\tau(k)}
- F_{i\sigma(j)\tau(k)}
- F_{\sigma(i)\sigma(j)k}
- F_{ijk} \,,\\
   F^{T_d;e}_{\sigma \tau;ik}&=F_{\sigma(i) i \tau (k)}-F_{i
   \sigma (i) \tau
   (k)}-F_{\sigma (i) i k}+F_{i
   \sigma (i) k}\, .
  \end{aligned}
\end{equation}
The leading-color action reads
\begin{equation}
  \label{eq_Tijkd_LC}
  \begin{aligned}
   \sum_{l,m,n}F_{lmn}\bm{T}_{lmn}^{\alpha\beta;\tilde{\alpha}\tilde{\beta}}\ket{\sigma}\!\!\bra{\sigma}&=
    \frac{1}{8}\sum_i^N F^{T_d;o}_{\sigma \sigma;ii}\left(\ket{\tilde{\sigma}\chi_{iqr}}\!\!\bra{\tilde{\sigma}\chi_{iqr}}-\ket{\tilde{\sigma}\chi_{irq}}\!\!\bra{\tilde{\sigma}\chi_{irq}} \right)\\
   & \quad +\frac{1}{8}\sum_i^N F^{T_d;e}_{\sigma \sigma;ii}\left(\ket{\tilde{\sigma}\chi_{irq}}\!\!\bra{\tilde{\sigma}\chi_{irq}}+\ket{\tilde{\sigma}\chi_{iqr}}\!\!\bra{\tilde{\sigma}\chi_{iqr}} \right)+\dots\,.
  \end{aligned}
\end{equation}
The physical intuition for this result is that we can pick any color dipole $[i \sigma(i)]$ and insert two gluons. There are two ways to do this, we can either end up with the dipoles 
$[iqr\sigma(i)]$, which is encapsulated in $\ket{\tilde{\sigma}\chi_{iqr}}\!\!\bra{\tilde{\sigma}\chi_{iqr}} $, or we can get the same structure with the two gluons interchanged. Furthermore, if we inserted the color structure $\bm{T}_{lmn}^{\alpha\beta;\tilde{\alpha}\tilde{\beta}}$ at the last step in the shower evolution, the odd contribution in \eqref{eq_Tijkd_LC} would immediately contract to zero. However, if $\bm{T}_{lmn}^{\alpha\beta;\tilde{\alpha}\tilde{\beta}}$ is inserted at an intermediate shower time, the odd contribution can be undone again by the shower evolution, leading to a non-vanishing result.

\subsection{Quadrupole color structure}
The last color structure we analyze is the quadrupole color structure $\bm{Q}_{lmno}^{\alpha\beta;\tilde{\alpha}\tilde{\beta}}$ that shows up in the double-emission anomalous dimension for clustering cross sections. We again have to understand $\bm{Q}_{lmno}^{\alpha\beta;\tilde{\alpha}\tilde{\beta}}$ as shorthand notation for an operator along the lines of \eqref{eq_double_emission_dipole_operator}. $\bm{Q}_{lmno}^{\alpha\beta;\tilde{\alpha}\tilde{\beta}}$ is built by acting with the operator $(T_{l,L}^\alpha T_{m,L}^\beta)_+$ from the left, and with its hermitian conjugate from the right.
Using equation \eqref{eq_TialphaTjbetaSymAction}, the action on a generic color-flow state is then given by
\begin{equation}\label{eq:cfQoff}
 \begin{aligned}
   &\sum_{l,m,n,o} F_{lmno}\bm{Q}_{lmno}^{\alpha\beta;\tilde{\alpha}\tilde{\beta}}  \ket{\sigma}\!\!\bra{\tau}=\\
   &\frac{1}{16}\sum_{i,j,k,l}^N\left[ F_{\sigma \tau ;i j k l}{}^{e e}\ket{\sigma_e^{ij}}\!\!\bra{\tau^{kl}_e}+F_{\sigma \tau ;i j k l}{}^{e o}\ket{\sigma_e^{ij}}\!\!\bra{\tau_o^{kl}}+F_{\sigma \tau ;i j k l}{}^{o e}\ket{\sigma_o^{ij}}\!\!\bra{\tau^{kl}_e} +F_{\sigma \tau ;i j k l}{}^{o o}\ket{\sigma_o^{ij}}\!\!\bra{\tau_o^{kl}}\right] ,
 \end{aligned}
\end{equation}
where we defined the states
\begin{equation}\label{eq:cfsigmae}
  \begin{aligned}
    \ket{\sigma_e^{ij}}&=\ket{\tilde{\sigma}\chi_{iq}\chi_{jr}}+\ket{\tilde{\sigma}\chi_{jr}\chi_{iq}}-\frac{2}{N_c}\left(\ket{\tilde{\sigma}\chi_{iq}}+\ket{\tilde{\sigma}\chi_{jr}}   \right)+\frac{2}{N_c^2}\ket{\tilde{\sigma}} \,,\\    
    \ket{\sigma_o^{ij}}&=\ket{\tilde{\sigma}\chi_{iq}\chi_{jr}}-\ket{\tilde{\sigma}\chi_{jr}\chi_{iq}}=\delta_{ij}\left(\ket{\tilde{\sigma}\chi_{irq}} -\ket{\tilde{\sigma}\chi_{iqr}} \right) ,
  \end{aligned}
\end{equation}
 where we introduced the kinematic functions
\begin{equation}\label{eq:cfFQee}
  \begin{aligned}
    &F_{\sigma \tau ;i j k l}{}^{e e}=F_{\sigma (i) j \tau (k) l}+F_{i \sigma (j) \tau (k) l}-F_{\sigma (i) \sigma (j) \tau (k) l}+F_{\sigma (i) j k \tau (l)}+F_{i \sigma (j) k \tau (l)}-F_{\sigma (i) \sigma (j) k \tau (l)}\\
    &\quad-F_{\sigma (i) j \tau (k) \tau (l)}-F_{i \sigma (j) \tau (k) \tau (l)}+F_{\sigma (i) \sigma (j) \tau (k) \tau (l)}-F_{\sigma (i) j k l}-F_{i \sigma (j) k l}+F_{\sigma (i) \sigma (j) k l}-F_{i j \tau (k) l}\\
    &\quad-F_{i j k \tau (l)}+F_{i j \tau (k) \tau (l)}+F_{i j k l} \,,\\
     &F_{\sigma \tau ;i j k l}{}^{e o}=F_{\sigma (i) j \tau (k) l}+F_{i \sigma (j) \tau (k) l}-F_{\sigma (i) \sigma (j) \tau (k) l}-F_{\sigma (i) j k \tau (l)}-F_{i \sigma (j) k \tau (l)}+F_{\sigma (i) \sigma (j) k \tau (l)}\\
     &\quad-F_{i j \tau (k) l}+F_{i j k \tau (l)} \,,\\
     &F_{\sigma \tau ;i j k l}{}^{o e}=F_{\sigma (i) j \tau (k) l}-F_{i \sigma (j) \tau (k) l}+F_{\sigma (i) j k \tau (l)}-F_{i \sigma (j) k \tau (l)}-F_{\sigma (i) j \tau (k) \tau (l)}+F_{i \sigma (j) \tau (k) \tau (l)}\\&
     \quad-F_{\sigma (i) j k l}+F_{i \sigma (j) k l} \,,\\
     &F_{\sigma \tau ;i j k l}{}^{o o}=F_{\sigma (i) j \tau (k) l}-F_{i \sigma (j) \tau (k) l}-F_{\sigma (i) j k \tau (l)}+F_{i \sigma (j) k \tau (l)}\, .
  \end{aligned}
\end{equation}
The leading-color contribution to the quadrupole term reads
\begin{equation}
 \label{eq_Q_LC}
 \begin{aligned}
   &\sum_{l,m,n,o} F_{lmno}\bm{Q}_{lmno}^{\alpha\beta;\tilde{\alpha}\tilde{\beta}} \!\! \ket{\sigma}\bra{\sigma}=\\
   &\quad\frac{1}{16}\sum_{i}^N\ket{\tilde{\sigma}\chi_{iqr}}\!\!\bra{\tilde{\sigma}\chi_{iqr}}\left( -F_{\sigma \sigma ;i i i i}{}^{o e}-F_{\sigma \sigma ;i i i i}{}^{e o}+F_{\sigma \sigma ;i i i i}{}^{e e}+F_{\sigma \sigma ;i i i i}{}^{o o} \right)\\
   &+\frac{1}{16}\sum_{i}^N\ket{\tilde{\sigma}\chi_{irq}}\!\!\bra{\tilde{\sigma}\chi_{irq}}\left( F_{\sigma \sigma ;i i i i}{}^{o e}+F_{\sigma \sigma ;i i i i}{}^{e o}+F_{\sigma \sigma ;i i i i}{}^{e e}+F_{\sigma \sigma ;i i i i}{}^{o o} \right)\\
   &+\frac{1}{4}\sum_{(ij)}^N\ket{\tilde{\sigma}\chi_{iq}\chi_{jr}}\!\!\bra{\tilde{\sigma}\chi_{iq}\chi_{jr}}F_{\sigma \sigma ;i j i j}{}^{e e} +\dots\, .
 \end{aligned}
\end{equation}
The first two lines of the right-hand side of \eqref{eq_Q_LC} represent the two ways to insert a color-connected gluon pair into the dipole $[i \sigma(i)]$, and the third line represents the independent insertion of one gluon into the $[i \sigma(i)]$ dipole, and the other into the $[j \sigma(j)]$ dipole.

This concludes our analysis of the color structures appearing in the two-loop anomalous dimensions for fixed-cone and clustering cross sections. Some of our results have been stated in \cite{Platzer:2020lbr} in a different notation. We have checked that our two-loop ingredients \eqref{eq_Tlmn_v_CF} and \eqref{eq_ifalphabcsymTmbTnc} are consistent with equations (A.7) and (A.2) in \cite{Platzer:2020lbr}. To our knowledge, we are the first to present the action of the double-emission color operators in the color-flow basis.  

\section{Lorentz transformation properties of the building blocks}
\label{sec:lorentz}
Throughout the paper the light-like vectors $n_a$ are normalized to unit energy in the lab frame, $n_a=(1,\vec n_a)$, i.e.\ $v\cdot n_a=1$ for $v=(1,\vec 0)$, and all results are expressed through the scalar products $n_{ab}=n_a\cdot n_b$ and $n_{va}=v\cdot n_a$, the angular measure $[d\Omega_q]$ defined in \eqref{eq:dOmega}, and, in the clustering case, the energy fraction $\xi$ of \eqref{eq_xi_definition}. In this appendix we collect the behavior of these building blocks under Lorentz transformations, which allows one to read off the frame (in)dependence of the results in the main text.

\paragraph{Rescaling of light-like vectors.} A Lorentz transformation maps a lab-normalized vector $n_a$ to a light-like vector $\Lambda n_a$ whose energy is no longer one; the correctly normalized direction in the new frame is $n_a'=\Lambda n_a/(v\cdot\Lambda n_a)$. Introducing $\tilde v=\Lambda^{-1}v$ and using the invariance of the scalar product, $v\cdot\Lambda n_a=n_{\tilde v a}$, this becomes $n_a'=\Lambda\,(n_a/n_{\tilde v a})$: the rescaled lab-frame vector $n_a/n_{\tilde v a}$ is carried by $\Lambda$ exactly onto the correctly normalized direction of the new frame. Since all our results are built from scalar products, the overall transformation $\Lambda$ drops out, and evaluating an expression in a different frame amounts to the rescaling
\begin{equation}\label{eq:rescaling}
 n_a\;\to\;\lambda_a\,n_a\,,\qquad \lambda_a=\frac{1}{n_{\tilde v a}}\,,\qquad n_{ab}\to\lambda_a\lambda_b\,n_{ab}\,,
\end{equation}
of all light-like vectors. In this language the lab-frame reference vector carries degree zero: $v$ is a genuine four-vector, normalized as $v^2=1$, rather than a direction with a conventional normalization. An expression built from scalar products is therefore frame independent if and only if it is homogeneous of degree zero in each light-like vector and free of $v$. An expression of degree zero that does involve $v$ is a Lorentz scalar, but depends on the choice of the lab frame, i.e., one needs to declare in which frame $v$ takes the form $v=(1,0,\dots)$; this is what is meant by the frame dependence of the \XS scheme results in Sections~\ref{sec:soft_function} and~\ref{sec:XSanomdim}. The degrees of homogeneity of the basic objects are
\begin{equation}\label{eq:degrees}
\begin{aligned}
 n_{ab}&: \;(1,1)\ \text{in}\ (n_a,n_b)\,, & n_{va}&:\;1\ \text{in}\ n_a\,,\ 0\ \text{in}\ v\,,\\
 W^q_{ij}=\frac{n_{ij}}{n_{iq}n_{jq}}&:\;-2\ \text{in}\ n_q\,,\ 0\ \text{in}\ n_i,n_j\,, &
 W^{qr}_{ij}=\frac{n_{ij}}{n_{iq}n_{qr}n_{jr}}&:\;-2\ \text{in}\ n_q\ \text{and}\ n_r\,,\ 0\ \text{in}\ n_i,n_j\,,
\end{aligned}
\end{equation}
and consequently every product of two dipole functions $W^q_{ab}W^r_{cd}$, and with it all two-emission kinematic functions $k_{ijk;qr}$, $l_{ijk;qr}$, $K^{(a,b,c)}_{ij;qr}$ , has degree $-2$ in $n_q$ and in $n_r$ and degree zero in the hard directions.

\paragraph{Angular measures.} The invariant phase space of a massless emission, $d^{d-1}q/(2E_q)=E_q^{d-3}dE_q\,d^{d-2}\Omega_q/2$, is frame independent, while the solid-angle element of lab-normalized directions has degree $d-2=2-2\epsilon$ in $n_q$. The factor $(n_{vq})^{2\epsilon}$ in the definition \eqref{eq:dOmega} of $[d\Omega_q]$ precisely compensates the $\epsilon$ dependence, so that
\begin{equation}\label{eq:measuredegree}
 [d\Omega_q]\;:\;\text{degree}\ +2\ \text{in}\ n_q\quad\text{for all } d\,,
\end{equation}
and the combinations $[d\Omega_q]\,W^q_{ij}$ and $[d\Omega_q][d\Omega_r]\,W^q_{ab}W^r_{cd}$ are Lorentz invariant in $d$ dimensions. This is what allows us to evaluate the angular integrals in any frame, e.g.\ in the dipole rest frame in Section~\ref{sec:soft_function}. By the same token, a term in an anomalous dimension multiplying $[d\Omega_q][d\Omega_r]$ is Lorentz invariant if and only if it has degree $-2$ in $n_q$ and $n_r$ and degree zero in all other vectors, and contains $v$ at most through the invariant combinations listed below. It is furthermore frame independent if it also does not depend on $v$.

\paragraph{Energies and the energy fraction.} In our convention for the normalized direction vectors, the four-vector $q=E_q n_q$ in the lab frame becomes $q'=\Lambda q =E_q' n_q'$ in a second frame. Thus, energies transform as
\begin{equation}\label{eq:energytrafo}
  E_q' = E_q\, n_{\tilde v q}\equiv E_q\, \tilde{v}\cdot n_q\,.
\end{equation}
For two emissions with momenta $q=E_qn_q$ and $r=E_rn_r$, the energy fractions $\xi'=E_q'/(E_q'+E_r')$ and $\barxi'=1-\xi'$ (and measures built from them) can be expressed in terms of the lab-frame $\xi=E_q/(E_q+E_r)$ and $\barxi=1-\xi$ as 
\begin{equation}\label{eq:xitrafo}
 \xi'=\frac{\xi\,n_{\tilde v q}}{\xi\,n_{\tilde v q}+\barxi\,n_{\tilde v r}}\,,\qquad \frac{d\xi'}{\xi'\barxi'}=\frac{d\xi}{\xi\barxi}\,,\qquad \delta(\xi')\,d\xi'=\delta(\xi)\,d\xi\,,\quad \delta(\barxi')\,d\xi'=\delta(\barxi)\,d\xi\,.
\end{equation}
The endpoints $\xi=0,1$ are frame independent, as is the measure $d\xi/(\xi\barxi)$, but the plus distributions of Section~\ref{sec:clustering}, which regulate the endpoints by subtracting the values at $\xi=0$ and $\xi=1$, are not: using \eqref{eq:xitrafo} one finds
\begin{equation}\label{eq:plustrafo}
\begin{aligned}
 \Big(\frac{1}{\xi'\barxi'}\Big)_+d\xi'&=\left[\Big(\frac{1}{\xi\barxi}\Big)_+ +\ln\!\Big(\frac{n_{\tilde v q}}{n_{\tilde v r}}\Big)\big(\delta(\xi)-\delta(\barxi)\big)\right]d\xi\,,\\
 \Big(\frac{1}{\xi'}\Big)_+d\xi'&=\left[\Big(\frac{1}{\xi}\Big)_+ +\ln\!\Big(\frac{n_{\tilde v q}}{n_{\tilde v r}}\Big)\delta(\xi)\right]d\xi\,,\qquad
 \Big(\frac{1}{\barxi'}\Big)_+d\xi'=\left[\Big(\frac{1}{\barxi}\Big)_+ -\ln\!\Big(\frac{n_{\tilde v q}}{n_{\tilde v r}}\Big)\delta(\barxi)\right]d\xi\,,
\end{aligned}
\end{equation}
which are the replacement rules \eqref{eq:framerules}. In contrast, the kernels that are singular at one endpoint only and involve the direction of a hard leg,
\begin{equation}\label{eq:kernelinv}
 \frac{n_{kr}\,d\xi}{\xi\,(\xi n_{kq}+\barxi n_{kr})}=\frac{n_k\cdot r}{n_k\cdot(q+r)}\,\frac{d\xi}{\xi\barxi}\,,
\end{equation}
are invariant together with their plus-distribution versions, since both factors on the right-hand side are, and the endpoint subtraction of the plus distribution uses the same kernel.

\paragraph{Invariant building blocks.} Summarizing, the following combinations are Lorentz invariant and free of the reference vector $v$:
\begin{itemize}
\item cross ratios $\dfrac{n_{ab}\,n_{cd}}{n_{ac}\,n_{bd}}$, such as the arguments of all logarithms in \eqref{eq:KbarFuncs}, \eqref{eq:Kbarnew}, \eqref{eq:MijqrLS} and \eqref{eq:KbarijklClust}--\eqref{eq:KbarAClust};
\item ratios of dipole functions with the same emission, $W^q_{ab}/W^q_{cd}$, and products of dipole functions with the angular measures, $[d\Omega_q]\,W^q_{ab}$, $[d\Omega_q][d\Omega_r]\,W^q_{ab}W^r_{cd}$, and $[d\Omega_q][d\Omega_r]$ times any of $k_{ijk;qr}$, $l_{ijk;qr}$, $K^{(a,b,c)}_{ij;qr}$, $W^{qr}_{ij}$;
\item the distributions $\delta(\xi)\,d\xi$, $\delta(\barxi)\,d\xi$, $d\xi/(\xi\barxi)$, $\big(n_{kr}/(\xi(\xi n_{kq}+\barxi n_{kr}))\big)_+d\xi$, as well as the combinations $\Big[\big(\frac{1}{\xi\barxi}\big)_+ +\frac{1}{2}\big(\delta(\xi)-\delta(\barxi)\big)\ln X\Big]\,d\xi$ for any ratio $X$ of scalar products of degree $+2$ in $n_q$, $-2$ in $n_r$ and $0$ in all other vectors: under \eqref{eq:rescaling} the shift $\ln(\lambda_r/\lambda_q)\big(\delta(\xi)-\delta(\barxi)\big)$ of the plus distribution cancels against the shift of $\frac{1}{2}\ln X$.
\end{itemize}
The following combinations are Lorentz invariant but frame dependent as they depend on the choice of $v$:
\begin{itemize}
\item $\dfrac{n_{va}\,n_{vb}}{n_{ab}}$ and in particular $\hat W^q_{ij}=n_{vq}^2W^q_{ij}$ of \eqref{eq:dipoledefinition}, $n_{vq}n_{vr}W^{qr}_{ij}$, the arguments $2n_{vq}^2W^q_{ij}$ of the logarithms in \eqref{eq:Gamma2XStripole}, \eqref{eq:Gamma2XSdipole} and \eqref{eq:modIm}, and the function $S_{ijqv}$;
\item $\ln\dfrac{n_{kq}\,n_{vr}}{n_{kr}\,n_{vq}}$ of \eqref{eq:Gamma2XStripole}, the frame factor $\dfrac{n_{vq}n_{vr}}{(\xi n_{vq}+\barxi n_{vr})^2}$ of \eqref{eq:framefactor};
\item the invariant measure $[d\Omega_q]/n_{vq}^2$ and the energies $v\cdot q$.
\end{itemize}
The \XS results \eqref{eq:Gamma2XStripole}, \eqref{eq:Gamma2XSdipole}, \eqref{eq:dKclustDefs} and \eqref{eq:framerules} are of the second type, while the \LS results \eqref{eq:KbarFuncs}, \eqref{eq:Kbarnew} and \eqref{eq:KbarijklClust}--\eqref{eq:KbarAClust} are built exclusively from the first type of building blocks, which is the precise meaning of the statement that the \LS scheme is frame independent.

\end{appendix}

\bibliographystyle{JHEP}
\bibliography{references}

\end{document}